\documentclass[aps,superscriptaddress,floatfix,nofootinbib,preprintnumbers,eqsecnum,twocolumn]{revtex4-1}
\usepackage[utf8]{inputenc}
\usepackage{lipsum, graphicx}
\usepackage{amssymb,amsmath,slashed,multirow,mathtools}
\usepackage{comment,color,placeins,amsmath,amsfonts,enumitem}
\usepackage{cancel}
\usepackage[utf8]{inputenc}

\usepackage{subfigure}

\usepackage{hyperref}
\usepackage[usenames,dvipsnames]{xcolor}
\hypersetup
{
  colorlinks,
  linkcolor={blue!80!black},
  citecolor={blue!70!black},
  urlcolor={blue!70!black}
}

\makeatletter
\renewcommand*\env@matrix[1][\arraystretch]{%
  \edef\arraystretch{#1}%
  \hskip -\arraycolsep
  \let\@ifnextchar\new@ifnextchar
  \array{*\c@MaxMatrixCols c}}
\makeatother

\def\beq{\begin{equation}}
\def\eeq{\end{equation}}

\def\bea{\begin{eqnarray}}
\def\eea{\end{eqnarray}}

\begin{document}
\title{Lighting Electroweak-Violating ALP-Lepton Interactions at $e^{+}e^{-}$ and $ep$ Colliders}
\author{Chih-Ting Lu}
\email{ctlu@njnu.edu.cn}
\affiliation{\mbox{Department of Physics and Institute of Theoretical Physics, Nanjing Normal University, Nanjing, 210023, China}}

\begin{abstract} 

Recently, Altmannshofer, Dror and Gori (2022) claimed there is a four-point interaction, $W$-$\ell$-$\nu$-$a$, in the electroweak-violating scenario of axion-like particle (ALP) and lepton interactions which plays a critical role in searching for ALPs from $\pi^{\pm}$, $K^{\pm}$ mesons and $W$ boson decays because of the novel energy enhancements. Inspired by this interesting finding, we first propose new t-channel processes, $e^{+}e^{-}\rightarrow\nu_e a\overline{\nu_e}$ and $e^{-}p\rightarrow\nu_e a j$, for electrophilic ALPs ($e$ALPs) which also involve $W$-$\ell$-$\nu$-$a$ four-point interaction and have obvious energy enhancement behaviors in their cross sections when the collision energy is increasing. 
On the other hand, heavier $e$ALPs mainly decay to a photon pair induced by the chiral anomaly instead of an electron-positron pair. Therefore, studies of these t-channel processes with a photon pair plus missing energy at $e^+ e^-$ and $ep$ colliders open a new door to search for $e$ALPs at high energy colliders. The proposed search strategies are not only aiming to generate a larger production rate of $e$ALPs, but also trying to distinguish electroweak-violating ALP-lepton interactions from electroweak-preserving ones. 

\end{abstract}

\maketitle

\FloatBarrier

\section{Introduction}
\label{sec:Introduction}

Axion-Like-Particles (ALPs) are the generic extension of the Standard Model (SM) with the type of pseudoscalars behaving like Nambu–Goldstone bosons. Compared with the traditional QCD axion which aims to explain the strong CP problem~\cite{Peccei:1977hh,Weinberg:1977ma,Wilczek:1977pj} and can serve as the dark matter (DM) candidate~\cite{Preskill:1982cy,Abbott:1982af,Dine:1982ah}, motivations of new physics models with ALPs are diverse. For example, ALPs may come from different spontaneous symmetry breaking patterns of global symmetries~\cite{Preskill:1982cy,Abbott:1982af,Dine:1982ah,Kim:1979if,Bagger:1994hh} or compactifications of very high energy string theories~\cite{Svrcek:2006yi,Arvanitaki:2009fg,Cicoli:2012sz}. ALPs may also affect the structure of the electroweak phase transition~\cite{Jeong:2018jqe,Im:2021xoy} and explain the hierarchy problem of the Higgs boson mass~\cite{Graham:2015cka}. ALPs can also be wave-like DM candidates like QCD axions~\cite{Arias:2012az}. In addition, QCD axions are restricted to a very low and narrow mass range, $10^{-12}$ eV $\lesssim m_a\lesssim 10^{-2}$ eV. However, the mass of ALPs can extend from almost massless to the electroweak scale or even higher. Therefore, we have to set up various search strategies to hunt ALPs in different mass regions.

In general, we can apply the effective field theory (EFT) approach to ALPs~\cite{Brivio:2017ije,Bauer:2017ris,Bauer:2018uxu,Ebadi:2019gij,Bauer:2020jbp,Bauer:2021mvw}. In such a framework, different coupling types between ALPs and SM particles are possible and each of them can be independent of the others. Hence, we can explore each coupling type at a time. Although searching for ALPs is a model-dependent issue, the mass of ALPs below about $10$ MeV scale has already suffered from severe constraints from cosmological and astrophysical observations~\cite{Raffelt:1990yz,Marsh:2015xka} as well as low energy beam dump experiments~\cite{Bjorken:1988as,Dobrich:2015jyk,Dobrich:2019dxc}. For ALPs heavier than the MeV scale, bounds from various meson decay channels~\cite{Izaguirre:2016dfi,Benson:2018vya,Gori:2020xvq,Altmannshofer:2022izm} and low energy $e^+ e^-$ colliders such as BaBar~\cite{BaBar:2011kau} and Belle II~\cite{Belle-II:2020jti} play important roles. Finally, we resort to high energy colliders, like LEP~\cite{OPAL:2002vhf,Mimasu:2014nea,Jaeckel:2015jla} and LHC~\cite{ATLAS:2014jdv,ATLAS:2015rsn,Knapen:2016moh} to explore ${\cal O}(1-100)$ GeV ALPs. Hunting for ALPs is not yet a monotonous task, and people are still chasing different new proposals.

Recently, the authors in Ref.~\cite{Altmannshofer:2022izm} proposed a new way to explore the ALP and lepton interactions. 
They emphasized that the less attention given to the $W$-$\ell$-$\nu$-$a$ four-point interaction for leptophilic ALPs ($\ell$ALPs)\footnote{The previous studies of $\ell$ALPs can be found in~\cite{Kirpichnikov:2020lws,Han:2020dwo,Chang:2021myh,Cheung:2021mol,Bertuzzo:2022fcm,Cheung:2022umw,Lucente:2022esm}.} can result in novel energy enhancement effects in the processes of $\ell$ALPs from charged mesons and $W$ boson decays in the electroweak-violating scenario. 
Indeed, they have shown this type of interaction can explore new parameter space and cannot be overlooked. 
Inspired by this work, we first propose how to explore $W$-$\ell$-$\nu$-$a$ four-point interaction for heavier electrophilic ALPs ($e$ALPs) and concretely display the energy enhancement behaviors at $e^+ e^-$ and $ep$ colliders. We point out the t-channel $e^{+}e^{-}\rightarrow\nu_e a\overline{\nu_e}$ and $e^{-}p\rightarrow\nu_e a j$ processes have obvious energy enhancements from momentum transferring of the initial state $e^{\pm}$ to the final states ALP and $\nu (\overline{\nu})$. This unique feature not only increases the $e$ALP production rates, but opens a door to solely explore $W$-$\ell$-$\nu$-$a$ four-point interaction such that we can distinguish electroweak-violating ALP-lepton interactions from electroweak-preserving ones.

On the other hand, the dominant decay mode of the heavier $e$ALP is a pair of photons, mainly from the chiral anomaly instead of a pair of electron-positron. As a result, the signal final states are a photon pair plus missing energy (a photon pair plus missing energy and a backward jet) at $e^+ e^-$ ($ep$) colliders.  
Most interestingly, when the produced $e$ALP is highly boosted, two photons are too collimated to pass the photon isolation criterion. In this situation, we group these two collimated, non-isolated photons inside a novel jet-like object called "photon-jet"~\cite{Dobrescu:2000jt,Toro:2012sv,Draper:2012xt,ATLAS:2012soa,Ellis:2012sd,Ellis:2012zp,Knapen:2015dap,Agrawal:2015dbf,Chang:2015sdy,Aparicio:2016iwr,Dasgupta:2016wxw,Domingo:2016unq,Chiang:2016eav,Dillon:2016tqp,Allanach:2017qbs,Chakraborty:2017mbz,ATLAS:2018dfo,Sheff:2020jyw,Wang:2021uyb,Ren:2021prq,Ai:2022qvs,CMS:2022fyt}. The photon-jet signature is clean and unique at both $e^+ e^-$ and $ep$ colliders such that it can help us to extract the signals from backgrounds. We find that searching for these two processes at $e^+ e^-$ and $ep$ colliders with signatures of both two isolated photons and a photon-jet for different mass ranges of $e$ALP can provide much stronger future bounds than the existing ones. Furthermore, these possible future bounds are hard to exceed by other channels when exploring $\ell$ALPs in collider experiments. Therefore, this could be another motivation to build $e^+ e^-$ and/or $ep$ colliders in the future.

We organize this paper as follows. In Sec.\,\ref{sec:rev}, a brief review on ALP-lepton interactions is provided and decay modes of the $\ell$ALP are discussed. In Sec.\,\ref{sec:kin}, we analytically and numerically display energy enhancement behaviors in $e^{+}e^{-}\rightarrow\nu_e a\overline{\nu_e}$ and $e^{-}p\rightarrow\nu_e a j$. In Sec.\,\ref{sec:filter}, we demonstrate signal-to-background analysis at $e^+ e^-$ and $ep$ colliders and predict possible future bounds as well as compare them with existing constraints. Finally, we summarize our findings along with some further discussions in Sec.\,\ref{sec:con}. Some kinematic distributions for both signals and SM backgrounds as well as supplemental materials are collected in Appendix~\ref{app:rec}.

\section{A Brief Review on ALP-lepton interactions}
\label{sec:rev}

Taking ALPs as pseudo Nambu-Goldstone bosons arising from the global Peccei-Quinn (PQ) symmetry~\cite{Peccei:1977hh}, $U(1)_{\text{PQ}}$, breaking, ALPs receive a shift symmetry, $a(x)\rightarrow a(x)+\text{const}$. Because of this shift symmetry, the ALP interactions with SM fermion pairs are characterized by the derivatively-coupled type. Here we focus on the $\ell$ALP and the Lagrangian of its interactions is in the form  
${\cal L}_{\ell\text{ALP}} = \partial_{\mu}a ~J^{\mu}_{ \text{PQ},\ell}$.  
The general lepton current can be written as~\cite{Altmannshofer:2022izm,Bertuzzo:2022fcm} 
\begin{equation} 
\label{eq:Jint}
J^{\mu}_{\text{PQ},\ell} = \frac{c^V_{\ell}}{2\Lambda}\overline{\ell}\gamma^{\mu}\ell + \frac{c^A_{\ell}}{2\Lambda}\overline{\ell}\gamma^{\mu}\gamma_5\ell + \frac{c_{\nu}}{2\Lambda}\overline{\nu_{\ell}}\gamma^{\mu} P_L \nu_{\ell} \,, 
\end{equation} 
where $\Lambda$ is the relevant new physics scale breaking $U(1)_{\text{PQ}}$ symmetry and $c^V_{\ell}$, $c^A_{\ell}$, $c_{\nu}$ are dimensionless couplings. Note that the electroweak invariance isn't imposed in Eq.~(\ref{eq:Jint}), and the condition $c_{\nu} = c^V_{\ell} -c^A_{\ell}$ doesn't need to be held. As pointed out in Ref.~\cite{Altmannshofer:2022izm}, the first and the third terms in the general lepton current can be generated independently from two electroweak invariant high-order operators $\left(\overline{HL}\right)\gamma^{\mu}\left(HL\right)$ and $\left(\overline{H^{\dagger}L}\right)\gamma^{\mu}\left(H^{\dagger}L\right)$. Here $L$ and $H$ are SM lepton and complex scalar $SU(2)$ doublets, respectively. After integration by parts of $\partial_{\mu}a ~J^{\mu}_{\text{PQ},\ell}$, the ${\cal L}_{\ell\text{ALP}}$ can be represented as~\cite{Altmannshofer:2022izm}  
\begin{align} 
& a ~\partial_{\mu}J^{\mu}_{\text{PQ},\ell} = i c^A_{\ell}\frac{m_{\ell}}{\Lambda}~a\overline{\ell}\gamma_5\ell \label{eq:int} \\
& + \frac{\alpha_{\text{em}}}{4\pi\Lambda} \bigg[  \frac{ c^V_{\ell} -c^A_{\ell} + c_{\nu}}{4 s^2 _W}~a W^{+}_{\mu\nu}\tilde W ^{-,\mu\nu} \notag \\ 
& + \frac{c^V_{\ell} - c^A_{\ell} (1 -4 s^2_W)}{2s _W c_W}~a F_{\mu\nu}\tilde{Z} ^{\mu\nu} - c^A_{\ell}~a F_{\mu\nu} \tilde{F}^{\mu\nu} + \notag \\ 
& \frac{c^V_{\ell} (1 -4 s^2_W) -c^A_{\ell} (1 -4 s^2_W +8 s^4_W)  + c_{\nu}}{8 s^2_W c^2_W}~a Z_{\mu\nu}\tilde{Z}^{\mu\nu}\bigg]  \notag \\ 
& + \frac{ig_W}{2\sqrt{2}\Lambda}(c^A_{\ell} - c^V_{\ell} + c_{\nu})~a (\bar\ell \gamma^{\mu} P _L \nu) W_{\mu}^{-} ~+~\text{h.c.}  \,, \notag 
\end{align} 
where $\alpha_{\text{EM}}$ is the fine structure constant, $g_W$ is the weak coupling constant, $s_W$ ($c_W$) is the sine (cosine) of the weak mixing angle.

The term in the first line of Eq.~(\ref{eq:int}) has been intensively studied for light $\ell$ALPs~\cite{Bjorken:1988as,Bauer:2021mvw,Kirpichnikov:2020lws,Chang:2021myh,Bertuzzo:2022fcm,Cheung:2022umw} and we label it as "$\boldsymbol{a\ell\ell}$". The $\ell$ALP can be produced from this interaction via the radiation of charged leptons. However, this term has a $m_{\ell}/\Lambda$ suppression, so we have to resort to high intensity experiments, like SLAC E137~\cite{Bjorken:1988as} and Jefferson Lab BDX~\cite{BDX:2016akw} to search for $\ell$ALPs. Terms on the second to the fourth lines of Eq.~(\ref{eq:int}) arise from the chiral anomaly and we label them as "$\boldsymbol{aVV'}$". These terms are not proportional to $m_{\ell}$, but they are suppressed by the factor $\alpha_{\text{em}}/4\pi$. The $\ell$ALP can be produced from its connections with gauge bosons via flavor-changing meson decays~\cite{Izaguirre:2016dfi,Gori:2020xvq,Bauer:2021mvw} as well as direct gauge boson fusion and associated gauge boson production processes~\cite{Mimasu:2014nea,Jaeckel:2015jla,Knapen:2016moh,Liu:2021lan}. The terms in the final line of Eq.~(\ref{eq:int}) were first shown in Ref.~\cite{Raffelt:1987yt} for stellar bounds on ALPs and recently have been studied for $\ell$ALP productions from charged mesons and $W$ boson decays in details~\cite{Altmannshofer:2022izm}. We label them as "$\boldsymbol{aW\ell\nu}$". 
This interaction vanishes when the general lepton current in Eq.~(\ref{eq:Jint}) preserves the electroweak symmetry. This four-point interaction $W$-$\ell$-$\nu$-$a$ is also not proportional to $m_{\ell}$, and has obvious $\left(\text{energy}/\Lambda\right)$ enhancement for some specific processes. The main point of this work is to first show this novel enhancement behavior is not only crucial for the $\ell$ALP productions from charged mesons and $W$ boson decays to constrain the light $\ell$ALP~\cite{Altmannshofer:2022izm}, but also important to search for heavier $\ell$ALPs from the t-channel processes, $\ell^{+}\ell^{-}\rightarrow\nu_{\ell} a\overline{\nu_{\ell}}$ and $\ell^{-}p\rightarrow\nu_{\ell} a j$ in the electroweak-violating scenario at $\ell^{+}\ell^{-}$ and $\ell p$ colliders, respectively.

Apart from the production of $\ell$ALPs, decay modes of the $\ell$ALP are also important to search for $\ell$ALPs in collider experiments. For mass of $\ell$ALP below the electroweak scale ($m_a\lesssim M_W$), the $\ell$ALP mainly decays to $\ell^{+}\ell^{-}$ and $\gamma\gamma$ in the form~\cite{Bauer:2017ris,Bauer:2018uxu,Chang:2021myh}  
\begin{equation}
\Gamma_{a\rightarrow\ell^{+}\ell^{-}} = \frac{(c^A_{\ell})^2 m^2_{\ell} m_a}{8\pi\Lambda^2}\sqrt{1-\frac{4m^2_{\ell}}{m^2_a}} ~,~~ 
\Gamma_{a\rightarrow\gamma\gamma} = \frac{g^2_{a\gamma\gamma}m^3_a}{64\pi} ~,
\end{equation}
where 
\begin{equation}
g_{a\gamma\gamma} = \frac{\alpha_{\text{em}}}{\pi}\frac{c^A_{\ell}}{\Lambda}\lvert 1 - {\cal F} (\frac{m^2_a}{4m^2_{\ell}})\rvert \,,
\end{equation} 
and the loop function ${\cal F} (z > 1) = \frac{1}{z}\text{arctan}^2\left(\frac{1}{\sqrt{1/z -1}}\right)$. Note the $\ell$ALP interaction with a photon pair comes from both chiral anomaly as shown in Eq.~(\ref{eq:int}) and one-loop triangle Feynman diagrams\footnote{Here only charged leptons are considered in the loop, and the contribution from $W$ boson is much suppressed and can be safely ignored.}. 

\begin{figure}[tb]
\centering{\includegraphics[width=0.4\textwidth]{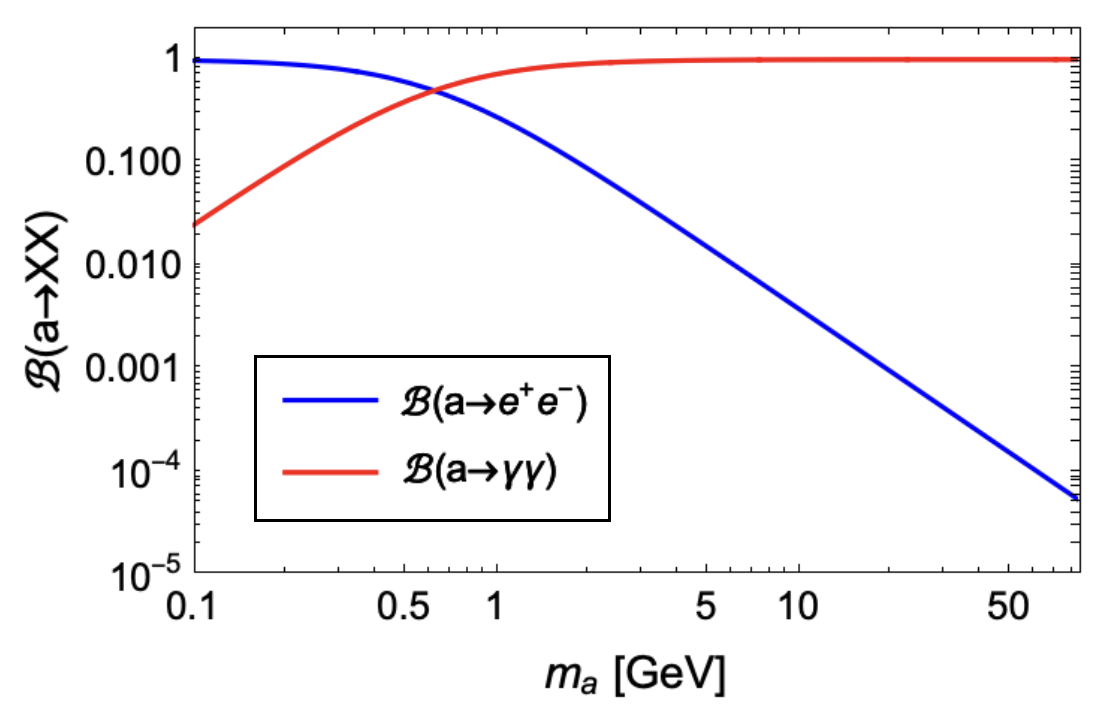}}
\caption{The decay branching ratios of $e$ALP below the electroweak scale ($m_a\lesssim M_W$). 
}
\label{fig:lALP_BR}
\end{figure}

Taking $e$ALP as an example, we show the branching ratios for $a\rightarrow e^{+}e^{-}$ and $a\rightarrow\gamma\gamma$ in Fig.~\ref{fig:lALP_BR}. The crossing point of ${\cal B}(a\rightarrow e^{+}e^{-})$ and ${\cal B}(a\rightarrow\gamma\gamma)$ is around $m_a = 0.6$ GeV. When $m_a\gg m_e$ as the case in Fig.~\ref{fig:lALP_BR}, ${\cal F} (z)\ll 1$ and the $e$ALP interaction with a photon pair is dominated from chiral anomaly. Since this chiral anomaly driven $\boldsymbol{a\gamma\gamma}$ interaction is almost independent of the charged lepton mass, we can expect the decay of heavier $\ell$ALP becomes photophilic~\cite{Chang:2021myh}.

\section{Energy enhancement behaviors in $e^{+}e^{-}\rightarrow\nu_e a\overline{\nu_e}$ and $e^{-}p\rightarrow\nu_e a j$}
\label{sec:kin}
 
\begin{figure}[tb]
\centering{\includegraphics[width=0.48\textwidth]{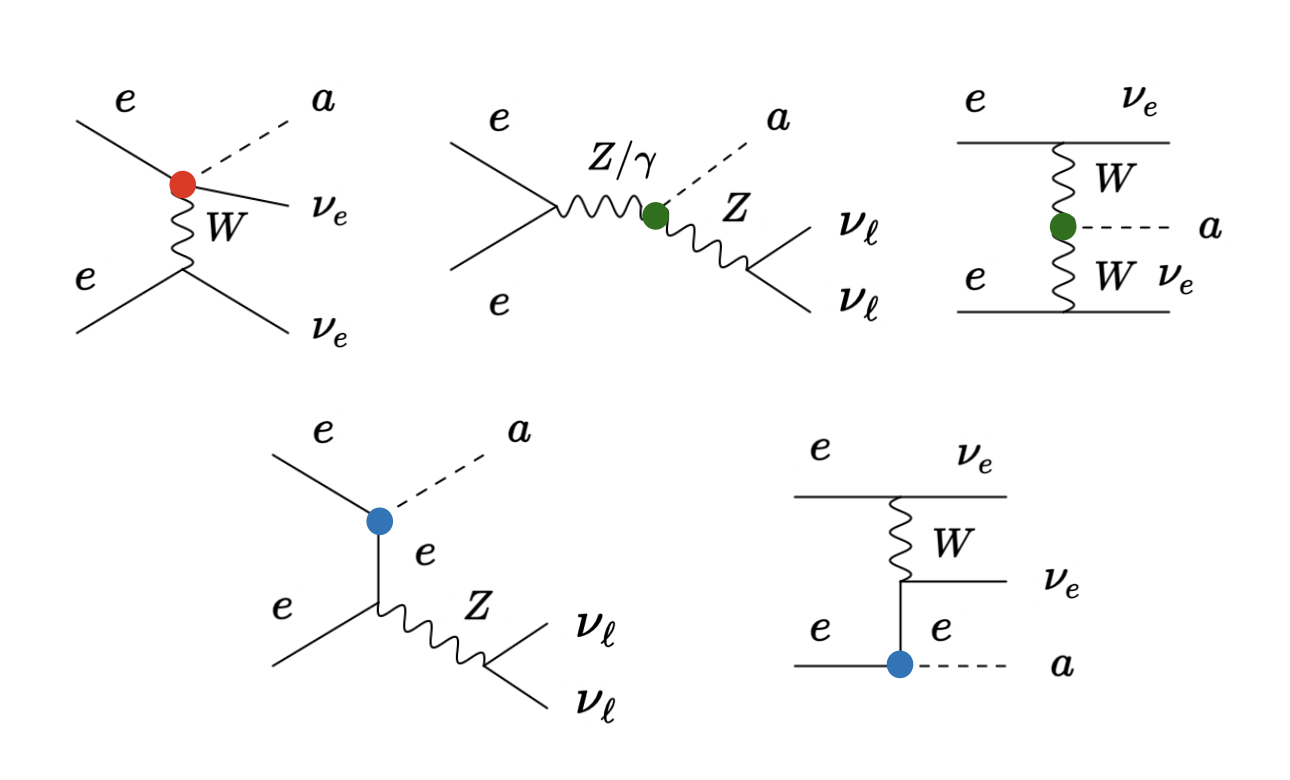}}
\caption{Feynman diagrams for $e^{+}e^{-}\rightarrow\nu_e a\overline{\nu_e}$. Here the color markers indicate red for $\boldsymbol{aW\ell\nu}$ interaction, green for $\boldsymbol{aVV'}$ interaction and blue for $\boldsymbol{a\ell\ell}$ interaction.  
}
\label{fig:Feyn1}
\end{figure}

\begin{figure}[tb]
\centering{\includegraphics[width=0.48\textwidth]{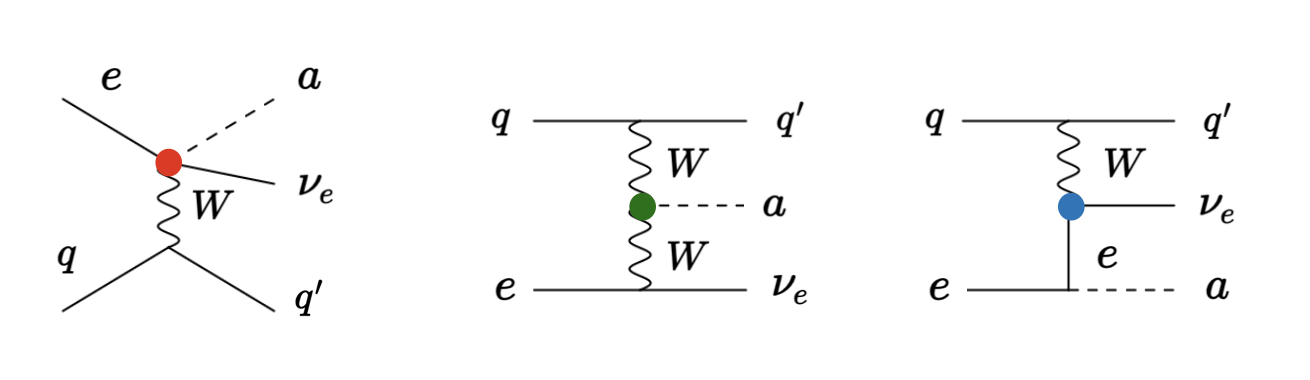}}
\caption{Feynman diagrams for $e^{-}p\rightarrow\nu_e a j$. The color markers are the same as in Fig.~\ref{fig:Feyn1}.
}
\label{fig:Feyn2}
\end{figure}

In this section, we concretely study the energy enhancement behaviors in $e^{+}e^{-}\rightarrow\nu_e a\overline{\nu_e}$ and $e^{-}p\rightarrow\nu_e a j$ via the $\boldsymbol{aW\ell\nu}$ interaction. Using $e^{+}e^{-}\rightarrow\nu_e a\overline{\nu_e}$ as an example, the relevant Feynman diagrams for this process are shown in Fig.~\ref{fig:Feyn1}. We will see numerical results later that the  contributions from $\boldsymbol{aVV'}$ and $\boldsymbol{a\ell\ell}$ interactions in this process are much smaller than the one from $\boldsymbol{aW\ell\nu}$ interaction. In order to simplify the analytical form for discussions and highlight the energy enhancement behaviors here, we will only show the amplitude square with the average (sum) over initial (final) polarization from the first diagram in Fig.~\ref{fig:Feyn1} for the process $e^{-}(p_1)e^{+}(p_2)\rightarrow\nu_e(q_1)a(q_2)\overline{\nu_e}(q_3)$,   
\begin{align} 
& \overline{\lvert{\cal M}\rvert ^2} = \label{eq:amplitude} \\ 
& \frac{g^4_W\left( c^A_{\ell}-c^V_{\ell}+c_{\nu}\right) ^2}{32\Lambda^2}\left(\frac{1}{k^2-M^2_W}+\frac{1}{k^{\prime 2}-M^2_W}\right) ^2 \notag \\ \notag
& \times\left( s-2m^2_e\right)\left[ s-m^2_a -2q_2\cdot (q_1 +q_3)\right] \notag  \,, \notag 
\end{align} 
where $s = \left( p_1 +p_2\right) ^2 = \left( q_1 + q_2 + q_3\right) ^2$, $k = p_2 -q_3$ and $k' = p_1 -q_1$. It's clear to see this amplitude square can be enhanced when the momentum transferring in this $t$-channel process is large enough and it is estimated to be smaller than 
\begin{equation}
\overline{\lvert{\cal M}\rvert ^2}~ < ~\frac{g^4_W\left( c^A_{\ell}-c^V_{\ell}+c_{\nu}\right) ^2}{16\Lambda^2\left( s-M^2_W\right) ^2}\left( s-2m^2_e\right)\left(\sqrt{s}-m_a\right) ^2.
\end{equation} 
On the other hand, the related Feynman diagrams for the process $e^{-}p\rightarrow\nu_e a j$ are shown in Fig.~\ref{fig:Feyn2}. Again, except for the $\boldsymbol{aW\ell\nu}$ interaction, contributions from other two types of $e$ALP interactions can be safely ignored and the analytical form for the amplitude square from the first diagram in Fig.~\ref{fig:Feyn2} is similar to Eq.~(\ref{eq:amplitude}) and we will not show it here.

Since $\boldsymbol{aW\ell\nu}$ interaction can only appear in an electroweak-violating scenario, we follow the strategy in Ref.~\cite{Altmannshofer:2022izm} to define two different scenarios : 
\begin{align} 
& \text{Electroweak Violating}~ (\textbf{EWV}) : c^V_{\ell} = c_{\nu} = 0, c^A_{\ell}\neq 0, \notag \\ 
& \text{Electroweak Preserving}~ (\textbf{EWP}) : c_{\nu} = 0, c^V_{\ell} = c^A_{\ell}\neq 0\,. 
\label{eq:EWV_EWP}
\end{align} 
Note the \textbf{EWV} (\textbf{EWP}) scenario is pure axial-vector (right-handed coupling) current in Eq.~(\ref{eq:Jint}) and the \textbf{EWV} scenario is usually studied ALP and SM fermion pair interactions in the literature.

\begin{figure}[tb]
\centering{\includegraphics[width=0.4\textwidth]{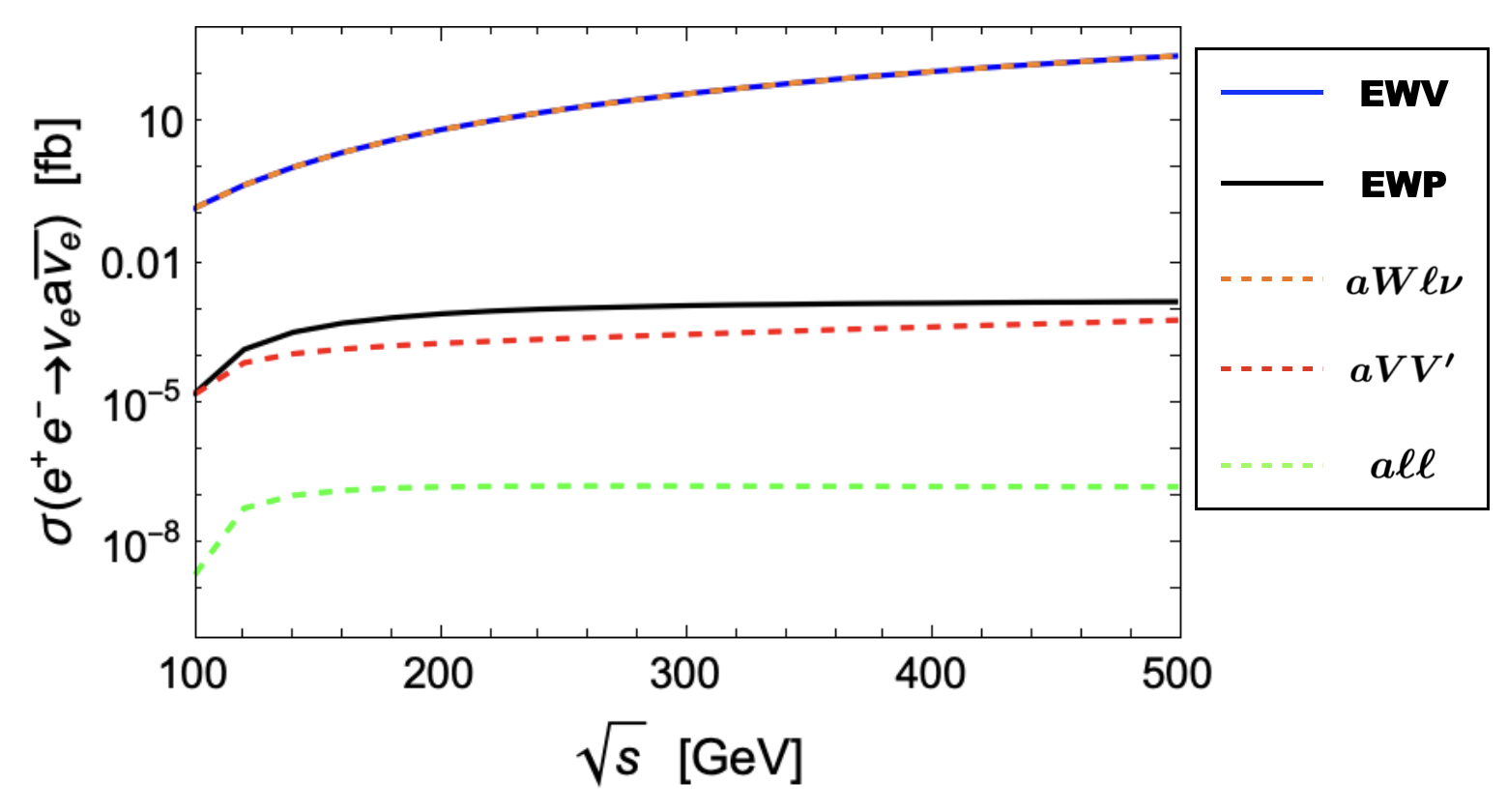}}
\caption{The energy enhancement behaviors of cross sections in $e^{+}e^{-}\rightarrow\nu_e a\overline{\nu_e}$ with $c^A_e / \Lambda = 0.01$ GeV$^{-1}$, $c^V_e=c_{\nu}=0$ (\textbf{EWV} : solid-blue line) and $c^A_e / \Lambda = c^V_e / \Lambda = 0.01$ GeV$^{-1}$, $c_{\nu}=0$ (\textbf{EWP} : solid-black line). The dashed-orange, dashed-red and dashed-green lines are contributions from $\boldsymbol{aW\ell\nu}$, $\boldsymbol{aVV'}$ and $\boldsymbol{a\ell\ell}$ interactions in \textbf{EWV} scenario, respectively. 
}
\label{fig:ALP_ee_Xsec}
\end{figure}

\begin{figure}[tb]
\centering{\includegraphics[width=0.4\textwidth]{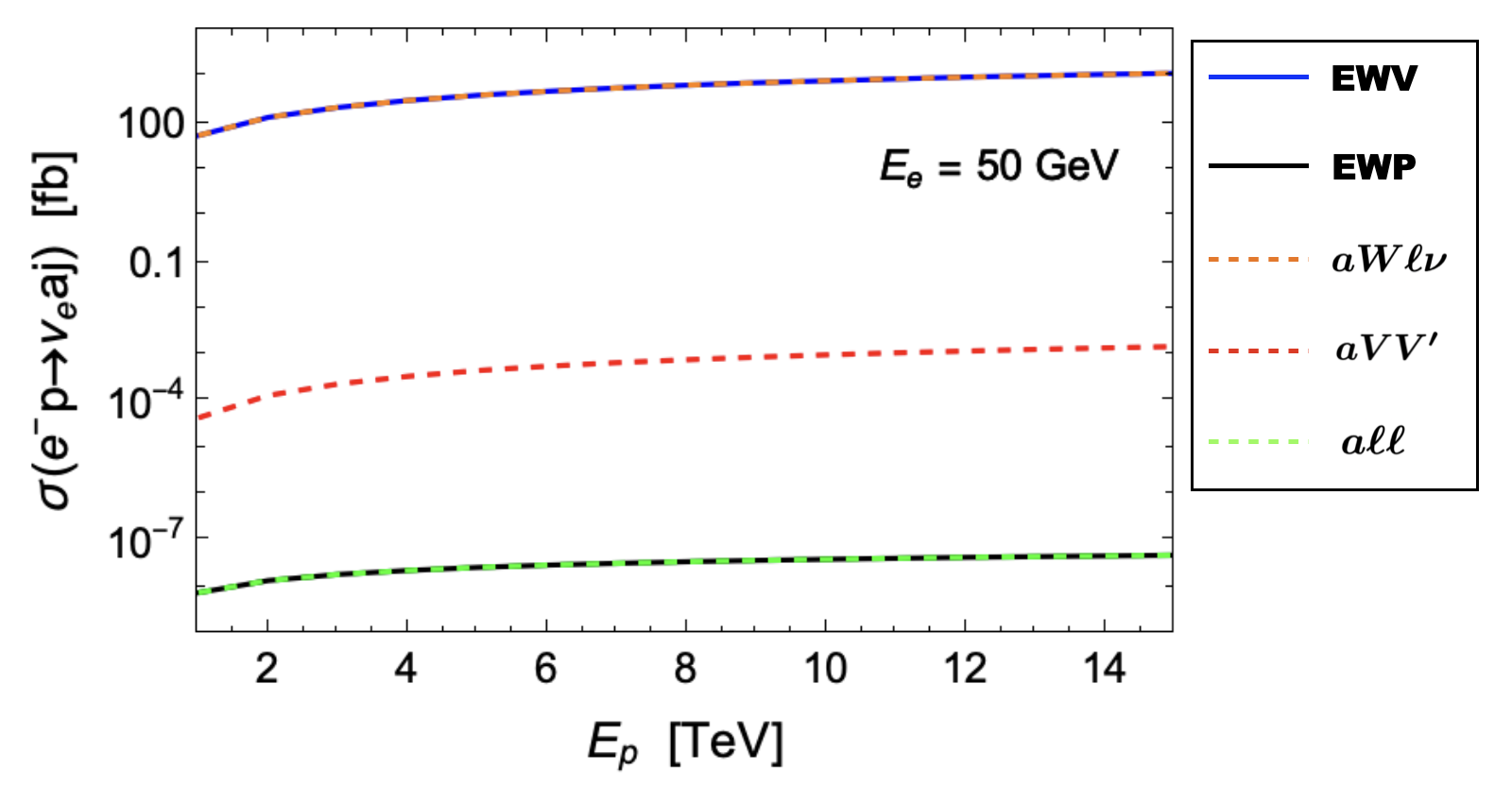}}
\caption{The same as Fig.~\ref{fig:ALP_ee_Xsec} except for the process $e^{-}p\rightarrow\nu_e a j$ with $E_e = 50$ GeV.
}
\label{fig:ALP_ep_Xsec}
\end{figure}

Although the energy enhancement behaviors are useful for exploring $e$ALPs from the $e^{+}e^{-}\rightarrow\nu_e a\overline{\nu_e}$ and $e^{-}p\rightarrow\nu_e a j$ processes at $e^{+}e^{-}$ and $ep$ colliders, these two processes still suffer from three-body phase space suppression. Therefore, it is important to numerically determine the cross-sections for varying center-of-mass energies. 
We use FeynRules~\cite{Alloul:2013bka} to implement ${\cal L}_{\ell\text{ALP}}$ in Eq.~(\ref{eq:int}) and apply Madgraph5\underline{\hspace{0.5em}}aMC@NLO~\cite{Alwall:2014hca} to calculate cross sections for these two processes. As benchmark points, we set $c^A_e / \Lambda = 0.01$ GeV$^{-1}$, $c^V_e=c_{\nu}=0$ ($c^A_e / \Lambda = c^V_e / \Lambda = 0.01$ GeV$^{-1}$, $c_{\nu}=0$) for the \textbf{EWV} (\textbf{EWP}) scenario. We vary the center-of-mass energy $\sqrt{s}=100-500$ GeV at $e^{+}e^{-}$ colliders and $E_p = 1-15$ TeV with fixed $E_e = 50$ GeV at $ep$ colliders and show their energy enhancement behaviors in Fig.~\ref{fig:ALP_ee_Xsec} and Fig.~\ref{fig:ALP_ep_Xsec}, respectively.

First of all, the cross sections in \textbf{EWV} scenario are more than four (nine) orders of magnitudes larger than the ones in \textbf{EWP} scenario of Fig.~\ref{fig:ALP_ee_Xsec} (Fig.~\ref{fig:ALP_ep_Xsec}). Therefore, these two processes are powerful to distinguish $e$ALPs in \textbf{EWV} scenario from  the \textbf{EWP} scenario. The main reason is that there is $\boldsymbol{aW\ell\nu}$ interaction in \textbf{EWV} scenario, but not in \textbf{EWP} scenario and the $\boldsymbol{aW\ell\nu}$ interaction contributes to almost the whole amount in cross sections of \textbf{EWV} scenario. In order to know quantitative contributions to cross sections from $\boldsymbol{aW\ell\nu}$, $\boldsymbol{aVV'}$ and $\boldsymbol{a\ell\ell}$ interactions of $e^{+}e^{-}\rightarrow\nu_e a\overline{\nu_e}$ and $e^{-}p\rightarrow\nu_e a j$ processes in \textbf{EWV} scenario, we label them in dashed lines of both Fig.~\ref{fig:ALP_ee_Xsec} and Fig.~\ref{fig:ALP_ep_Xsec}. 
The subleading contribution in \textbf{EWV} scenario comes from $\boldsymbol{aVV'}$ interaction, but it's about five (six) orders of magnitude smaller than the one from $\boldsymbol{aW\ell\nu}$ interaction in Fig.~\ref{fig:ALP_ee_Xsec} (Fig.~\ref{fig:ALP_ep_Xsec}). Contributions from $\boldsymbol{a\ell\ell}$ interaction are very tiny in the \textbf{EWV} scenario for both of these two processes and can be totally ignored. Note for \textbf{EWP} scenario, both $\boldsymbol{aW\ell\nu}$ and $\boldsymbol{aWW}$ interactions vanish in $e^{-}p\rightarrow\nu_e a j$ such that it overlaps with the dashed line of $\boldsymbol{a\ell\ell}$ interaction. Therefore, $e^{+}e^{-}\rightarrow\nu_e a\overline{\nu_e}$ and $e^{-}p\rightarrow\nu_e a j$ processes at $e^{+}e^{-}$ and $ep$ colliders are ideal to explore the novel $\boldsymbol{aW\ell\nu}$ interaction and distinguish $e$ALPs in \textbf{EWV} scenario from \textbf{EWP} scenario. 
We will focus on searching for $e$ALPs in \textbf{EWV} scenario at $e^{+}e^{-}$ and $ep$ colliders in the next section.

\section{The study at $e^{+}e^{-}$ and $ep$ colliders}
\label{sec:filter} 

In this section, we further study the signal-to-background analysis at $e^{+}e^{-}$ colliders in Sec.~\ref{sec:CEPC} and $ep$ colliders in Sec.~\ref{sec:LHeC}. Our goal is to predict the future bounds for $e$ALPs at $e^{+}e^{-}$ and $ep$ colliders and compare them with existing constraints which are shown in Sec.~\ref{sec:Summary}.

\subsection{Exploring $e^{+}e^{-}\rightarrow\nu_e a\overline{\nu_e}$ at CEPC}
\label{sec:CEPC} 

\begin{figure}[tb]
\centering{\includegraphics[width=0.48\textwidth]{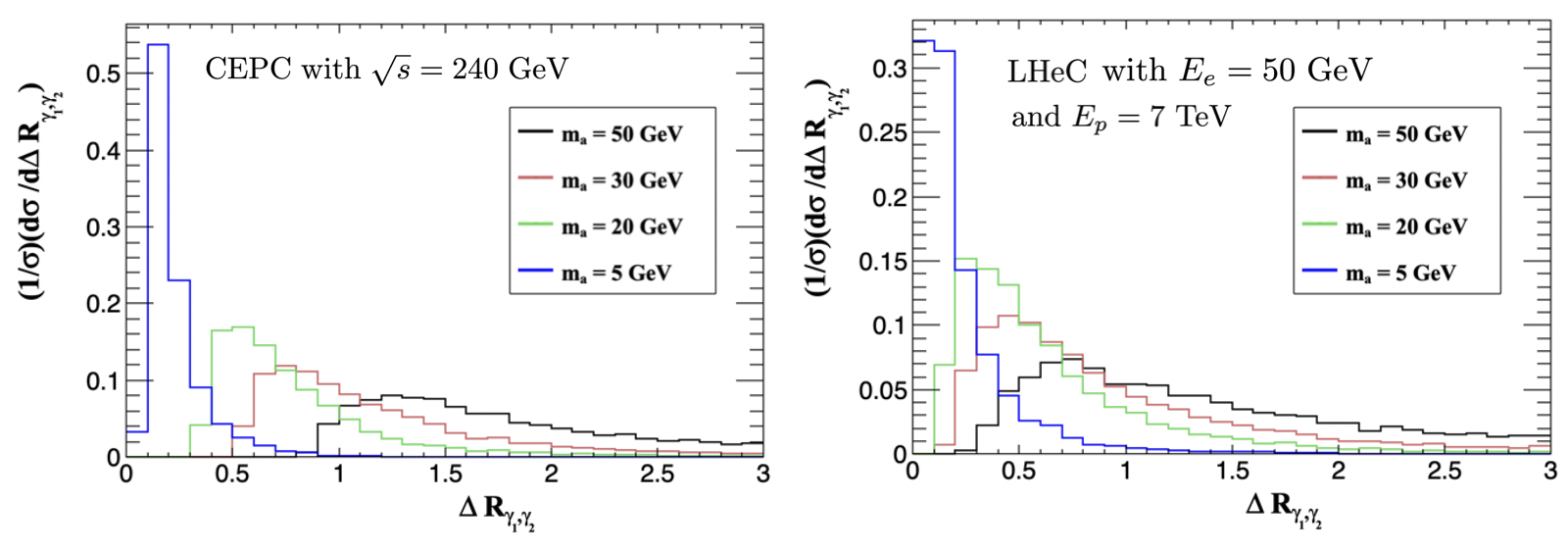}}
\caption{Distributions of the opening angle between two photons, $\Delta R_{\gamma_1\gamma_2}$, from $e^{+}e^{-}\rightarrow\nu_e (a\rightarrow\gamma\gamma)\overline{\nu_e}$ at CEPC with $\sqrt{s}=240$ GeV (left panel) and $e^{-}p\rightarrow\nu_e (a\rightarrow\gamma\gamma) j$ at LHeC with $E_e = 50$ GeV and $E_p = 7$ TeV (right panel) in parton-level. Four benchmark $e$ALP mass values, $m_a = 50$, $30$, $20$, $5$ GeV, are shown for comparison.
}
\label{fig:dR_aa}
\end{figure}

There are several proposed $e^{+}e^{-}$ colliders which can explore $e^{+}e^{-}\rightarrow\nu_e a\overline{\nu_e}$, like CEPC~\cite{CEPCStudyGroup:2018ghi}, ILC~\cite{TLEPDesignStudyWorkingGroup:2013myl}, CLIC~\cite{Baer:2013cma}, FCC-ee~\cite{CLICDetector:2013tfe}. Here we choose the CEPC with $\sqrt{s}=240$ GeV as a concrete example to study how to extract the $e$ALP signal from relevant SM backgrounds. According to Fig.~\ref{fig:lALP_BR}, the $e$ALP dominantly decays to $\gamma\gamma$ when $m_a\gtrsim 1$ GeV. Hence, we focus on $a\rightarrow\gamma\gamma$ decay mode in this work. Most interestingly, the $e$ALP becomes highly boosted at CEPC when it is light enough, so two photons in the final state may be too collimated to pass the photon isolation criterion at detectors. Take a cone size $R=0.5$ as the photon isolation criterion at CEPC, we can find that two photons cannot be isolated to each other at detectors when $m_a \lesssim 20$ GeV (parton-level) in the left panel of Fig.~\ref{fig:dR_aa}. For this kind of collimated, non-isolated photons, we can group them as a special signature "photon-jet" ($J_{\gamma}$) which is a non-QCD jet-like structure and deposits most of its energy in the electromagnetic calorimeter (ECAL) instead of hadron calorimeter (HCAL) as well as leaves less charged tracks compared with the QCD jet. Therefore, we classify the signal signatures as (1) two isolated photons plus missing energy (${\:/\!\!\!\! E}$) for $m_a\gtrsim 20$ GeV, (2) a $J_{\gamma}$ plus ${\:/\!\!\!\! E}$ for $m_a \lesssim 20$ GeV. 

\begin{table}[ht!]
\begin{center}\begin{tabular}{|c|c|c|c|}\hline cut flow in $\sigma$ [fb] & ~signal~ & ~$\nu_{\ell}\overline{\nu_{\ell}}\gamma\gamma$~ & ~$\nu_{\ell}\overline{\nu_{\ell}}(h\rightarrow\gamma\gamma)$~ \\ 
\hline Generator & $0.11$ & $263.60$ & $7.67\times 10^{-2}$ \\
\hline cut-(1) & $7.10\times 10^{-2}$ & $32.23$ & $5.47\times 10^{-2}$ \\
\hline cut-(2) & $7.02\times 10^{-2}$ & $30.24$ & $5.40\times 10^{-2}$ \\
\hline cut-(3) & $6.57\times 10^{-2}$ & $20.60$ & $3.68\times 10^{-5}$ \\
\hline cut-(4) & $5.97\times 10^{-2}$ & $11.93$ & $3.68\times 10^{-5}$ \\ 
\hline cut-(5) & $5.03\times 10^{-2}$ & $7.54$ & ~$3.84\times 10^{-6}$~ \\
\hline cut-(6) & $5.00\times 10^{-2}$ & $2.36$ & ~$1.53\times 10^{-6}$~ \\
\hline cut-(7) & $4.95\times 10^{-2}$ & $0.62$ & ~$0$~ \\
\hline \end{tabular} \caption{The cut-flow table for $e^{+}e^{-}\rightarrow\nu_e (a\rightarrow\gamma\gamma)\overline{\nu_e}$ and relevant SM backgrounds with signature of two isolated photons plus ${\:/\!\!\!\! E}$. The benchmark point $m_a = 50$ GeV with $c^A_e / \Lambda = 1$ TeV$^{-1}$ for signal is chosen. Each event selection has been mentioned in the main text. The "Generator" means the cross sections in parton-level calculated by Madgraph5\underline{\hspace{0.5em}}aMC@NLO.}
\label{tab:CF_ee_1}
\end{center}
\end{table}

For the first signal signature, the relevant SM backgrounds are $e^{+}e^{-}\rightarrow \gamma\gamma\nu_{\ell}\overline{\nu_{\ell}}$ and $e^{+}e^{-}\rightarrow\nu_{\ell}\overline{\nu_{\ell}}h\rightarrow\nu_{\ell}\overline{\nu_{\ell}} (\gamma\gamma)$.  
The benchmark point $m_a = 50$ GeV with $c^A_e / \Lambda = 1$ TeV$^{-1}$ for signal is chosen. We use Madgraph5\underline{\hspace{0.5em}}aMC@NLO to generate Monte Carlo samples for both signal and background processes and pass them to Pythia8~\cite{Sjostrand:2007gs} for the QED showering effect. The pre-selection cuts ($E_{\gamma} > 5$ GeV and $\lvert\eta_{\gamma}\rvert < 3.0$) at parton-level for both the signal and backgrounds are imposed. 
The CEPC template in Delphes3~\cite{deFavereau:2013fsa} is applied for the fast detector simulation where the photon isolation criterion is consistent with Ref.~\cite{Cobal:2020hmk,Ahmed:2022ude}. We require the following event selections to identify the signal signature and suppress background events :  
\begin{itemize} 
\item (1) $N(\gamma)\geqslant 2$ with $30$ GeV $< E_{\gamma_1} < 90$ GeV, $E_{\gamma_2} > 15$ GeV, $\lvert\eta_{\gamma_1}\rvert < 1.5$ and $\lvert\eta_{\gamma_2}\rvert < 2.0$, 
\item (2) $\lvert\cos\theta_{\gamma}\rvert < 0.95$ and $\left(E_{\gamma_1}+E_{\gamma_2}\right) /E_{\text{beam}} < 1.8$,  
\item (3) ${\:/\!\!\!\! E} > 120$ GeV and $\lvert\eta_{{\:/\!\!\!\! E}}\rvert < 2.0$, 
\item (4) Veto $85$ GeV $< M_{\:/\!\!\!\! E} < 95$ GeV,
\item (5) $\Delta\phi_{\gamma_1, {\:/\!\!\!\! E}} > 2.5$ and $\Delta\phi_{\gamma_2, {\:/\!\!\!\! E}} > 1.8$, 
\item (6) $2.2 < {\:/\!\!\!\! E}/M_{\gamma_1\gamma_2} < 3.6$, 
\item (7) $\lvert M_{\gamma_1\gamma_2}-m_a\rvert < 3$ GeV,  
\end{itemize} 
where $E_{\gamma_1}$, $E_{\gamma_2}$ ($\eta_{\gamma_1}$, $\eta_{\gamma_2}$) are the energy (pseudorapidity) of leading and subleading energetic photons, $\theta_{\gamma}$ is the polar angle of photon relative to the positron beam direct, $E_{\text{beam}}$ is the beam energy,  $M_{\gamma_1\gamma_2}$ is the invariant mass of a photon pair, $\Delta\phi_{\gamma_i, {\:/\!\!\!\! E}}$ is the azimuthal angle between the i-th photon and ${\:/\!\!\!\! E}$.   
The cut-flow table including signal and backgrounds for each event selection is presented in Table.~\ref{tab:CF_ee_1} and some kinematic distributions are shown in Fig.~\ref{fig:ee_50} of Appendix~\ref{app:rec}.

In signal events, two isolated photons and ${\:/\!\!\!\! E}$ distribute in the central region and we set $E_{\gamma_1} > 30$ GeV, $E_{\gamma_2} > 15$ GeV and ${\:/\!\!\!\! E} > 120$ GeV as the trigger for our candidate events. Moreover, we select $E_{\gamma_1} < 90$ GeV to further reduce the large background from $e^{+}e^{-}\rightarrow \gamma\gamma\nu_{\ell}\overline{\nu_{\ell}}$. Except for the aforementioned SM backgrounds, there are also potential QED-generated backgrounds, including $e^+e^-\rightarrow\gamma\gamma (\gamma)$ and radiative Bhabha scattering with one or more unobserved electrons. These backgrounds tend to produce photons near the beam direction. Therefore, they can be effectively managed by restricting the polar angles of the two photons and ensuring consistency with a photon pair recoiling from a massive object. We introduce two additional event selection criteria in cut-(2) to further suppress the QED-generated backgrounds, which are more stringent than those used in previous searches at LEP~\cite{OPAL:1997zll,OPAL:1998jka,OPAL:1999tmg}. However, as shown in Table.~\ref{tab:CF_ee_1}, cut-(2) only marginally reduce the signal and SM background events.

The neutrino pair from these two SM backgrounds are largely produced from the $Z$ boson decay, so we veto the $Z$ boson mass window, $85$ GeV $< M_{\:/\!\!\!\! E} < 95$ GeV, to suppress these background events. On the other hand, because two isolated photons are well-separated from ${\:/\!\!\!\! E}$, we apply this feature to reduce some events from $e^{+}e^{-}\rightarrow \gamma\gamma\nu_{\ell}\overline{\nu_{\ell}}$. Especially, the distribution of $\Delta\phi_{\gamma_2, {\:/\!\!\!\! E}}$ is not so large in both $e^{+}e^{-}\rightarrow \gamma\gamma\nu_{\ell}\overline{\nu_{\ell}}$ and $e^{+}e^{-}\rightarrow\nu_{\ell}\overline{\nu_{\ell}}h\rightarrow\nu_{\ell}\overline{\nu_{\ell}} (\gamma\gamma)$ compared to the signal which can be observed in Fig.~\ref{fig:ee_50}. The background events from $e^{+}e^{-}\rightarrow\nu_{\ell}\overline{\nu_{\ell}}h\rightarrow\nu_{\ell}\overline{\nu_{\ell}} (\gamma\gamma)$ are killed up to this step. Furthermore, we also apply the ratio, ${\:/\!\!\!\! E}/M_{\gamma_1\gamma_2}$, to further reduce a few background events. Finally, the $e$ALP mass window selection is very powerful to keep most signal events but highly reduce these two backgrounds\footnote{The chosen $e$ALP mass window reflects the energy measurement precision of the detector, and cut-(7) can be further refined in an actual experimental "bump search" analysis. However, conducting such an analysis is beyond the scope of this work.}.

We consider the most optimistic integrated luminosity of CEPC proposal~\cite{CEPCStudyGroup:2018ghi}, ${\cal L} = 5 ab^{-1}$, and define the signal significance $Z$ as~\cite{Cowan:2010js} 
\begin{equation}
Z = \sqrt{2\cdot\left( (N_s + N_b)\cdot ln(1+N_s/N_b)-N_s\right)},
\end{equation} 
where $N_s$ and $N_b$ are the expected signal and background event numbers. 
Here the systematic uncertainties are not taken into account in our simple analysis since the CEPC is still a future collider.
After all of these event selections in Table.~\ref{tab:CF_ee_1}, we find the signal significance can reach $Z=4.40$ for our benchmark point of ${\cal L} = 5 ab^{-1}$ and it's detectable in the future.

For the second signal signature, the only important SM background is $e^{+}e^{-}\rightarrow\nu_{\ell}\overline{\nu_{\ell}}\gamma$\footnote{As expected, it is hard to produce a single QCD jet plus large ${\:/\!\!\!\! E}$ at the clean $e^+ e^-$ colliders. Besides, the fake photon-jet candidate from instruments is beyond the scope of this work, and will not be considered here.}. Here we take the signal benchmark point as $m_a = 5$ GeV with $c^A_e / \Lambda = 1$ TeV$^{-1}$\footnote{The pre-selection cuts in parton-level are similar as before, except for $E_{\gamma} > 1$ GeV ($E_{\gamma} > 10$ GeV) for the signal (background).}. 
The $J_{\gamma}$ candidate is defined as the following. We apply the Cambridge/Aachen ($C/A$) jet clustering algorithm~\cite{Dokshitzer:1997in,Wobisch:1998wt} with a cone size $R = 0.4$ similar to previous photon-jet works~\cite{Ellis:2012sd,Ellis:2012zp,Allanach:2017qbs,Chakraborty:2017mbz,Sheff:2020jyw,Wang:2021uyb,Ren:2021prq} to group collimated photons. Then we require the hadronic energy fraction, $\theta_J$, of $J_{\gamma}$ to satisfy $log\theta_J < -2$ and it is defined as 
\begin{equation}
\theta_J = \frac{E_{J,\text{HCAL}}}{E_J}, 
\end{equation}
where $E_J$ is the total energy of the target jet and $E_{J,\text{HCAL}}$ is the recorded energy from this jet in HCAL. Note the crucial difference for the $J_{\gamma}$ signature at hadron colliders and $e^+ e^-$ colliders is the pile-up effect. The pile-up effect in hadron colliders makes it more challenging to distinguish $J_{\gamma}$ from the single photon and the QCD jet. However, since there are no pile-up collisions and underlying events at $e^+ e^-$ colliders, the $J_{\gamma}$ signature becomes very unique and almost no SM background can mimic it after considering some jet substructure observables. For this reason, we can choose a more stringent $J_{\gamma}$ criterion, $log\theta_J < -2$, than those previous works of $J_{\gamma}$ at LHC~\cite{Ellis:2012sd,Ellis:2012zp,Chakraborty:2017mbz,Wang:2021uyb,Ren:2021prq}.  

\begin{figure}[tb]
\centering{\includegraphics[width=0.4\textwidth]{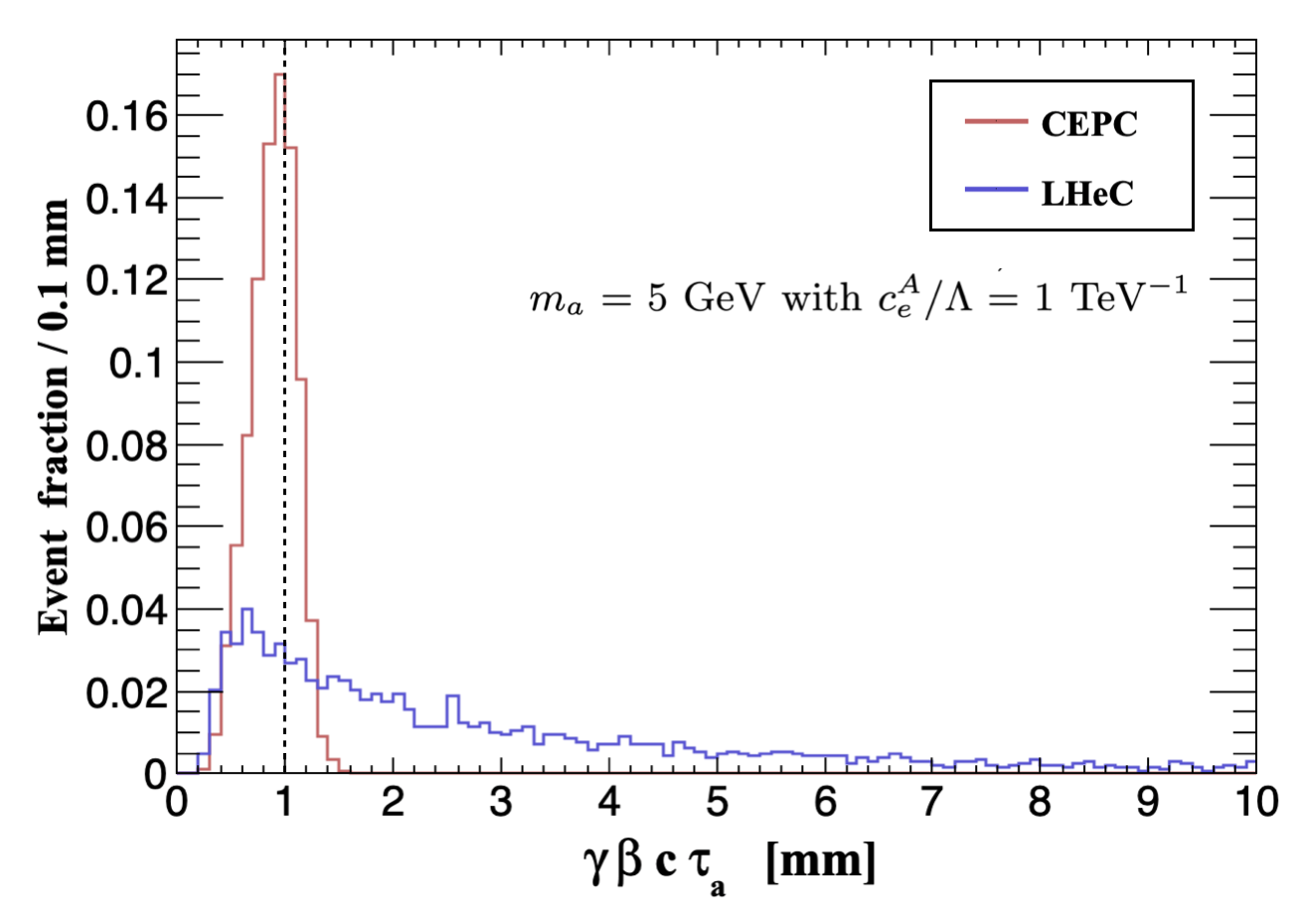}}
\caption{Distributions of the $e$ALP lab frame decay length from $e^{+}e^{-}\rightarrow\nu_e (a\rightarrow\gamma\gamma)\overline{\nu_e}$ at CEPC with $\sqrt{s}=240$ GeV (red line) and $e^{-}p\rightarrow\nu_e (a\rightarrow\gamma\gamma) j$ at LHeC with $E_e = 50$ GeV and $E_p = 7$ TeV (blue line) in detector-level. The benchmark point $m_a = 5$ GeV with $c^A_e / \Lambda = 1$ TeV$^{-1}$ is considered.
}
\label{fig:ALP_DecLen}
\end{figure}

On the other hand, the $e$ALP decay width for this benchmark point is $\Gamma_a = 3.90\times 10^{-12}$ GeV, which may be small enough for $e$ALP to become a long-lived particle (LLP) on the scale of collider experiments. In order to classify $e$ALPs belong to the prompt decay or the LLP at CEPC, we require the $e$ALP lab frame decay length $\gamma\beta c\tau_a < 1$ mm as a criterion of the prompt decay where $\gamma$ is the Lorentz factor, $\beta$ is the ratio of relative velocity to $c$, and $c$ is the speed of light. The distribution of $e$ALP lab frame decay length for this benchmark point at CEPC (detector-level) can be found in the red line of Fig.~\ref{fig:ALP_DecLen}. The physical size in radius of proposed detectors on CEPC can be summarized as (1) $16$ mm $\leqslant R_{\text{vertex}}\leqslant 60$ mm, (2) $0.15$ m $\leqslant R_{\text{ECAL}}\leqslant 1.81$ m, (3) $2.30$ m $\leqslant R_{\text{HCAL}}\leqslant 3.34$ m, (4) $2.30$ m $\leqslant R_{\text{muon}}\leqslant 3.34$ m~\cite{CEPCStudyGroup:2018ghi,Zhang:2021orr}. Therefore, we simply consider the $e$ALP lab frame decay length within $10^{-3}$ m $\leqslant\gamma\beta c\tau_a\leqslant 1.8$ m as a detectable LLP with a displaced $J_{\gamma}$ signature at CEPC.

\begin{table}[ht!]
\begin{center}\begin{tabular}{|c|c|c|}\hline cut flow in $\sigma$ [fb] & ~signal~ & ~$\nu_{\ell}\overline{\nu_{\ell}}\gamma$~ \\ 
\hline Generator & $0.16$ & $4266.80$ \\ 
\hline $\gamma\beta c\tau_a < 1$ mm & $0.10$ & $-$ \\ 
\hline cut-(1) & $8.49\times 10^{-2}$ & $520.29$ \\
\hline cut-(2) & $8.49\times 10^{-2}$ & $520.29$ \\
\hline cut-(3) & $8.10\times 10^{-2}$ & $387.77$ \\
\hline cut-(4) & $7.92\times 10^{-2}$ & $373.39$ \\ 
\hline cut-(5) & $7.70\times 10^{-2}$ & ~$0$~ \\
\hline cut-(6) & $7.67\times 10^{-2}$ & ~$0$~ \\
\hline \end{tabular} \caption{Similar to Table~\ref{tab:CF_ee_1}, but for the signal benchmark point $m_a = 5$ GeV with $c^A_e / \Lambda = 1$ TeV$^{-1}$ as well as the signature of a $J_{\gamma}$ candidate plus ${\:/\!\!\!\! E}$.}
\label{tab:CF_ee_2}
\end{center}
\end{table}

For the $e$ALP prompt decay, we set up event selections to identify the signal signature and suppress the background events in the following :  
\begin{itemize} 
\item (1) $N(J_{\gamma})\geqslant 1$ with $log\theta_J < -2$, $30$ GeV $< E_{J_{\gamma}} < 100$ GeV, and $\lvert\eta_{J_{\gamma}}\rvert < 1.5$, 
\item (2) ${\:/\!\!\!\! E} > 140$ GeV and $\lvert\eta_{{\:/\!\!\!\! E}}\rvert < 1.5$,    
\item (3) Veto $75$ GeV $< M_{\:/\!\!\!\! E} < 105$ GeV, 
\item (4) ${\:/\!\!\!\! E}/E_{J_{\gamma}} > 1.5$,
\item (5) $\tau_1 (J_{\gamma}) > 0.03$, 
\item (6) $\lvert M_{J_{\gamma}}-m_a\rvert < 1$ GeV,  
\end{itemize}  
where $\tau_1$ is the case of $N=1$ in the general N-subjettiness, $\tau_N$~\cite{Thaler:2010tr,Thaler:2011gf}. The N-subjettiness is a set of simple but powerful jet substructure observables to effectively count how many subjets inside a candidate jet. The $\tau_1$ can be represented as 
\begin{equation}
\tau_1 = \frac{\sum_k P_{T_k}\cdot\Delta R_{1,k}}{\sum_k P_{T_k}\cdot R} 
\end{equation}   
where $k$ runs over all constituents from both ECAL and HCAL cells of the jet, $P_{T_k}$ is the transverse momentum for the $k$-th constituent, $\Delta R_{1,k}$ is the opening angle between the first subjet and the $k$-th constituent of the jet. Hence, $\tau_1$ is nothing but a description of the energy distribution inside a candidate jet.
We can expect the single photon inside $J_{\gamma}$ is almost no structure, but $J_{\gamma}$ for the signal is a two-prong structure. Therefore, $\tau_1$ can be used to distinguish the signal from the background in this analysis. 
The cut-flow table of signal and background for each event selection are displayed in Table.~\ref{tab:CF_ee_2} and some kinematic distributions are also shown in Fig.~\ref{fig:ee_5} of Appendix~\ref{app:rec}.

We take a $J_{\gamma}$ candidate with $E_{J_{\gamma}} > 30$ GeV and ${\:/\!\!\!\! E} > 140$ GeV as the trigger and both of them are distributed in the central region. The event selection, $\lvert\cos\theta_{J_{\gamma}}\rvert < 0.95$ and $E_{J_{\gamma}} /E_{\text{beam}} < 1.8$, is applied between cut-(1) and cut-(2) to eliminate potential QED-generated backgrounds. However, it's worth noting that events from both the signal and $e^{+}e^{-}\rightarrow\nu_{\ell}\overline{\nu_{\ell}}\gamma$ can completely pass this event selection. Hence, we have not included it in Table~\ref{tab:CF_ee_2}. The event selections $E_{J_{\gamma}} < 100$ GeV and $Z$ boson mass window ($75$ GeV $< {\:/\!\!\!\! E} < 105$ GeV) veto as well as ${\:/\!\!\!\! E}/E_{J_{\gamma}} > 1.5$ are used to keep most signal events but remove large parts of background events. Note $J_{\gamma}$ and the sum of missing pieces almost share half of the whole center-of-mass energy and scatter back-to-back, so cut-(2) keeps the same event rate as cut-(1). As mentioned before, the jet substructure observables are the most powerful variables to distinguish $J_{\gamma}$ from the single photon in this analysis. Here we apply two of them in this work : N-subjettiness and jet mass. 
The jet mass of a single photon is close to zero because the photon is massless. However, the jet mass of signal peaks at $m_a$. Similarly, we apply $\tau_1 (J_{\gamma})$ to show the different feature between the single photon and the signal.  
After involving these two jet substructure observables, the background has been totally killed. The key reason to distinguish the signal from the single-photon background is that $e^+ e^-$ colliders are free from the pile-up effect and other contamination from QCD events. With ${\cal L} = 5 ab^{-1}$, there are $384$ signal events left for this benchmark point after all event selections.

For the $e$ALP as a LLP, detecting the novel displaced $J_{\gamma}$ signature is highly dependent on the design of detectors and it is beyond the scope of this paper. Hence, we simply apply cut-(1), cut-(2), cut-(5), cut-(6) in the previous text and assume this kind of displaced $J_{\gamma}$ signature after the above event selections is background-free at CEPC. In this approach, we find there are $38$ signal events left for this benchmark point with ${\cal L} = 5 ab^{-1}$ which are much smaller than numbers of the prompt decay one. This is because only high energy regions of $J_{\gamma}$ can make the $e$ALP to be a LLP for this benchmark point. However, once $\Lambda$ is increasing or $m_a$ is decreasing, the tendency of survival events will be greatly changed.

\subsection{Exploring $e^{-}p\rightarrow\nu_e a j$ at LHeC} 
\label{sec:LHeC}

For $ep$ colliders, there are two proposed experiments, LHeC~\cite{LHeC:2020van} and FCC-he~\cite{FCC:2018byv}. Here we choose the LHeC with $E_e = 50$ GeV and $E_p = 7$ TeV as a practical example to study the $e^{-}p\rightarrow\nu_e a j$ process. Again, we focus on the $a\rightarrow\gamma\gamma$ decay mode and use Madgraph5\underline{\hspace{0.5em}}aMC@NLO for event generation as well as Pythia8 for QCD, QED showering and hadronization effects. The pre-selection cuts ($P_T^{\gamma} > 5$ GeV, $P_T^j > 20$ GeV, $\lvert\eta_{\gamma}\rvert < 4.5$ and $\lvert\eta_j\rvert < 5.0$) at parton-level for both the signal and backgrounds are imposed.
A cone size $R = 0.4$ for the photon isolation criterion at LHeC is considered, and two photons cannot be isolated to each other at detectors when $m_a \lesssim 30$ GeV (parton-level) as shown in the right panel of Fig.~\ref{fig:dR_aa}. Since the produced $e$ALPs at LHeC are more boosted than the ones at CEPC, we can expect the opening angle between two photons from the $e$ALP decay at LHeC is smaller than the one at CEPC. On the other hand, unlike the LHC, the pile-up effect and multiple interactions at LHeC are tiny and can be safely ignored~\cite{Hesari:2018ssq,Antusch:2019eiz,Andre:2022xeh}. This is a key point that the $J_{\gamma}$ candidate will not suffer from serious pile-up pollution at LHeC and it helps us to easily distinguish $J_{\gamma}$ from the single-photon and the QCD jet rather than the situation at LHC. 

\begin{table}[ht!]
\begin{center}\begin{tabular}{|c|c|c|c|c|}\hline cut flow in $\sigma$ [fb] & ~signal~ & ~$\nu_e\gamma\gamma j$~ & ~$\nu_e (h\rightarrow\gamma\gamma)j$~ & ~$\nu_e\gamma jj$~ \\ 
\hline Generator & $4.31$ & $110.89$ & $0.10$ & $1462.10$ \\
\hline cut-(1) & $2.85$ & $13.16$ & $7.91\times 10^{-2}$ & $0.24$ \\
\hline cut-(2) & $2.49$ & $11.82$ & $6.90\times 10^{-2}$ & $0.11$ \\
\hline cut-(3) & $2.28$ & $10.71$ & $5.55\times 10^{-2}$ & $0.10$ \\
\hline cut-(4) & $1.99$ & $5.69$ & $3.18\times 10^{-2}$ & $6.85\times 10^{-2}$ \\ 
\hline cut-(5) & $1.87$ & $3.04$ & $6.93\times 10^{-3}$ & $3.75\times 10^{-2}$ \\
\hline cut-(6) & $1.82$ & $2.82$ & $6.88\times 10^{-3}$ & $3.32\times 10^{-2}$ \\ 
\hline cut-(7) & $1.77$ & $0.22$ & $1.04\times 10^{-6}$ & $2.66\times 10^{-3}$ \\ 
\hline \end{tabular} \caption{The cut-flow table for $e^{-}p\rightarrow\nu_e (a\rightarrow\gamma\gamma) j$ and relevant SM backgrounds with the signature of two isolated photons, a backward jet plus ${\:/\!\!\!\! E_T}$. The benchmark point $m_a = 50$ GeV with $c^A_e / \Lambda = 1$ TeV$^{-1}$ for signal is chosen. All related event selections have been mentioned in the main text.}
\label{tab:CF_ep_1}
\end{center}
\end{table}

For the signature with two isolated photons, a backward jet plus missing transverse energy (${\:/\!\!\!\! E_T}$), the relevant SM backgrounds are $e^{-}p\rightarrow\nu_e\gamma\gamma j$, $e^{-}p\rightarrow\nu_e hj\rightarrow\nu_e (\gamma\gamma)j$ and $e^{-}p\rightarrow\nu_e\gamma jj$ where one of QCD jets fakes to photon in the final process. Here we apply the rate of jet faking to photon as $P_{j\rightarrow\gamma}=5\times 10^{-4}$~\cite{ATLAS:2017muo}. 
We still choose the benchmark point $m_a = 50$ GeV with $c^A_e / \Lambda = 1$ TeV$^{-1}$ for signal-to-background analysis. For the fast detector simulation, the LHeC template in Delphes3 is used and the photon isolation criterion is consistent with Ref.~\cite{Gutierrez-Rodriguez:2020gsi}. The following event selections to identify the signal signature and suppress the background events are required, 
\begin{itemize}
\item (1) $N(\gamma)\geqslant 2$ with $p_T^{\gamma_1} > 30$ GeV, $p_T^{\gamma_2} > 10$ GeV, $-3.0 < \eta_{\gamma_1} < 0.0$ and $-3.5 < \eta_{\gamma_2} < 0.5$, 
\item (2) $N(j)\geqslant 1$ with $p_T^{j_1} > 20$ GeV and $-5.0 < \eta_{j_1} < -1.0$, 
\item (3) ${\:/\!\!\!\! E_T} > 20$ GeV and $2.0 < \eta_{\:/\!\!\!\! E_T} < 5.0$,   
\item (4) $\Delta\phi_{\gamma_1, {\:/\!\!\!\! E_T}} > 1.5$ and $\Delta\phi_{\gamma_2, {\:/\!\!\!\! E_T}} > 1.0$, 
\item (5) $\left({\:/\!\!\!\! E_T}+p_T^{j_1}\right) /M_{\gamma\gamma} > 1.0$, 
\item (6) $\left({\:/\!\!\!\! E_T}+p_T^{j_1}\right) /\left( p_T^{\gamma_1}+p_T^{\gamma_2}\right) < 1.2$, 
\item (7) $\lvert M_{\gamma_1\gamma_1}-m_a\rvert < 3$ GeV.
\end{itemize} 
The cut-flow table for signal and backgrounds from the above event selections is presented in Table.~\ref{tab:CF_ep_1} and some kinematic distributions are shown in Fig.~\ref{fig:ep_50} of Appendix~\ref{app:rec}.

Since $E_p$ is much larger than $E_e$ at LHeC, two isolated photons are distributed in the relatively backward region, ${\:/\!\!\!\! E_T}$ and jet are in the forward and backward regions. $p_T^{\gamma_1} > 30$ GeV, $p_T^{\gamma_2} > 10$ GeV, ${\:/\!\!\!\! E_T} > 20$ GeV and $p_T^{j_1} > 20$ GeV are required as the trigger for this signature. According to the geometric shape of this signal signature and different energy partitions for each object, we choose cut-(4), cut-(5) and cut-(6) to separate the signal from SM backgrounds. Finally, the $e$ALP mass window selection is the most stringent one which kills large parts of SM backgrounds. The most optimistic integrated luminosity of LHeC proposal~\cite{LHeC:2020van}, ${\cal L} = 1 ab^{-1}$, is used in this analysis. After involving all event selections, the signal significance is $Z=71.97$. This means we can explore much smaller $c^A_e / \Lambda$ for $m_a = 50$ GeV at LHeC. 

\begin{table}[ht!]
\begin{center}\begin{tabular}{|c|c|c|c|}\hline cut flow in $\sigma$ [fb] & ~signal~ & ~$\nu_e\gamma j$~ & ~$\nu_e jj$~  \\ 
\hline Generator & $5.47$ & $3596.30$ & $42514.00$ \\ 
\hline $\gamma\beta c\tau_a < 1$ mm & $1.30$ & $-$ & $-$ \\ 
\hline cut-(1) & $0.86$ & $1457.20$ & $326.89$ \\
\hline cut-(2) & $0.83$ & $795.35$ & $84.22$ \\
\hline cut-(3) & $0.75$ & $724.11$ & $79.08$ \\
\hline cut-(4) & $0.67$ & $646.21$ & $56.63$ \\ 
\hline cut-(5) & $0.58$ & $1.00$ & $3.53$ \\
\hline cut-(6) & $0.57$ & $0.38$ & $0.77$ \\
\hline \end{tabular} \caption{Similar to Table.~\ref{tab:CF_ep_1}, but for the signal benchmark point $m_a = 5$ GeV with $c^A_e / \Lambda = 1$ TeV$^{-1}$ and the signature of a $J_{\gamma}$ candidate, a backward jet plus ${\:/\!\!\!\! E_T}$.}
\label{tab:CF_ep_2}
\end{center}
\end{table}

For the $e$ALP prompt decay with a $J_{\gamma}$, a backward jet plus ${\:/\!\!\!\! E_T}$ signature, the relevant SM backgrounds are $e^{-}p\rightarrow\nu_e\gamma j$ and $e^{-}p\rightarrow\nu_e jj$ where the single photon and the QCD jet can mimic $J_{\gamma}$\footnote{Similar to $J_{\gamma}$ analysis at CEPC, the fake photon-jet candidate from instruments is not considered in our LHeC analysis.}. 
Here we also choose the $e$ALP lab frame decay length $\gamma\beta c\tau_a < 1$ mm as a criterion of the prompt decay at LHeC. Again, we choose the benchmark point $m_a = 5$ GeV with $c^A_e / \Lambda = 1$ TeV$^{-1}$ for signal-to-background analysis\footnote{The pre-selection cuts in parton-level are similar as before, except for $P_T^{\gamma} > 1$ GeV ($P_T^{\gamma} > 10$ GeV) for the signal (background).}. We use the same definition of $J_{\gamma}$ as the previous CEPC analysis. As shown before, the N-subjettiness and the jet mass observables for the $J_{\gamma}$ candidate are useful to pick out the signal from the single photon and the QCD jet. We set up event selections to identify the signal signature and suppress the background events in the following :  
\begin{itemize} 
\item (1) $N(J_{\gamma})\geqslant 1$ with $log\theta_J < -2$, $p_T^{J_{\gamma}} > 30$ GeV and $-3.5 < \eta_{\gamma_1} < 0.5$, 
\item (2) $N(j)\geqslant 1$ with $p_T^{j_1} > 20$ GeV and $-5.0 < \eta_{j_1} < -0.5$, 
\item (3) ${\:/\!\!\!\! E_T} > 20$ GeV and $1.5 < \eta_{\:/\!\!\!\! E_T} < 5.0$,  
\item (4) $\left({\:/\!\!\!\! E_T}+p_T^{j_1}\right) / p_T^{J_{\gamma}} < 1.2$,
\item (5) $0.03 < \tau_1 < 0.25$, $\tau_2/\tau_1 < 0.3$, $\tau_3/\tau_1 < 0.2$ and $\tau_3/\tau_2 < 0.4$, 
\item (6) $\lvert M_{J_{\gamma}}-m_a\rvert < 2$ GeV, 
\end{itemize} 
where the definition of the N-subjettiness observables, $\tau_2$ and $\tau_3$, can be found in Ref.~\cite{Thaler:2010tr,Thaler:2011gf}.  
The cut-flow table of signal and background for each event selection are displayed in Table.~\ref{tab:CF_ep_2} and some kinematic distributions are also shown in Fig.~\ref{fig:ep_5} of Appendix~\ref{app:rec}.

First of all, since the $e$ALP at LHeC is more boosted than the one at CEPC, its lab frame decay length can extend to a longer distance as shown in Fig.~\ref{fig:ALP_DecLen} such that the fraction of the $e$ALP prompt decay at LHeC becomes smaller than the one at CEPC. We then applied $J_{\gamma}$ selection criterion to pick out the $J_{\gamma}$ candidate from the backward QCD jet. A $J_{\gamma}$ with $p_T^{J_{\gamma}} > 30$ GeV, a jet with $p_T^{j_1} > 20$ GeV and ${\:/\!\!\!\! E_T} > 20$ GeV are required as the trigger for this signature. The ratio $\left({\:/\!\!\!\! E_T}+p_T^{j_1}\right) / p_T^{J_{\gamma}}$ is used to separate the signal from backgrounds. Furthermore, the powerful jet substructure observables, N-subjettiness and jet mass, are used to further distinguish the signal $J_{\gamma}$ from the single photon and the QCD jet as shown in Fig.~\ref{fig:ep_5}. Note there is still a small possibility for the backward QCD jet to be identified as a $J_{\gamma}$ candidate. On the other hand, unlike the environment in $e^+ e^-$ colliders, there is still some QCD contamination in this process at LHeC, though we don't need to consider the pile-up effect. It indicates SM backgrounds are much suppressed but cannot be ignored. With the help of these event selections, the signal significance can reach $Z=15.76$.

For the $e$ALP as a LLP, we first consider the physical size in radius of proposed detectors on LHeC. According to Ref.~\cite{LHeCStudyGroup:2012zhm}, we can summarize those parts relevant to this study as (1) $31$ mm $\leqslant R_{\text{vertex}}\leqslant 462$ mm, (2) $0.48$ m $\leqslant R_{\text{ECAL}}\leqslant 0.88$ m, (3) $1.2$ m $\leqslant R_{\text{HCAL}}\leqslant 2.6$ m. Therefore, we simply consider the $e$ALP lab frame decay length within $10^{-3}$ m $\leqslant\gamma\beta c\tau_a\leqslant 0.85$ m as a detectable LLP with a displaced $J_{\gamma}$ signature at LHeC. We further require cut-(1), cut-(2), cut-(3), cut-(5), cut-(6) in the previous text and assume this displaced $J_{\gamma}$ signature at LHeC is background-free after involving these event selections. Finally, there are $2047$ survival signal events for this benchmark point with ${\cal L} = 1 ab^{-1}$.

\subsection{Main results and existing bounds} 
\label{sec:Summary}

\begin{figure*}[ht!]
\includegraphics[width=0.8\textwidth]{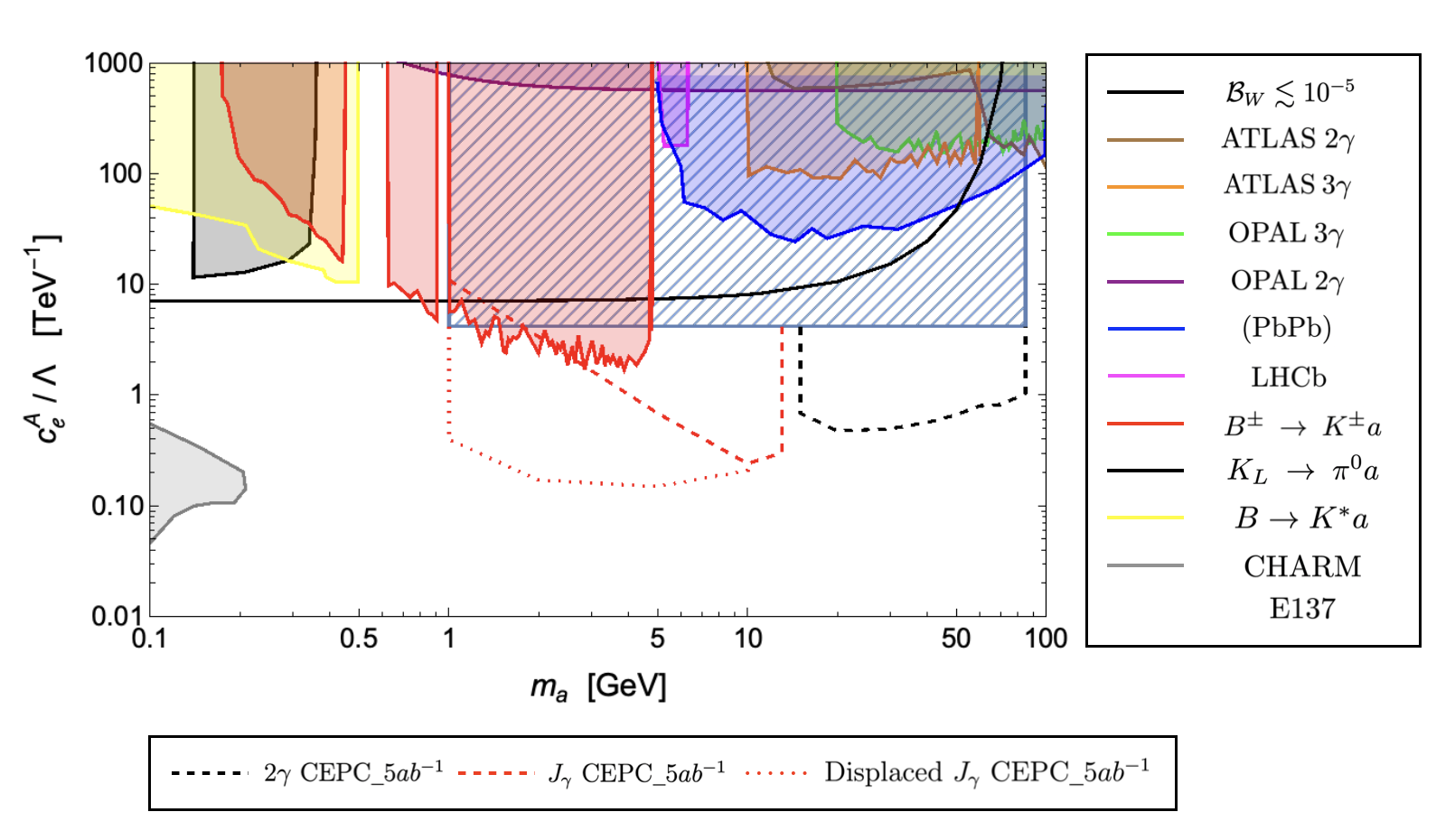}
\caption{The future bounds on $c^A_e/\Lambda$ of $e$ALPs from CEPC with ${\cal L} = 5 ab^{-1}$ within $95\%$ CL or $10$ survival events for background-free cases (dashed lines for the $e$ALP prompt decay and dotted lines for the $e$ALP as a LLP) as well as existing bounds (bulk regions). When adopting the EFT domain of validity with $g_{\ast}\sim O(1)$, the requirement $\frac{\Lambda}{c^A_e}\gtrsim 240$ GeV is indicated by the diagonal hatched region. Here we label "$2\gamma$" and "$J_{\gamma}$" to identify two kinds of signatures at CEPC and LHeC. ${\cal B}_W\lesssim 10^{-5}$ represents ${\cal B} (W^{\pm}\rightarrow\ell^{\pm}\nu a) < 10^{-5}$~\cite{Altmannshofer:2022izm} (solid-black line). Some collider bounds are in order : ATLAS $2\gamma$~\cite{ATLAS:2014jdv,Jaeckel:2015jla,Knapen:2016moh} (brown bulk), ATLAS $3\gamma$~\cite{ATLAS:2015rsn,Knapen:2016moh} (orange bulk), OPAL $3\gamma$~\cite{OPAL:2002vhf,Knapen:2016moh} (green bulk), OPAL $2\gamma$~\cite{OPAL:2002vhf,Knapen:2016moh}, ATLAS/CMS (PbPb)~\cite{dEnterria:2021ljz} (blue bulk) and LHCb~\cite{Benson:2018vya,CidVidal:2018blh} (magenta bulk). For light $e$ALPs, $B^{\pm}\rightarrow K^{\pm}a\rightarrow K^{\pm}(\gamma\gamma)$ from BaBar~\cite{BaBar:2021ich} (red bulk), $K_L\rightarrow\pi^0 a\rightarrow\pi^0 (e^+ e^-)$ from KTeV~\cite{KTeV:2003sls} (black bulk) and $B\rightarrow K^{\ast}a\rightarrow K^{\ast}(e^+ e^-)$ from LHCb~\cite{LHCb:2015ycz,Altmannshofer:2022izm} (yellow bulk) are involved. Finally, we also include the bounds from CHARM~\cite{CHARM:1985anb,Altmannshofer:2022izm} and SLAC E137~\cite{Bjorken:1988as} (gray bulk).}
\label{fig:summary_CEPC}
\end{figure*} 

\begin{figure*}[ht!]
\includegraphics[width=0.8\textwidth]{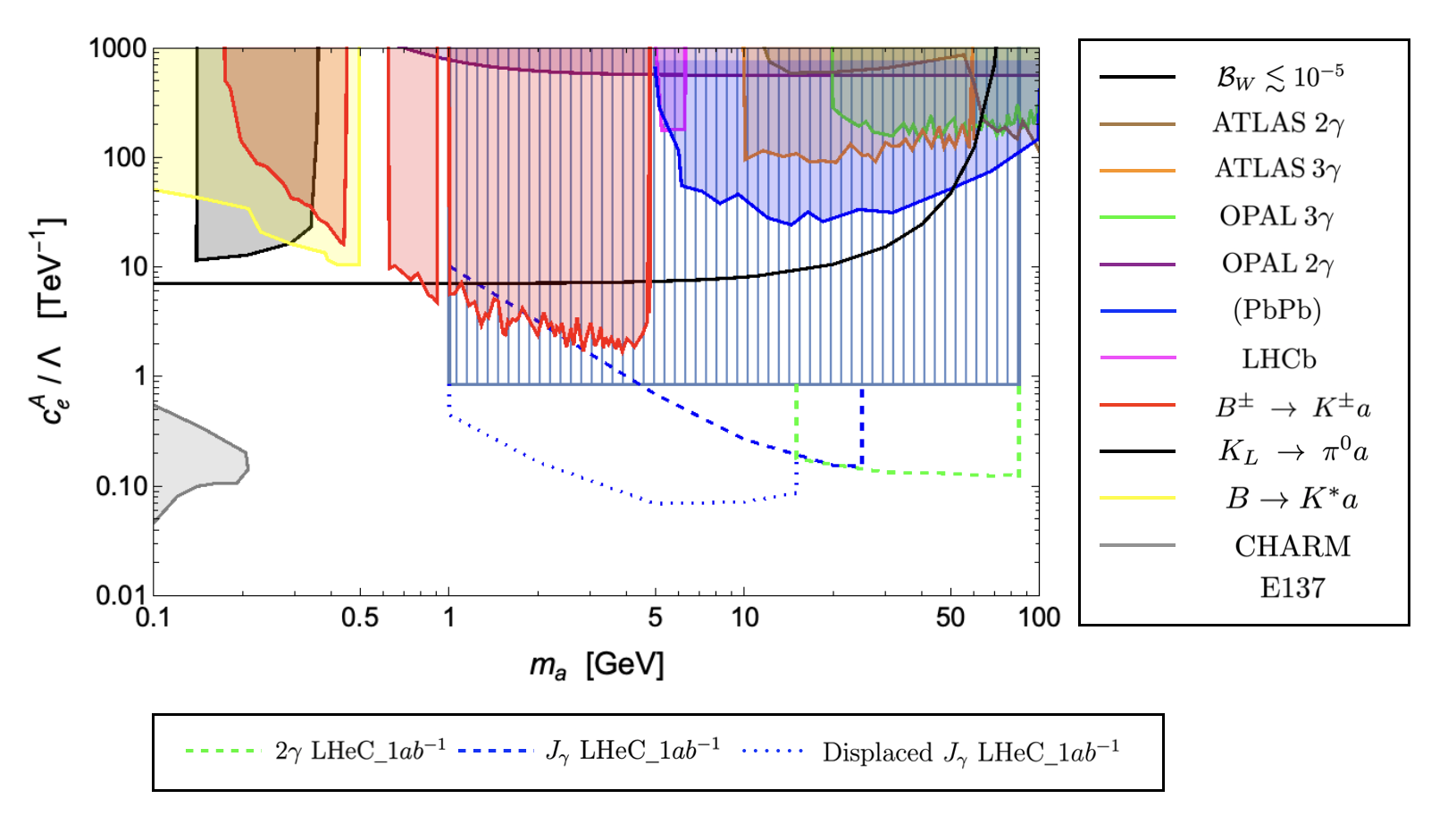}
\caption{The future bounds on $c^A_e/\Lambda$ of $e$ALPs from LHeC with ${\cal L} = 1 ab^{-1}$ within $95\%$ CL or $10$ survival events for background-free cases (dashed lines for the $e$ALP prompt decay and dotted lines for the $e$ALP as a LLP) as well as existing bounds (bulk regions). The requirement $\frac{\Lambda}{c^A_e}\gtrsim 1.18$ TeV based on the EFT domain of validity with $g_{\ast}\sim O(1)$ is indicated by the vertical hatched region. The labels for existing bounds are the same as Fig.~\ref{fig:summary_CEPC}.}
\label{fig:summary_LHeC}
\end{figure*} 

Based on the search strategies in Sec.~\ref{sec:CEPC} and~\ref{sec:LHeC}, we extend the study of signal benchmark points to a wide range of $m_a$ and try to find possible future bounds within $95\%$ confidence level (CL) ($Z=1.96$). On the other hand, in order to conservatively show the signal significance of the case without the survival background event after all event selections or the case of background-free assumption, we require at least $10$ signal events left. Moreover, only signal efficiency larger than $10\%$ after these event selections is considered here. We summarize these possible future bounds from CEPC with ${\cal L} = 5 ab^{-1}$ in Fig.~\ref{fig:summary_CEPC} and LHeC with ${\cal L} = 1 ab^{-1}$ in Fig.~\ref{fig:summary_LHeC}. Note some of our event selections for two isolated photons plus ${\:/\!\!\!\! E}$ at CEPC are sensitive to $m_a$, so we dynamically optimize event selections for different $m_a$ to suit each case. Some benchmark examples are listed in Table.~\ref{tab:other} of Appendix~\ref{app:rec}. We also restrict ourselves to $1$ GeV $\leqslant m_a \lesssim M_W$ for $e$ALPs below the electroweak scale.

The lower bound comes from technical issues of $J_{\gamma}$ analysis. As we have mentioned, the N-subjettiness and jet mass observables are efficient to separate the signal from SM backgrounds. However, when $m_a < 1$ GeV, the above two observables are no longer powerful for discrimination. In this situation, sophisticated jet substructure observables and/or machine learning techniques are required~\cite{Ellis:2012zp,Wang:2021uyb,Ren:2021prq,Ai:2022qvs}, but they are beyond the scope of this work. In addition, when the opening angle between two photons $\Delta R_{\gamma\gamma} \lesssim 0.04$, the trigger at ECAL is hard to distinguish a $J_{\gamma}$ from a single photon since this is close to the size of a standard single photon energy cluster at ECAL~\cite{ATLAS:2018dfo}.
We can clearly find the efficiency of two isolated photons decreases when $m_a\lesssim 20$ GeV because two photons become closer to each other and cannot pass the photon isolation criterion. Similarly, it's hard to group two photons inside a $J_{\gamma}$ candidate for $m_a\gtrsim 20 (30)$ at CEPC (LHeC). Hence, the analysis of signatures with two isolated photons and a $J_{\gamma}$ are complementary to each other for the middle $m_a$ of $e$ALP searches.

Some existing bounds are also shown in both Fig.~\ref{fig:summary_CEPC} and Fig.~\ref{fig:summary_LHeC} for comparison. First of all, the precise $W$ boson width measurements can indirectly test $\ell$ALPs with $m_a < M_W$ in the \textbf{EWV} scenario as pointed out in Ref.~\cite{Altmannshofer:2022izm} and we conservatively choose ${\cal B} (W^{\pm}\rightarrow\ell^{\pm}\nu a) < 10^{-5}$ as a benchmark value and mark it as black-solid line in both Fig.~\ref{fig:summary_CEPC} and Fig.~\ref{fig:summary_LHeC}. For heavier $e$ALPs ($m_a\gtrsim 5$ GeV), the ATLAS $2\gamma$~\cite{ATLAS:2014jdv,Jaeckel:2015jla,Knapen:2016moh} (brown bulk), ATLAS $3\gamma$~\cite{ATLAS:2015rsn,Knapen:2016moh} (orange bulk), OPAL $3\gamma$~\cite{OPAL:2002vhf,Knapen:2016moh} (green bulk), ATLAS/CMS (PbPb)~\cite{dEnterria:2021ljz} (blue bulk) as well as LHCb~\cite{Benson:2018vya,CidVidal:2018blh} (magenta bulk) can already bite some parameter space in the upper-right corner. Among these constraints, the one from ATLAS/CMS (PbPb) is the strongest and extends to a wide $m_a$ range. The bound from OPAL $2\gamma$~\cite{OPAL:2002vhf,Knapen:2016moh} can also extend to $0.5$ GeV $\lesssim m_a\lesssim 100$ GeV, but it's weaker than other constraints. All of the above bounds rely on the $\boldsymbol{aVV'}$ interaction. 
For lighter $e$ALPs ($m_a < 5$ GeV), $B^{\pm}\rightarrow K^{\pm}a\rightarrow K^{\pm}(\gamma\gamma)$ from BaBar~\cite{BaBar:2021ich} (red bulk), $K_L\rightarrow\pi^0 a\rightarrow\pi^0 (e^+ e^-)$ from KTeV~\cite{KTeV:2003sls} (black bulk) as well as $B\rightarrow K^{\ast}a\rightarrow K^{\ast}(e^+ e^-)$ from LHCb~\cite{LHCb:2015ycz,Altmannshofer:2022izm} (yellow bulk) can already exclude large parameter space. These strong constraints come from FCNC processes with $\boldsymbol{aVV'}$ interaction. Finally, we also show the bounds from CHARM~\cite{CHARM:1985anb,Altmannshofer:2022izm} and SLAC E137~\cite{Bjorken:1988as} (gray bulk) which come from the $\boldsymbol{a\ell\ell}$ interaction in the left-bottom region of both Fig.~\ref{fig:summary_CEPC} and Fig.~\ref{fig:summary_LHeC}. Note that all of the above bounds have been rescaled according to our definition of ALP-letpon interactions in Eq.~(\ref{eq:int}) and $e$ALP decay branching ratios in Fig.~\ref{fig:lALP_BR}. Some other bounds from Belle II~\cite{Belle-II:2020jti}, $K\rightarrow\pi a\rightarrow\pi (\gamma\gamma)$~\cite{Izaguirre:2016dfi}, and so on are so weak that we don't include them here.

We examine the EFT domain of validity for the four-point interaction, $\boldsymbol{aW\ell\nu}$, at CEPC and LHeC. Consistent with previous approaches in DM EFT searches in collider experiments~\cite{Racco:2015dxa,DeSimone:2016fbz}, we impose the condition $\sqrt{s} < M_{\text{cut}}=g_{\ast}M_{\ast}$. Here, $M_{\text{cut}}$ denotes the EFT cutoff scale, $g_{\ast}$ represents the coupling strength of the relevant new physics theory, and $M_{\ast}$ is the typical EFT interaction scale. It is important to note that apart from the requirement that $g_{\ast} < 4\pi$ to maintain perturbativity, we may not possess precise information regarding the specific value of $g_{\ast}$. This value could potentially be quite small, resulting in a significantly higher energy scale for $M_{\ast}$ at a fixed $\sqrt{s}$ in collider experiments. However, for order one couplings, the above condition can be translated into the requirement:
\begin{equation}
\sqrt{s}\lesssim\frac{\Lambda}{c^A_e}.
\label{eq:EFT_valid}
\end{equation}
For the case of CEPC ($\sqrt{s} = 240$ GeV), our anticipated future CEPC bounds comfortably satisfy $\frac{\Lambda}{c^A_e}\gtrsim 240$ GeV. We have highlighted this as the diagonal hatched region in Fig.~\ref{fig:summary_CEPC} to emphasize that this region may remain unexcluded at CEPC when adopting an EFT approach with $g_{\ast}\sim O(1)$. Similarly, for the case of LHeC ($\sqrt{s}\lesssim 2\sqrt{E_e E_p}\approx 1.18$ TeV), our expected future LHeC bounds comfortably fulfill $\frac{\Lambda}{c^A_e}\gtrsim 1.18$ TeV, as indicated by the vertical hatched region in Fig.~\ref{fig:summary_LHeC}. A significant portion of the anticipated LHeC bounds may not fall within the exclusion criteria for the EFT domain of validity, assuming $g_{\ast}\sim O(1)$\footnote{If the four-point interaction, $\boldsymbol{aW\ell\nu}$, originates from dimension-7 operators, such as $\partial_{\mu}a\left(\overline{HL}\right)\gamma^{\mu}\left(HL\right)$ and $\partial_{\mu}a\left(\overline{H^{\dagger}L}\right)\gamma^{\mu}\left(H^{\dagger}L\right)$ discussed in Sec.~\ref{sec:rev}, the EFT domain of validity in Eq.~(\ref{eq:EFT_valid}) needs to be modified to $\sqrt{s}\lesssim (\frac{\Lambda}{v})^2\frac{\Lambda}{c^A_e}$, where $v$ corresponds to the electroweak scale. Here we have omitted the order one coupling factors in this expression.}.

On the other hand, we observed that the efficiencies of the same event selections applied to both $e^{+}e^{-}\rightarrow\nu_e a\overline{\nu_e}$ and $e^{-}p\rightarrow\nu_e a j$ in \textbf{EWP} scenario are reduced. This reduction occurs because only Feynman diagrams involving $\boldsymbol{a\ell\ell}$ and/or $\boldsymbol{aVV'}$ interactions are included in this scenario. The kinematic distributions resulting from these Feynman diagrams differ from those of the $\boldsymbol{aW\ell\nu}$ interaction. Moreover, the production cross section of $e^{+}e^{-}\rightarrow\nu_e a\overline{\nu_e}$ ($e^{-}p\rightarrow\nu_e a j$) in \textbf{EWP} scenario is approximately four (nine) orders of magnitude smaller than in \textbf{EWV} scenario. Consequently, the expected future bounds for the same processes of $e$ALPs in \textbf{EWP} scenario do not meet the criteria for the EFT domain of validity. Therefore, for this analysis, we must resort to the UV-complete model approach instead of the EFT approach in \textbf{EWP} scenario.

All in all, compared with these existing bounds, our proposals to search for $e$ALPs via $e^{+}e^{-}\rightarrow\nu_e a\overline{\nu_e}$ and $e^{-}p\rightarrow\nu_e a j$ at $e^+ e^-$ and $ep$ colliders are still attractive and much stronger than some existing bounds. The possible future bounds of $c^A_e/\Lambda$ can reach to less than about $0.1-1.0$ TeV$^{-1}$ which open a new door to explore $e$ALPs in \textbf{EWV} scenario below the electroweak scale.

\section{Discussions and Conclusion}
\label{sec:con} 

The ALP is a well-motivated, postulated new particle beyond the SM. Searching for couplings among ALPs and various SM particles in different ALP mass ranges is an important mission to explore its particle nature. For ALP-lepton interactions, the recent work in Ref.~\cite{Altmannshofer:2022izm} has shown the less discussed four-point interaction, $W$-$\ell$-$\nu$-$a$, in electroweak-violating (EWV) scenario plays an important role to explore leptophilic ALPs for some charged current interaction processes. Especially, for concreteness, they have applied this kind of interaction to search for electronphilic ALPs ($e$ALPs) from $\pi^{\pm}$, $K^{\pm}$ mesons and $W$ boson decays and find the novel energy enhancement effect in these decay modes can largely increase the $e$ALP production rate.

In this work, we further explore heavier $e$ALPs via $W$-$\ell$-$\nu$-$a$ four-point interaction in collider experiments. New t-channel processes, $e^{+}e^{-}\rightarrow\nu_e a\overline{\nu_e}$ and $e^{-}p\rightarrow\nu_e a j$ to search for $e$ALPs at $e^+ e^-$ and $ep$ colliders are proposed. In the EWV scenario, there are obvious energy enhancement behaviors in production cross sections for these two processes as shown in Fig.~\ref{fig:ALP_ee_Xsec} and~\ref{fig:ALP_ep_Xsec} such that they can be used as new channels not only to search for $e$ALPs but also to distinguish EWV ALP-lepton interactions from electroweak-preserving ones at high energy colliders.

For $m_a\gtrsim 1$ GeV, the $e$ALPs dominantly decay to a photon pair induced by the chiral anomaly as shown in Fig.~\ref{fig:lALP_BR}. It is noteworthy that two photons from the $e$ALP decay at colliders may be too collimated to pass the photon isolation criterion when $e$ALP is highly boosted. We apply photon-jet analysis to this situation. Therefore, we study two kinds of signal signatures, two isolated photons and a photon-jet, at $e^+ e^-$ and $ep$ colliders. Taking CEPC with ${\cal L} = 5 ab^{-1}$ and LHeC with ${\cal L} = 1 ab^{-1}$ as two examples, we find the possible future bounds of $c^A_e/\Lambda$ can be lower than about $0.1-1.0$ TeV$^{-1}$ for $1$ GeV $\leqslant m_a\lesssim M_W$ which is much stronger than existing bounds as shown in both Fig.~\ref{fig:summary_CEPC} and Fig.~\ref{fig:summary_LHeC}.

Before closing, we discussed some possible extensions of this work. First, we don't consider the polarized $e^+$, $e^-$ at $e^+ e^-$ and $ep$ colliders. The signal significance should be enhanced once the specific $e^+$, $e^-$ polarization is chosen. Moreover, boosted decision tree (BDT)~\cite{Roe:2004na}, convolutional neural network (CNN)~\cite{Ayyar:2020ijy} and other advanced machine learning techniques can further improve the simple analysis here, but the main features to explore these two new $e$ALP production channels have been emphasized in this work. On the other hand, for light $e$ALPs with $c^A_e/\Lambda\lesssim 1$ TeV$^{-1}$, they become long-lived particles inside colliders. A more elaborate detector simulation and analysis for this novel displaced photon-jet signature beyond this work is needed. Finally, the extension to muonphilic ALPs at $\mu^+\mu^-$ colliders is left for another work~\cite{Lu:2023ryd}.

\acknowledgments{We thank Lei Wu, Shu-Yu Ho, Zeren Simon Wang and Yue-Lin Sming Tsai for helpful discussions.
}

\appendix

\section{Some kinematic distributions and supplemental information}
\label{app:rec}

\begin{figure*}[ht!]
\includegraphics[width=0.35\textwidth]{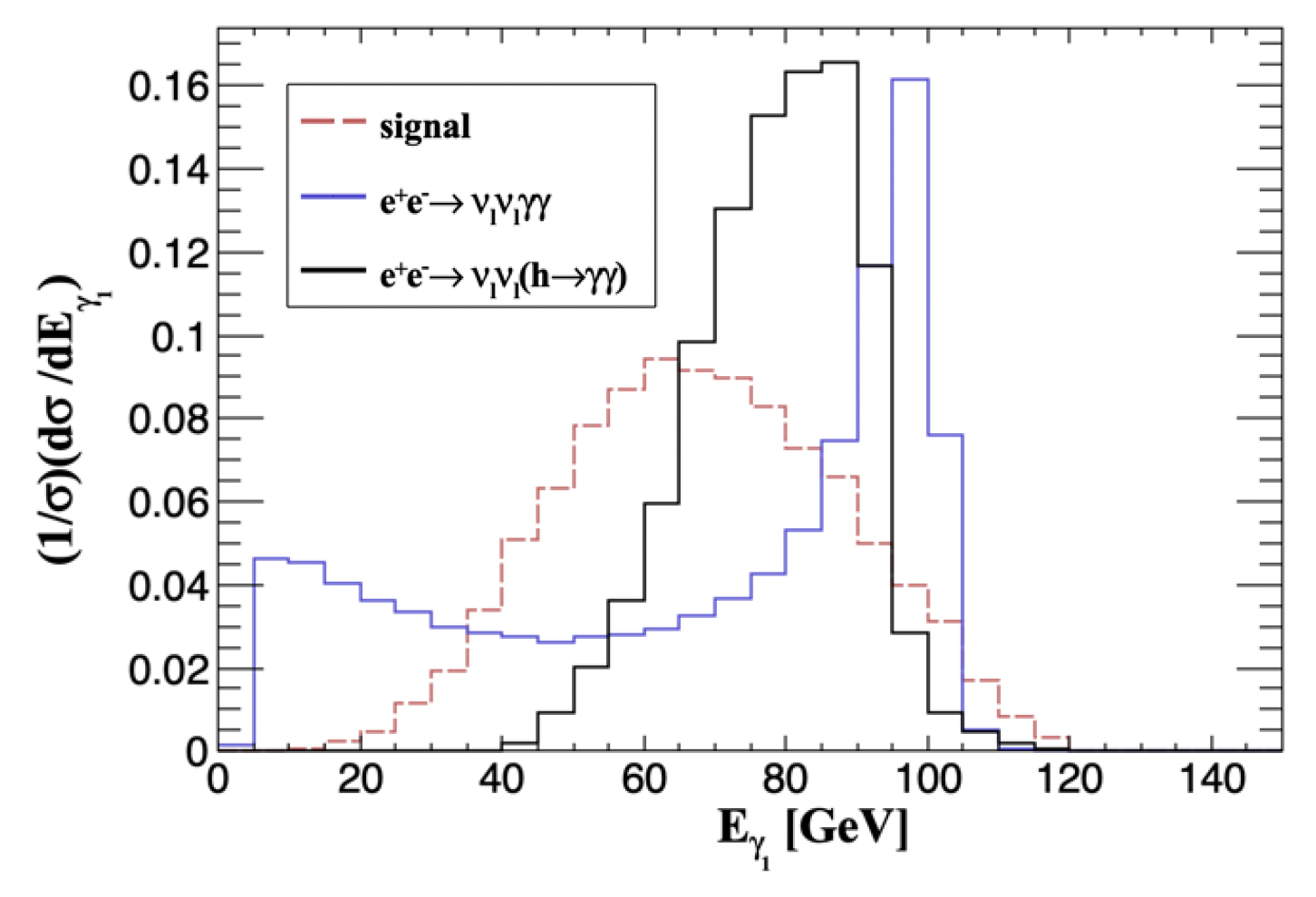}\qquad
\includegraphics[width=0.35\textwidth]{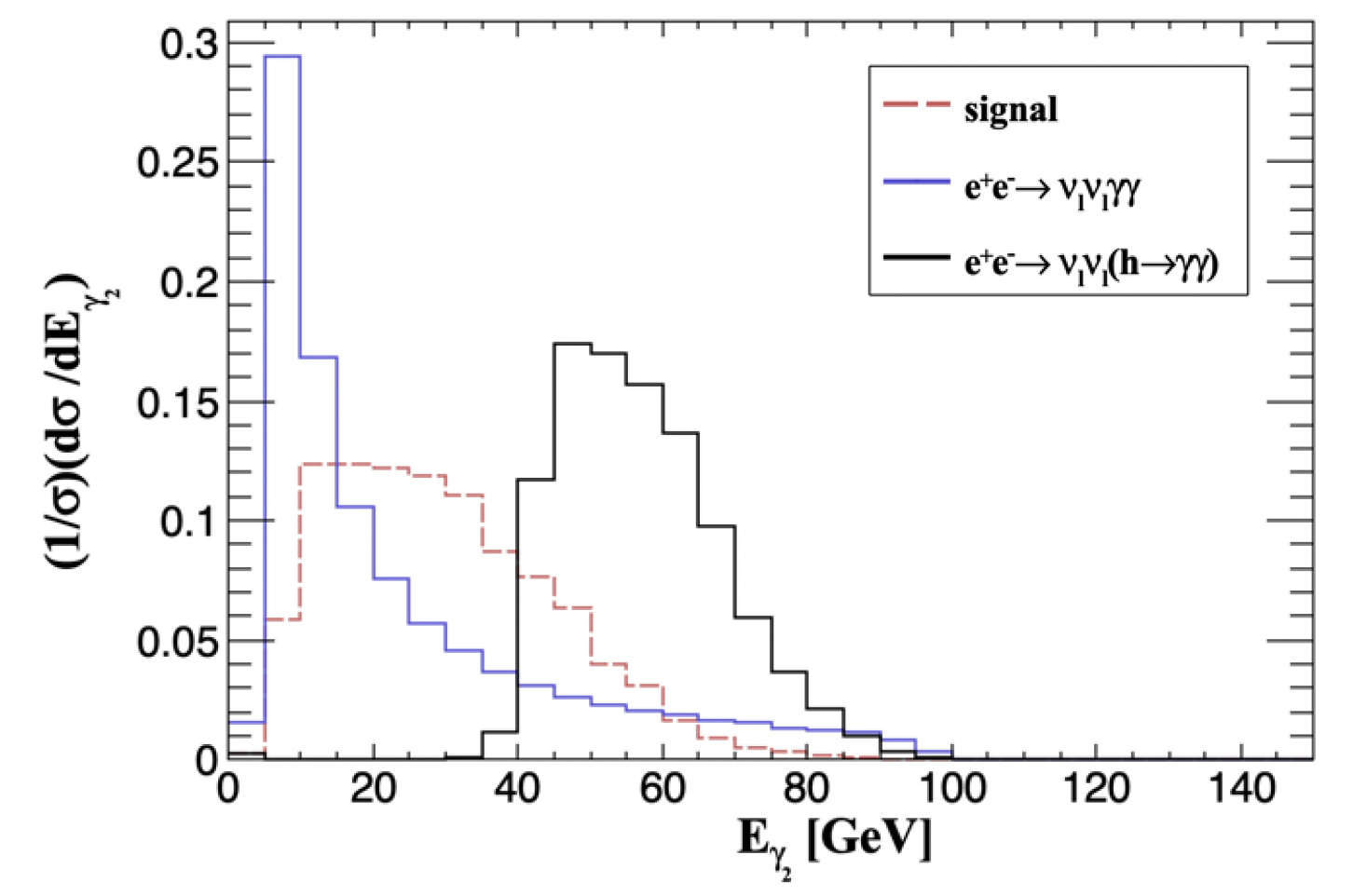}\qquad
\includegraphics[width=0.35\textwidth]{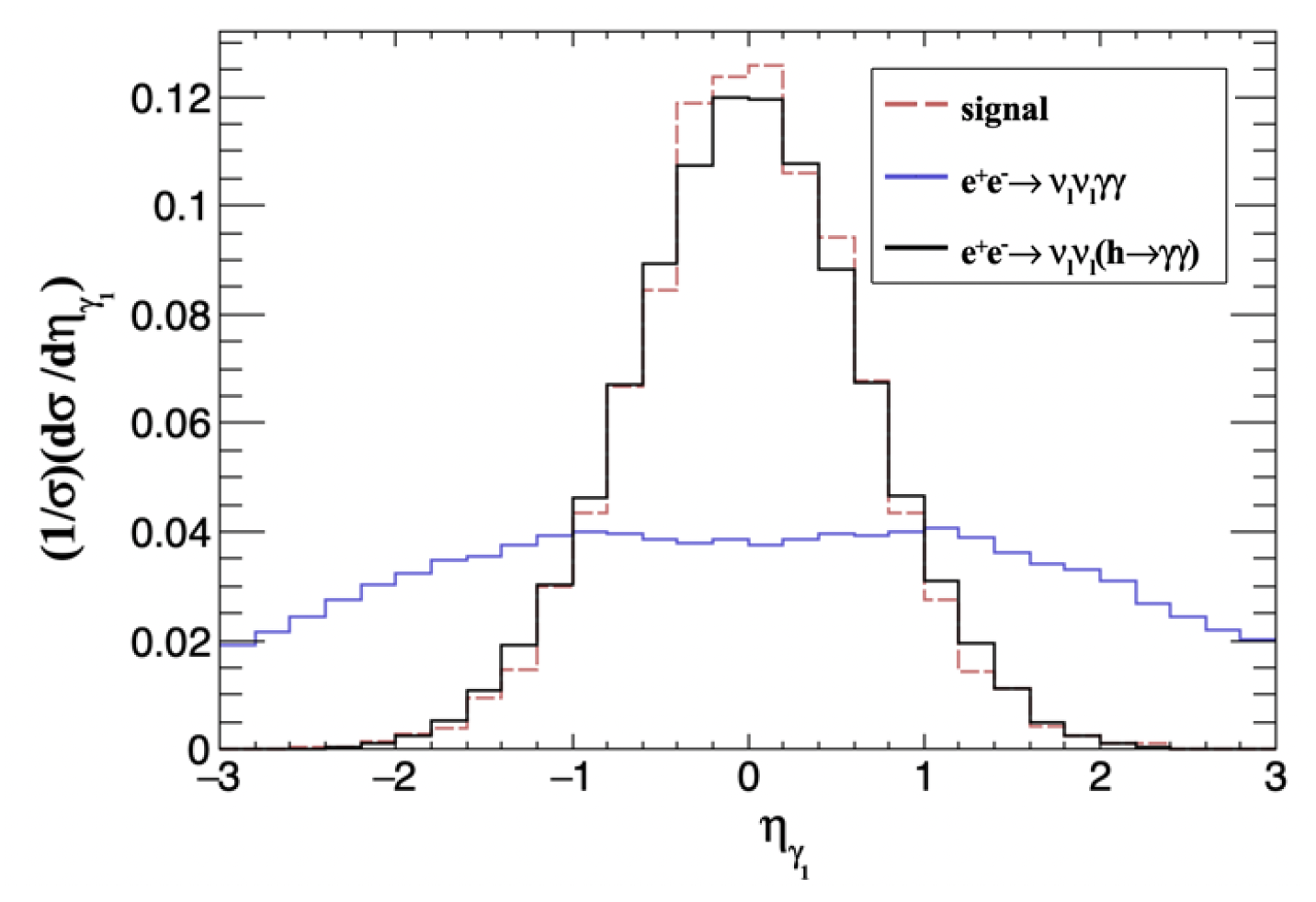}\qquad
\includegraphics[width=0.35\textwidth]{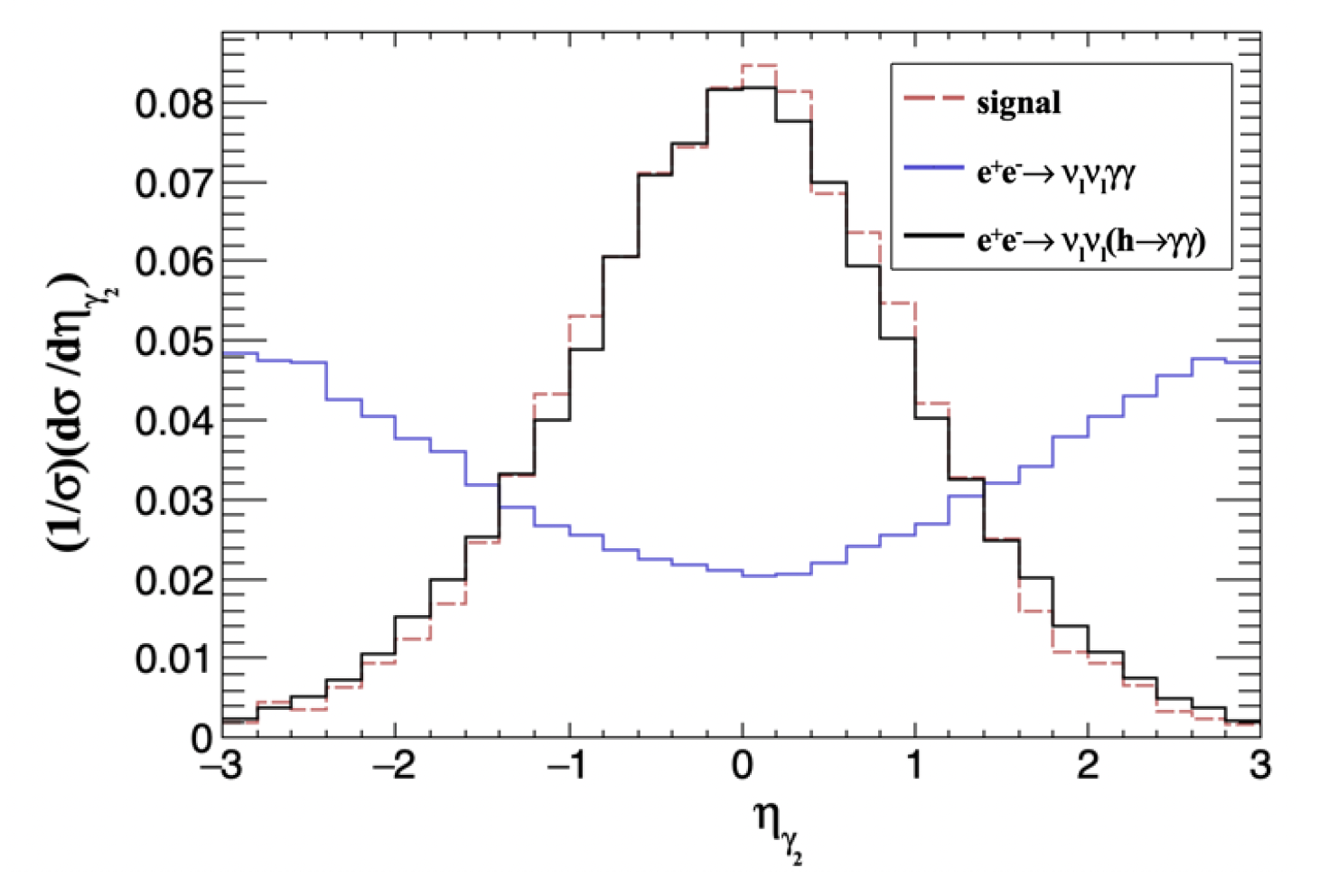}\qquad
\includegraphics[width=0.35\textwidth]{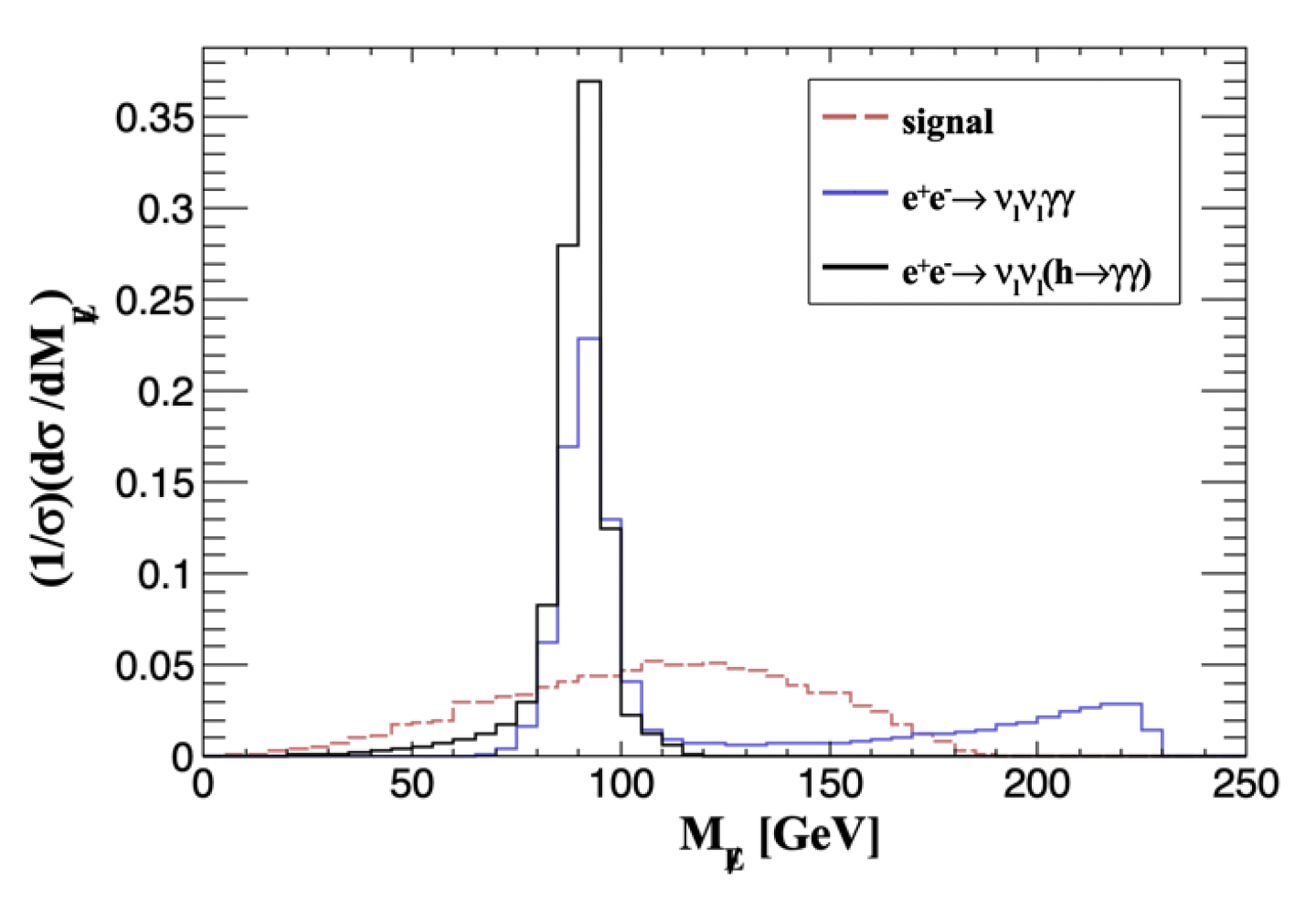}\qquad
\includegraphics[width=0.35\textwidth]{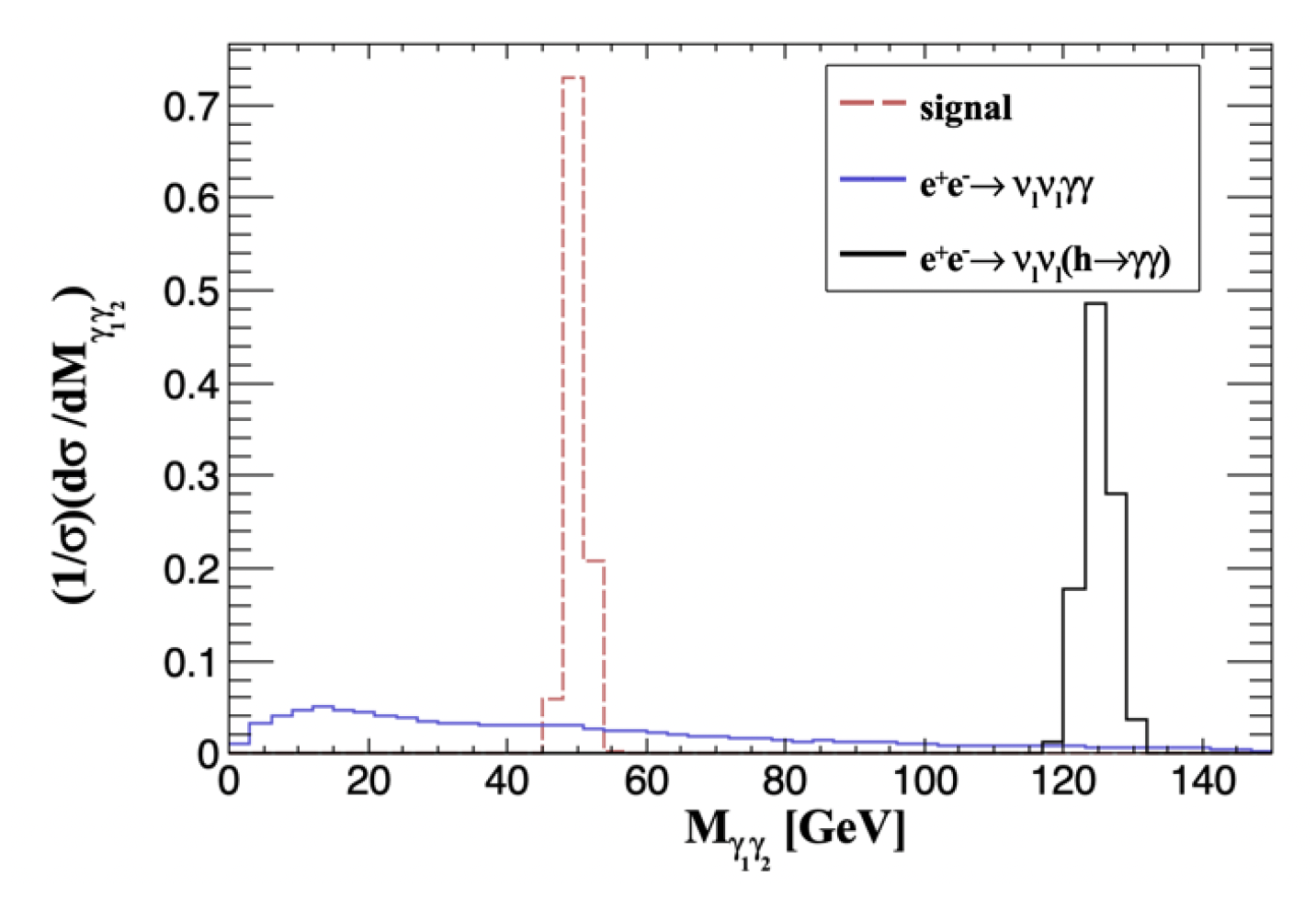}\qquad
\includegraphics[width=0.35\textwidth]{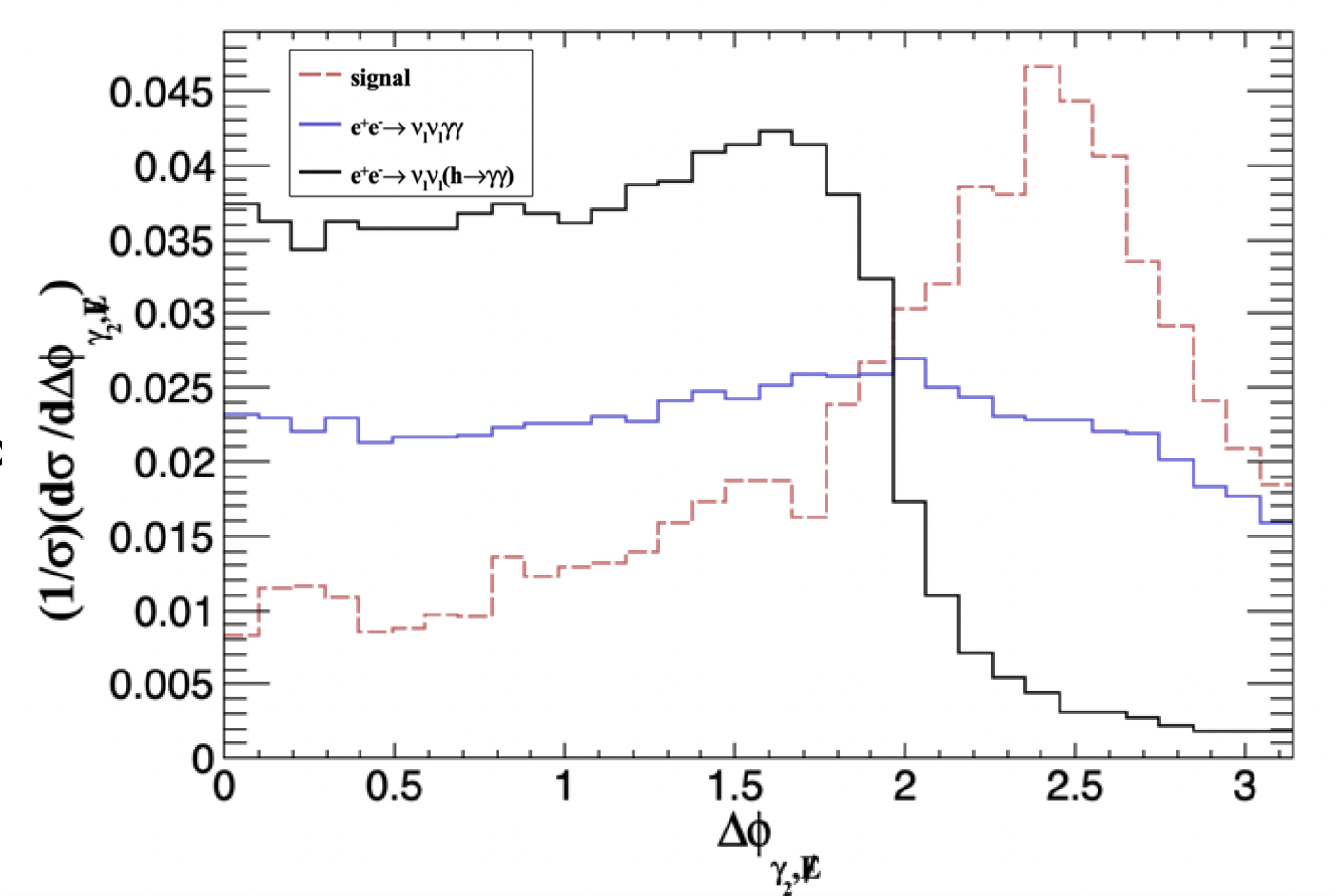}\qquad
\includegraphics[width=0.35\textwidth]{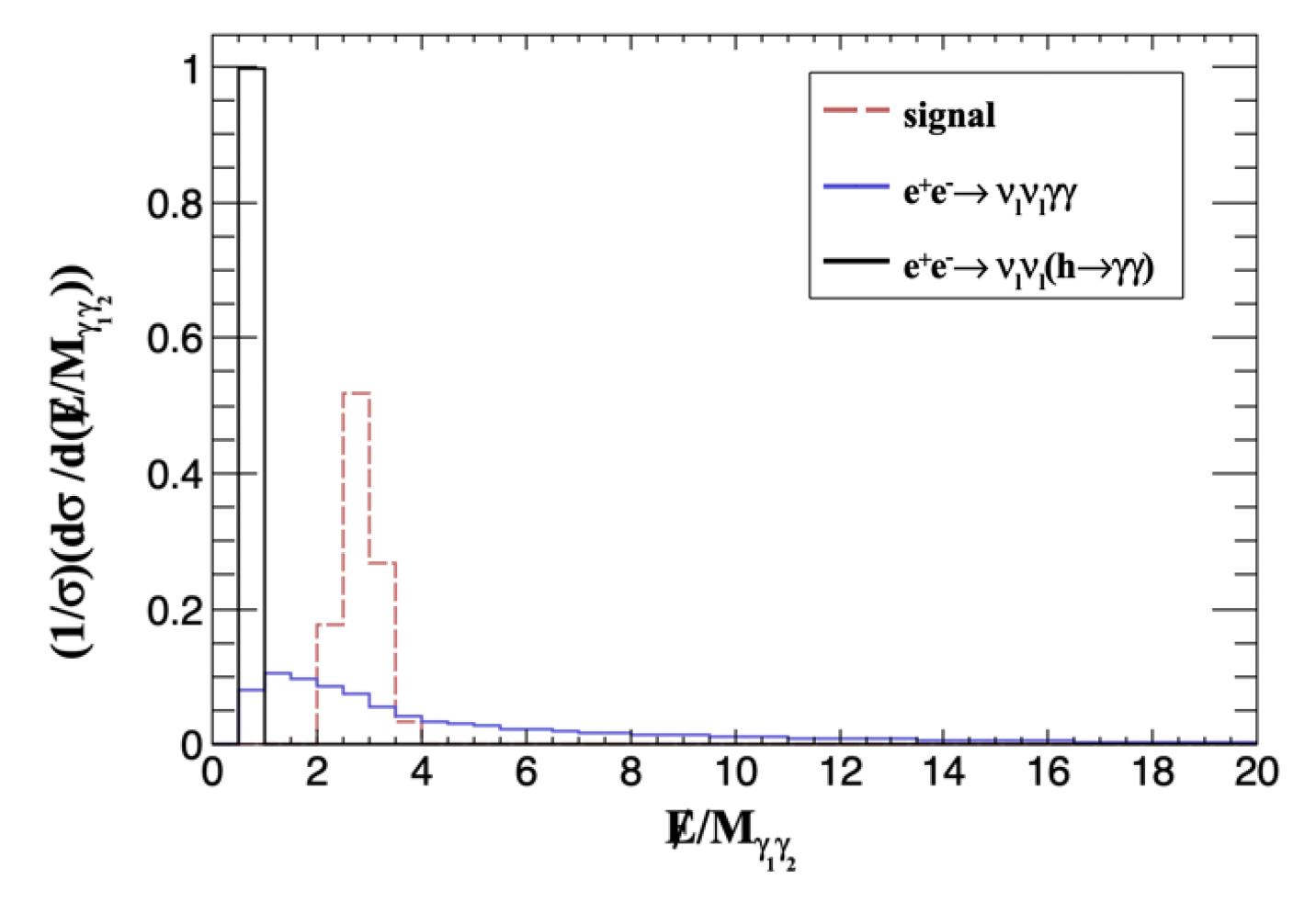}
\caption{Some signal and background kinematic distributions for the signature of two isolated photons plus ${\:/\!\!\!\! E}$ at CEPC for $m_a = 50$ GeV with $c^A_e / \Lambda = 1$ TeV$^{-1}$.}
\label{fig:ee_50}
\end{figure*} 

\begin{figure*}[ht!]
\includegraphics[width=0.35\textwidth]{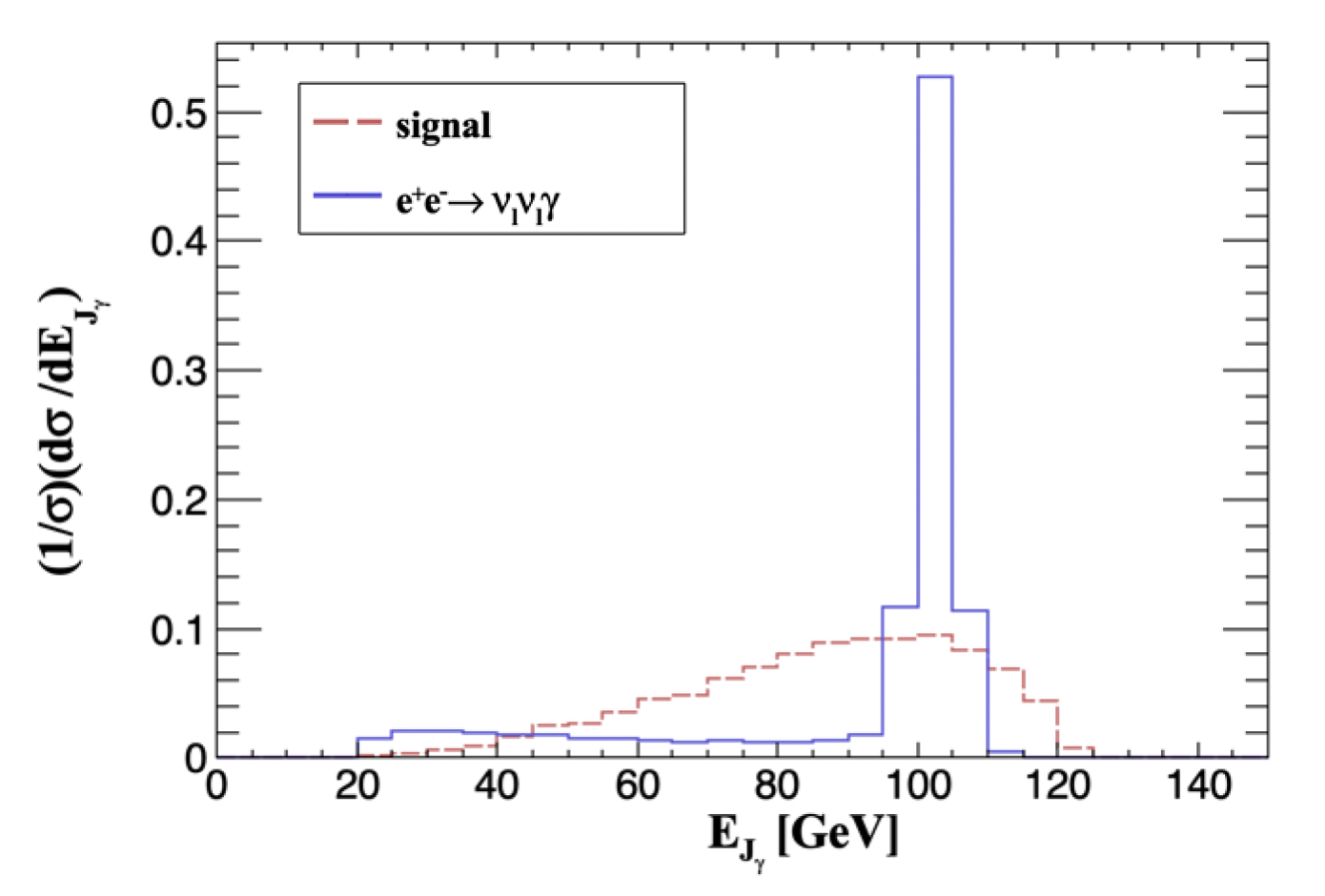}\qquad
\includegraphics[width=0.35\textwidth]{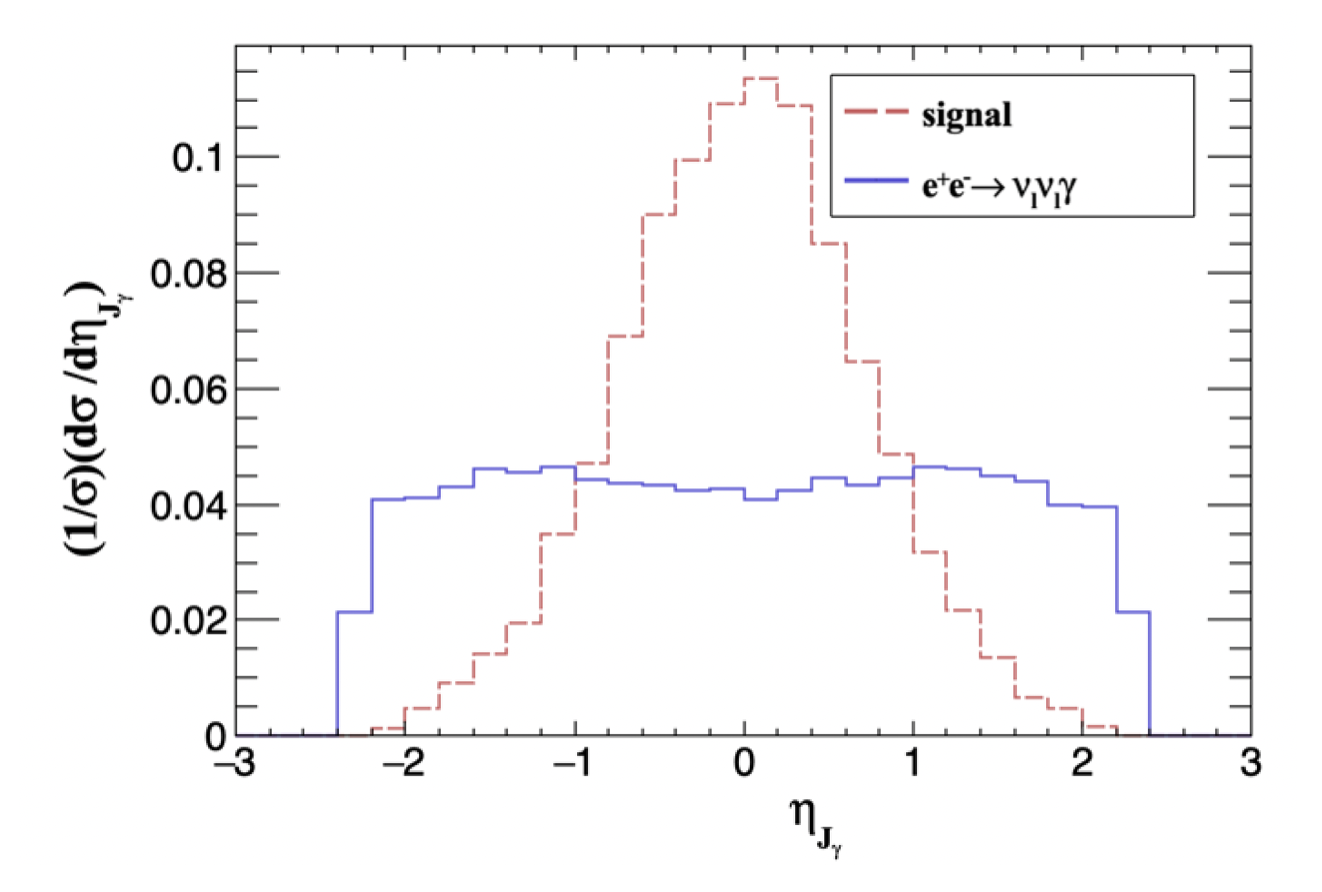}\qquad
\includegraphics[width=0.35\textwidth]{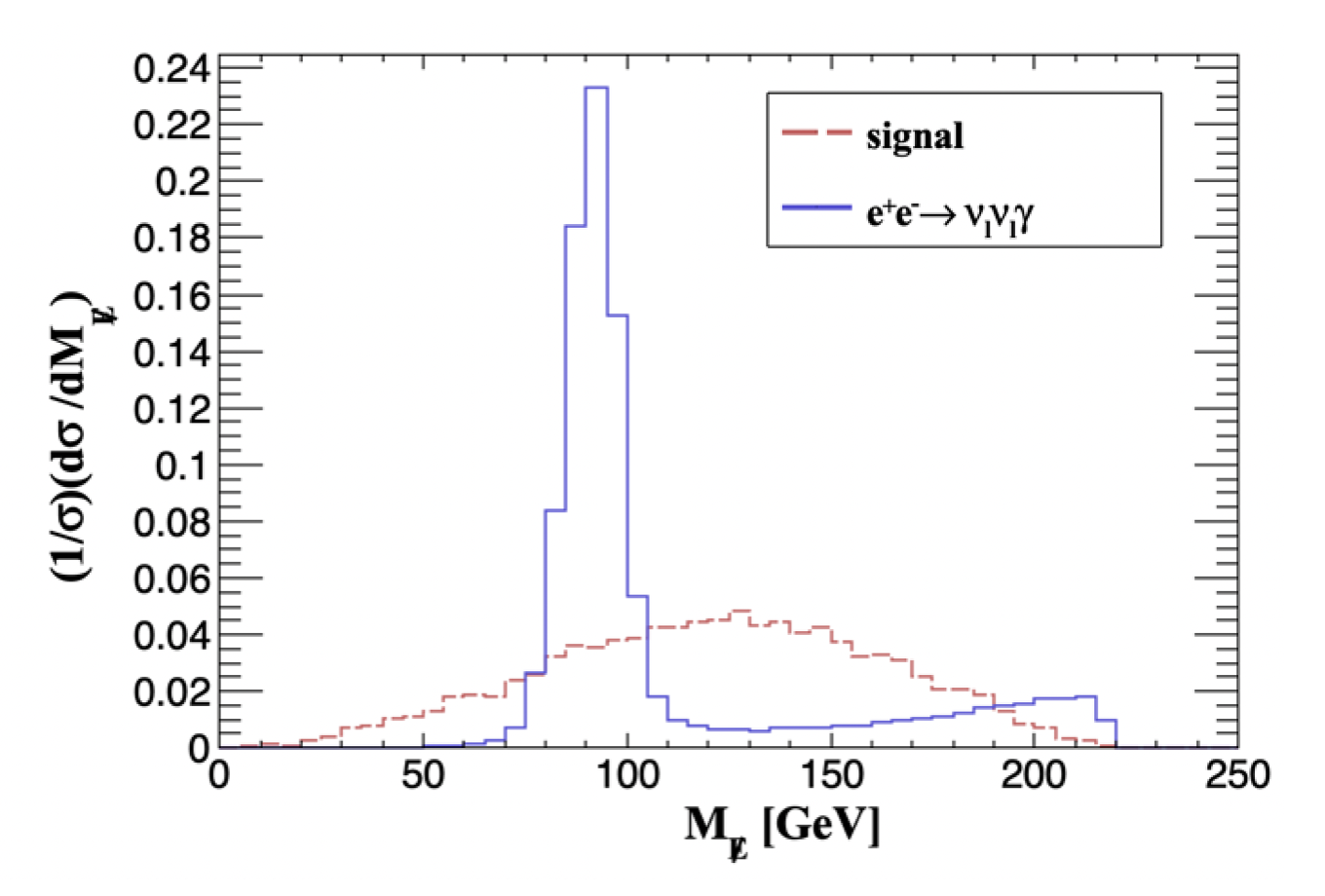}\qquad
\includegraphics[width=0.35\textwidth]{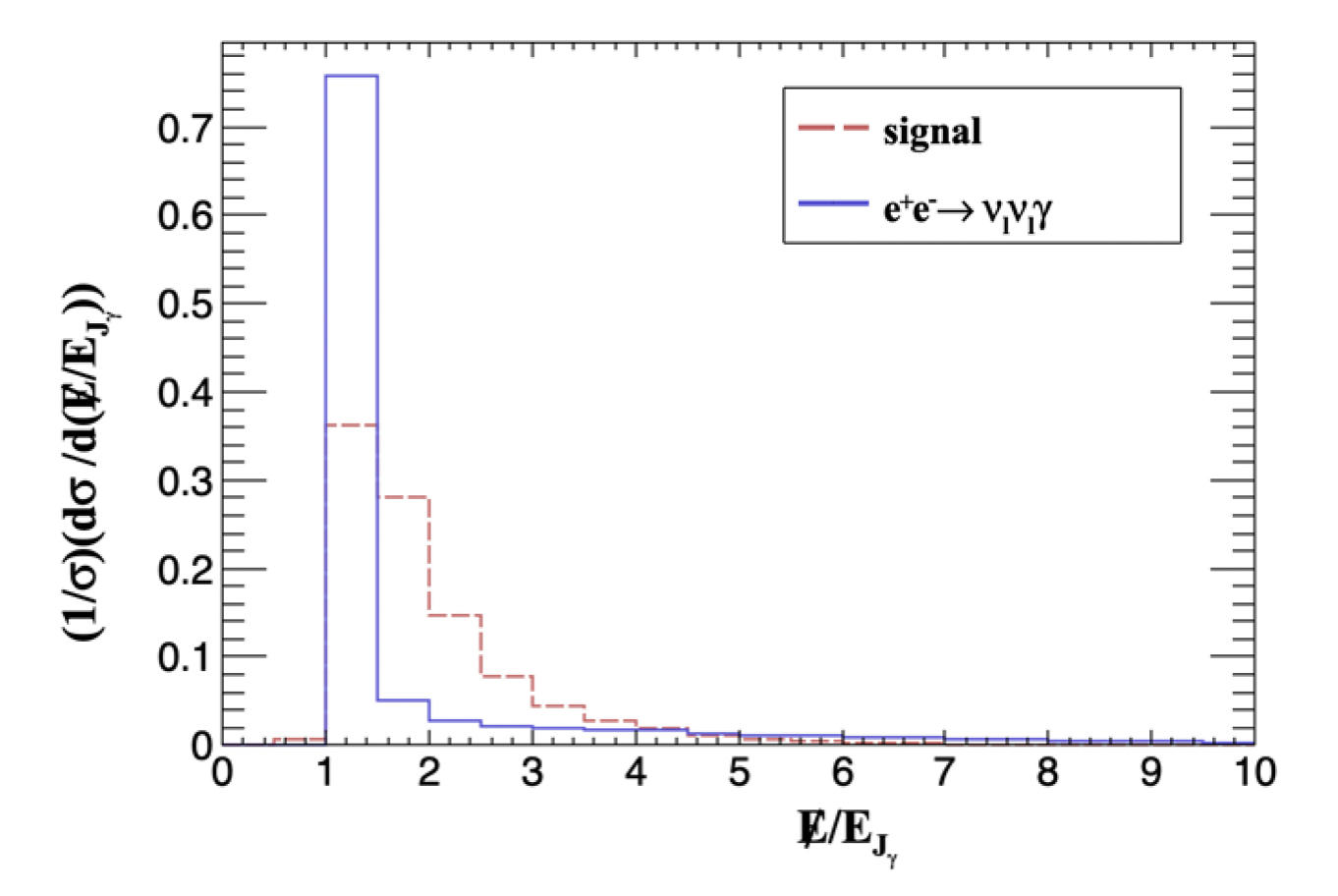}\qquad
\includegraphics[width=0.35\textwidth]{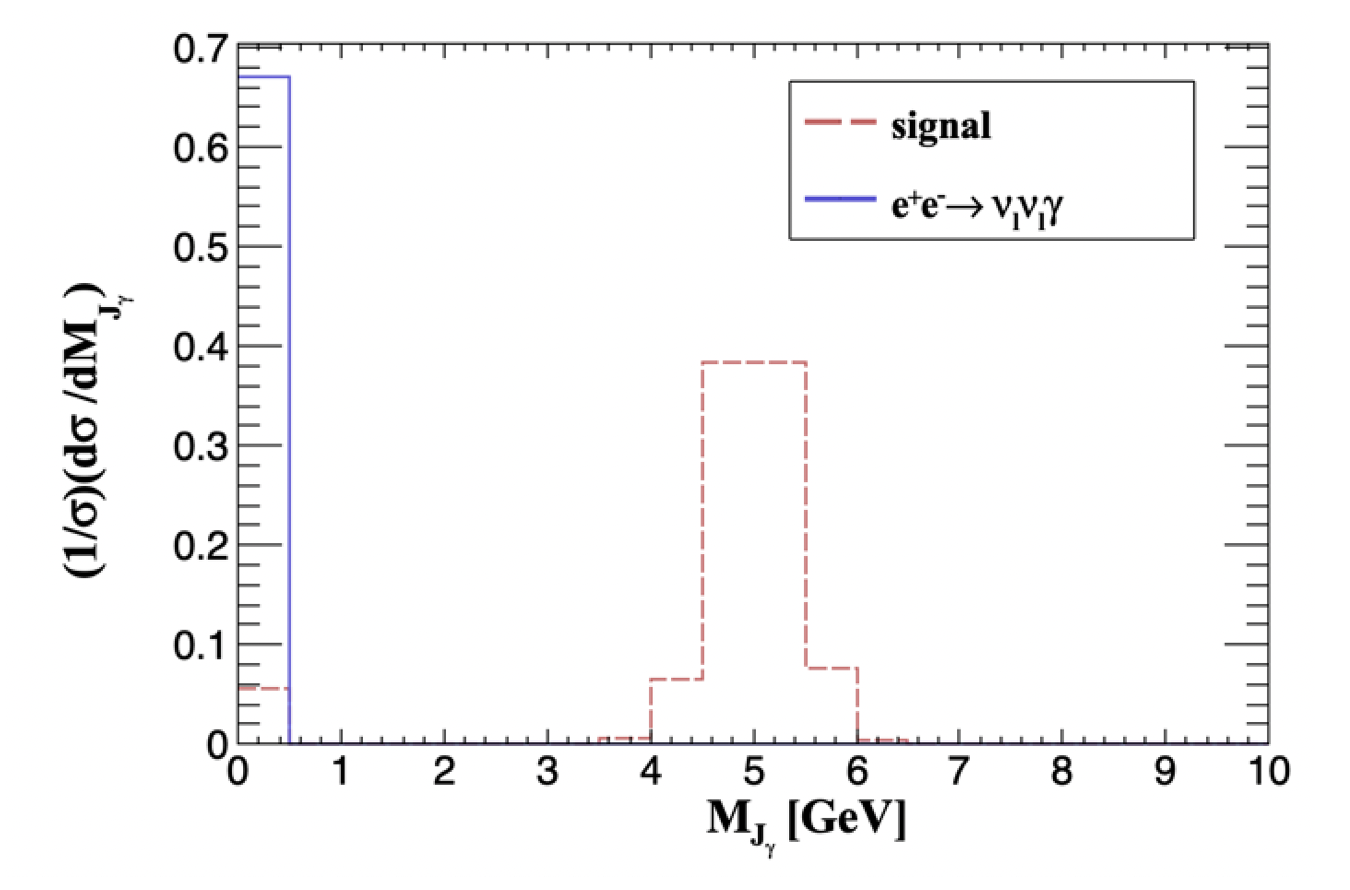}\qquad
\includegraphics[width=0.35\textwidth]{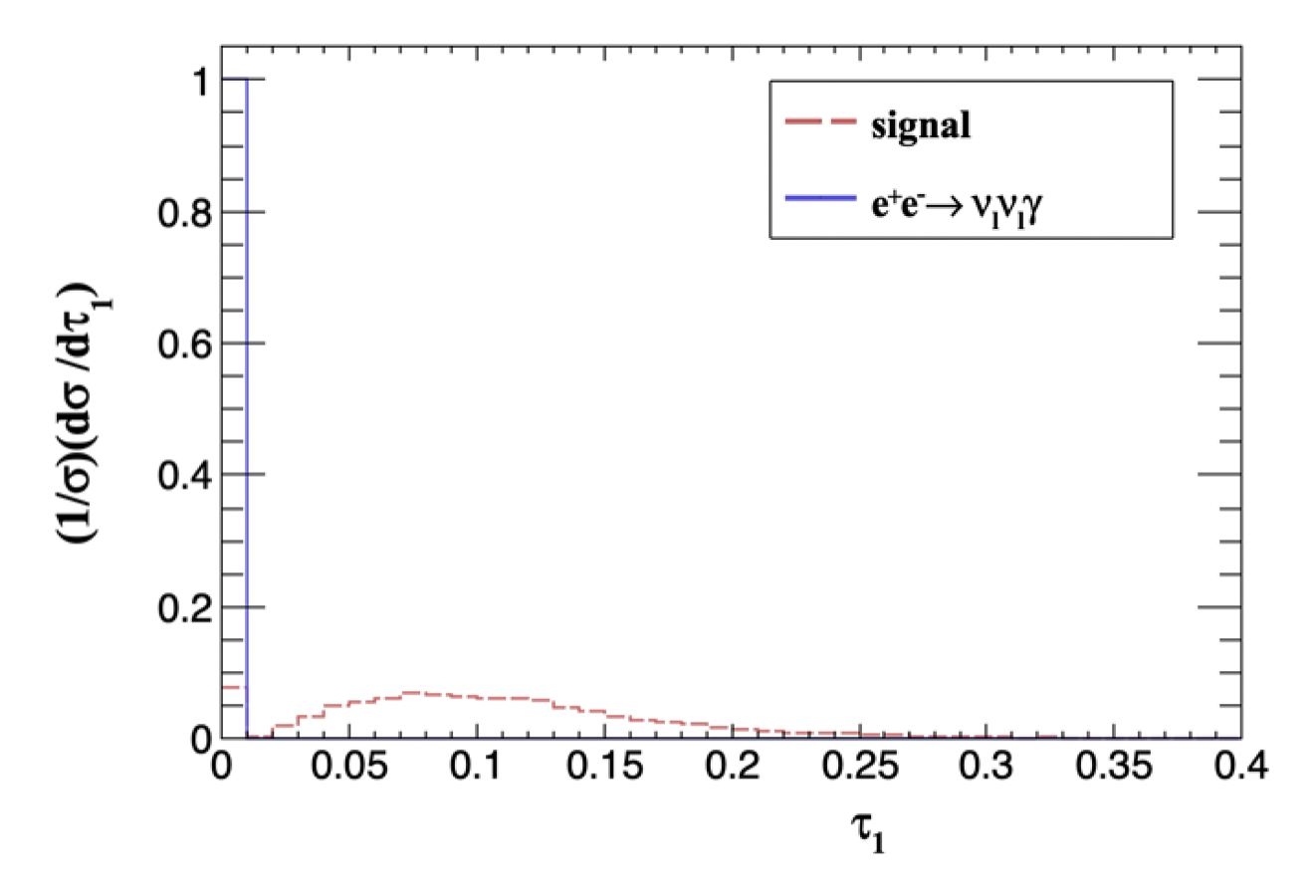}
\caption{Some signal and background kinematic distributions for the signature of a $J_{\gamma}$ plus ${\:/\!\!\!\! E}$  at CEPC for $m_a = 5$ GeV with $c^A_e / \Lambda = 1$ TeV$^{-1}$.}
\label{fig:ee_5}
\end{figure*} 

\begin{figure*}[ht!]
\includegraphics[width=0.35\textwidth]{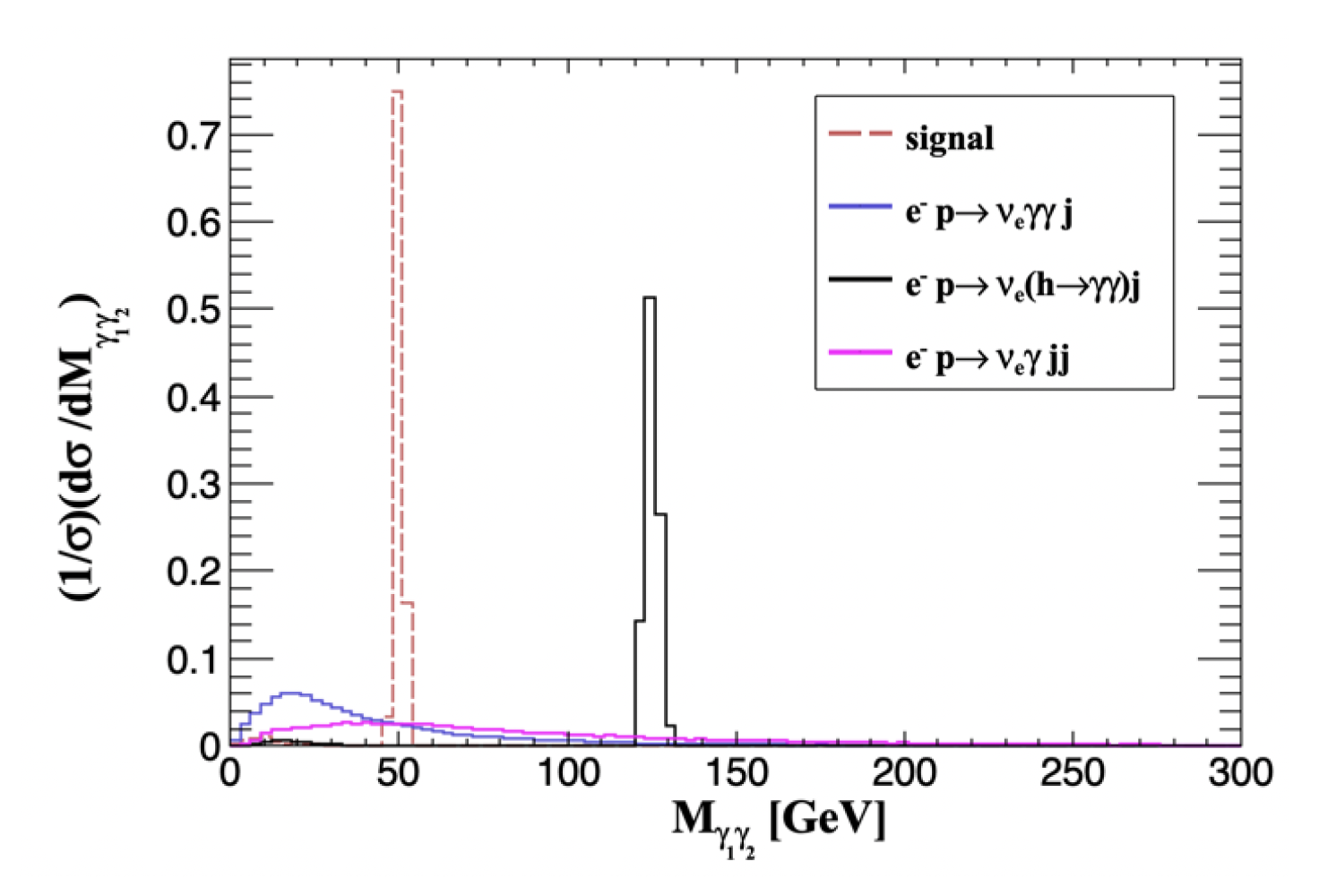}\qquad
\includegraphics[width=0.35\textwidth]{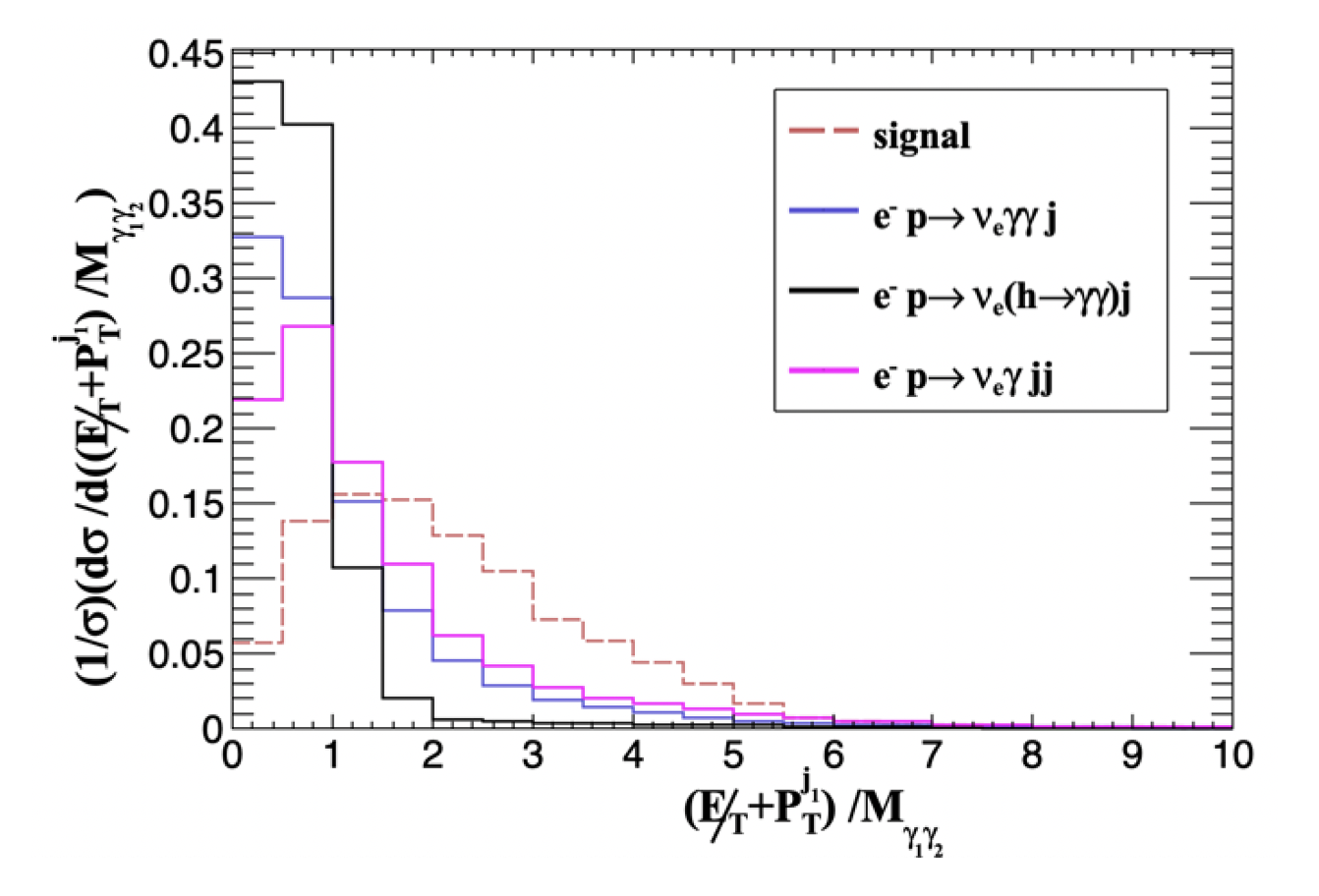}\qquad
\includegraphics[width=0.35\textwidth]{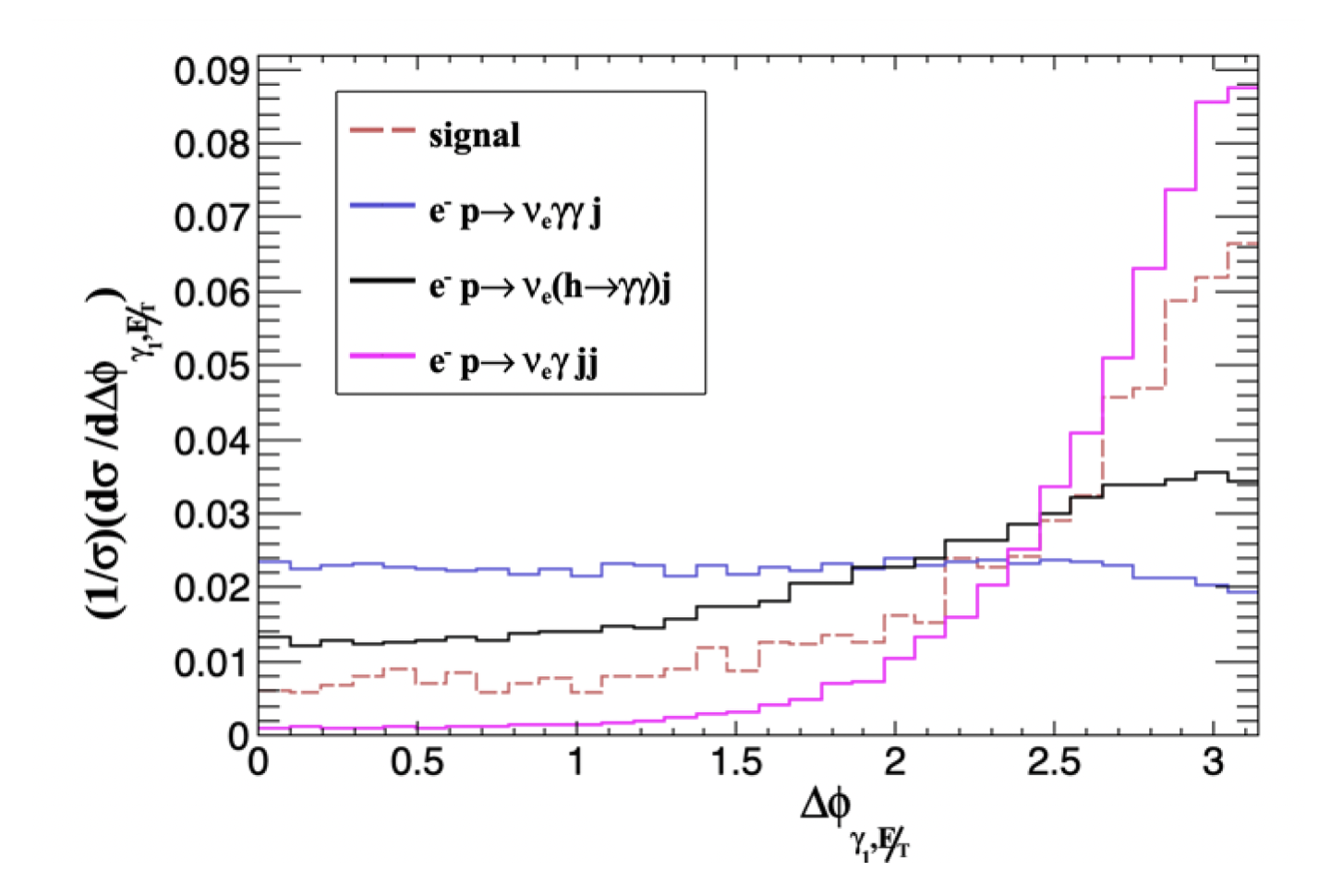}\qquad
\includegraphics[width=0.33\textwidth]{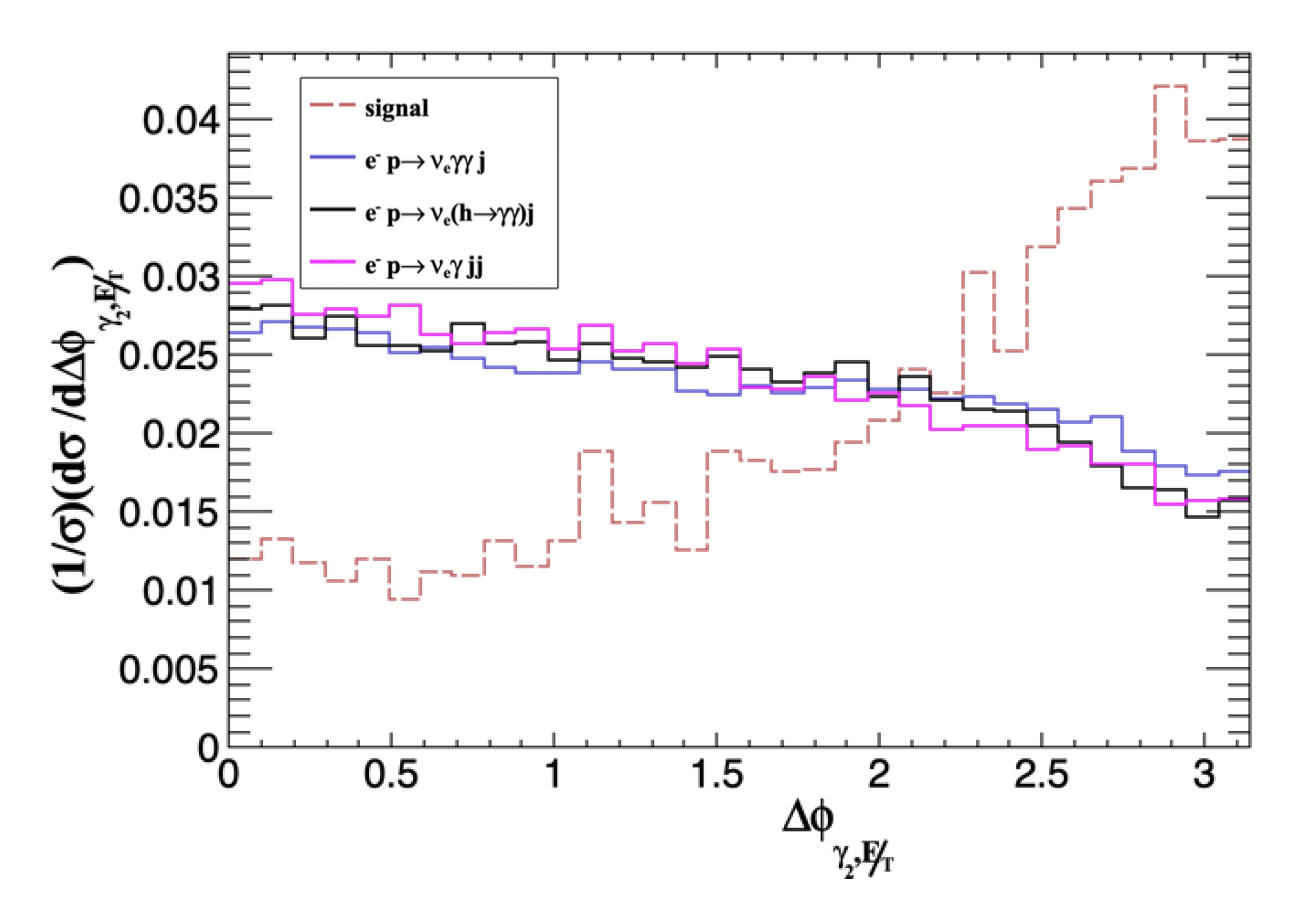}
\caption{Some signal and background kinematic distributions for the signature of two isolated photons, a backward jet plus ${\:/\!\!\!\! E_T}$ at LHeC for $m_a = 50$ GeV with $c^A_e / \Lambda = 1$ TeV$^{-1}$.}
\label{fig:ep_50}
\end{figure*} 

\begin{figure*}[ht!]
\includegraphics[width=0.35\textwidth]{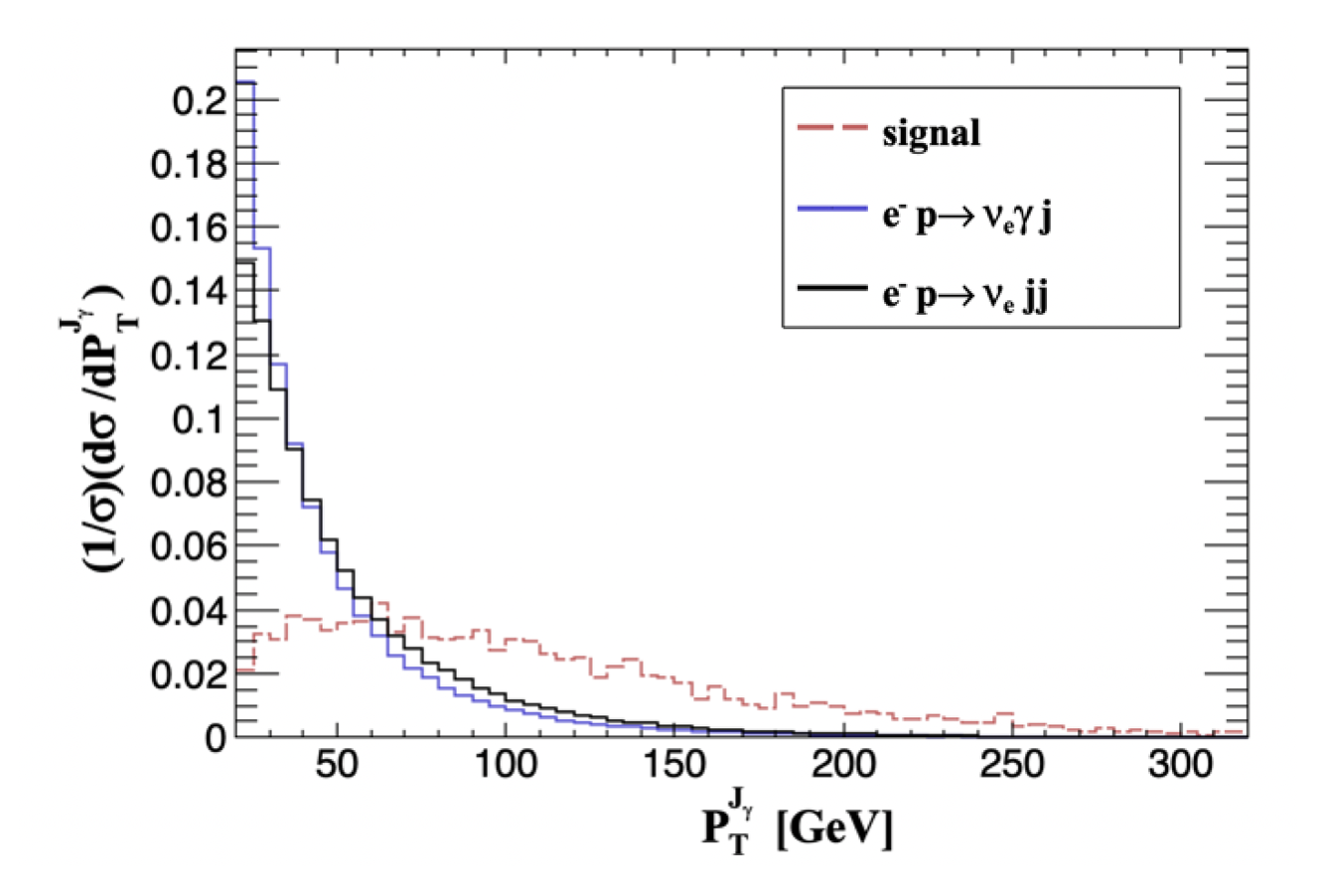}\qquad
\includegraphics[width=0.35\textwidth]{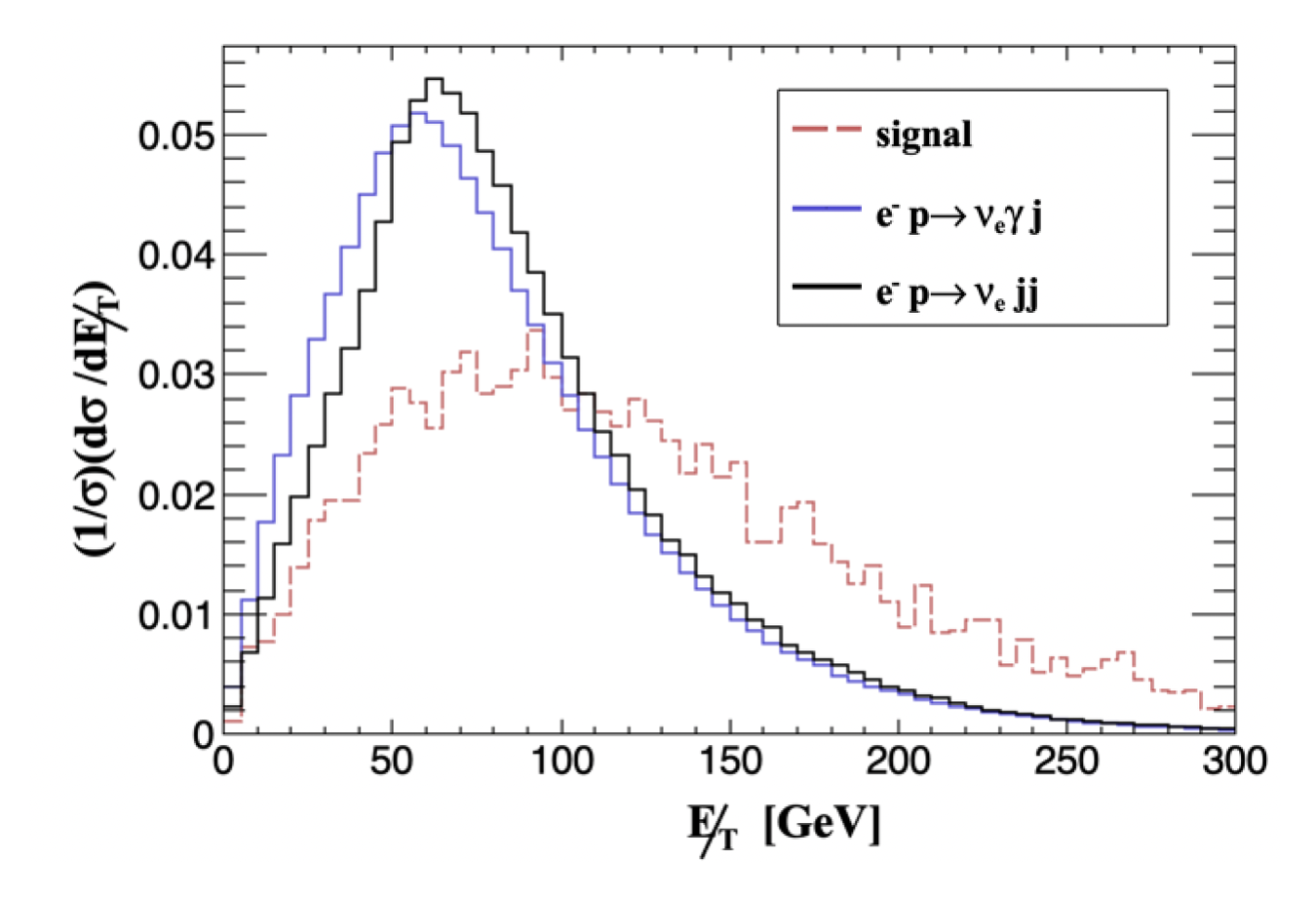}\qquad
\includegraphics[width=0.35\textwidth]{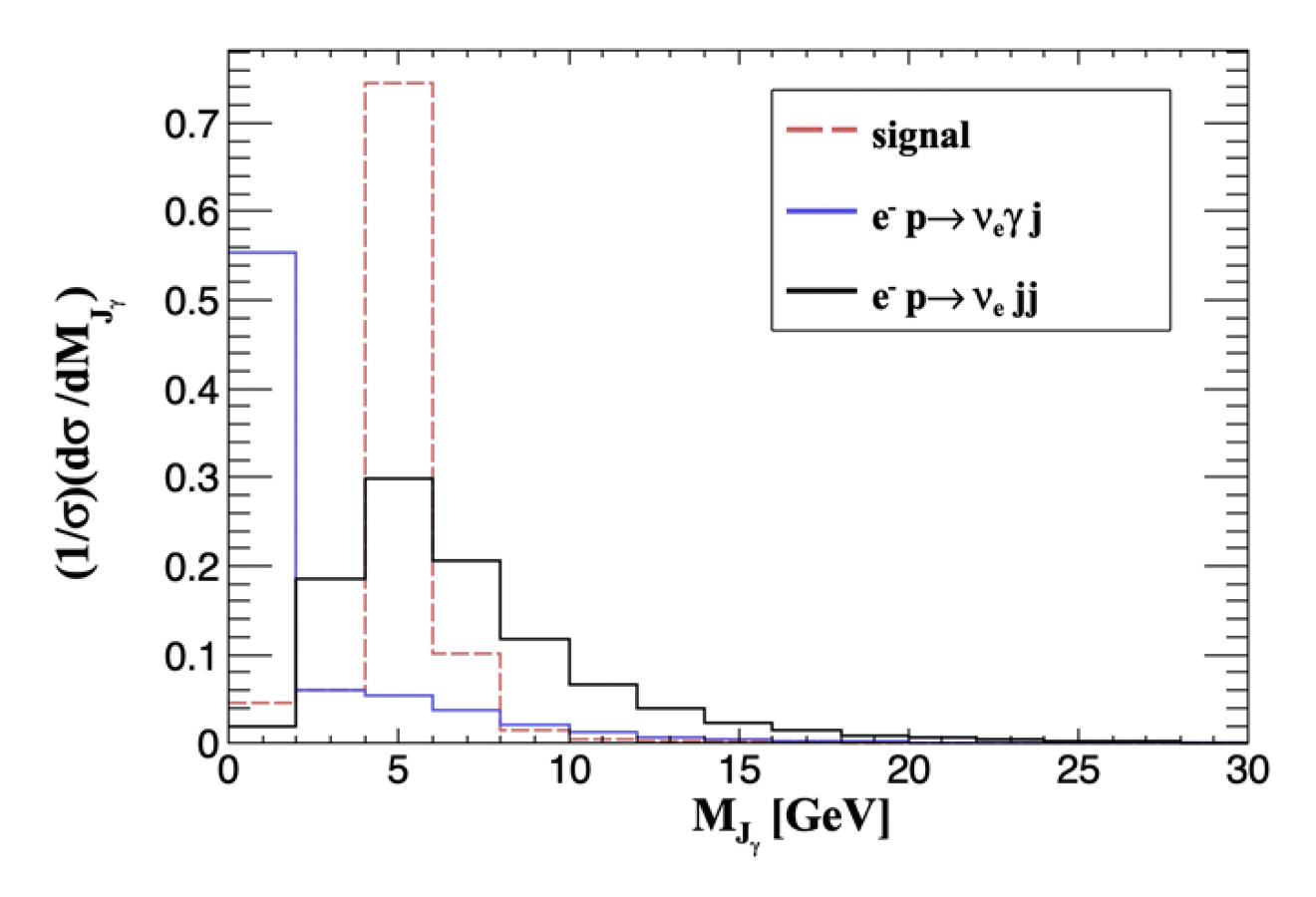}\qquad
\includegraphics[width=0.35\textwidth]{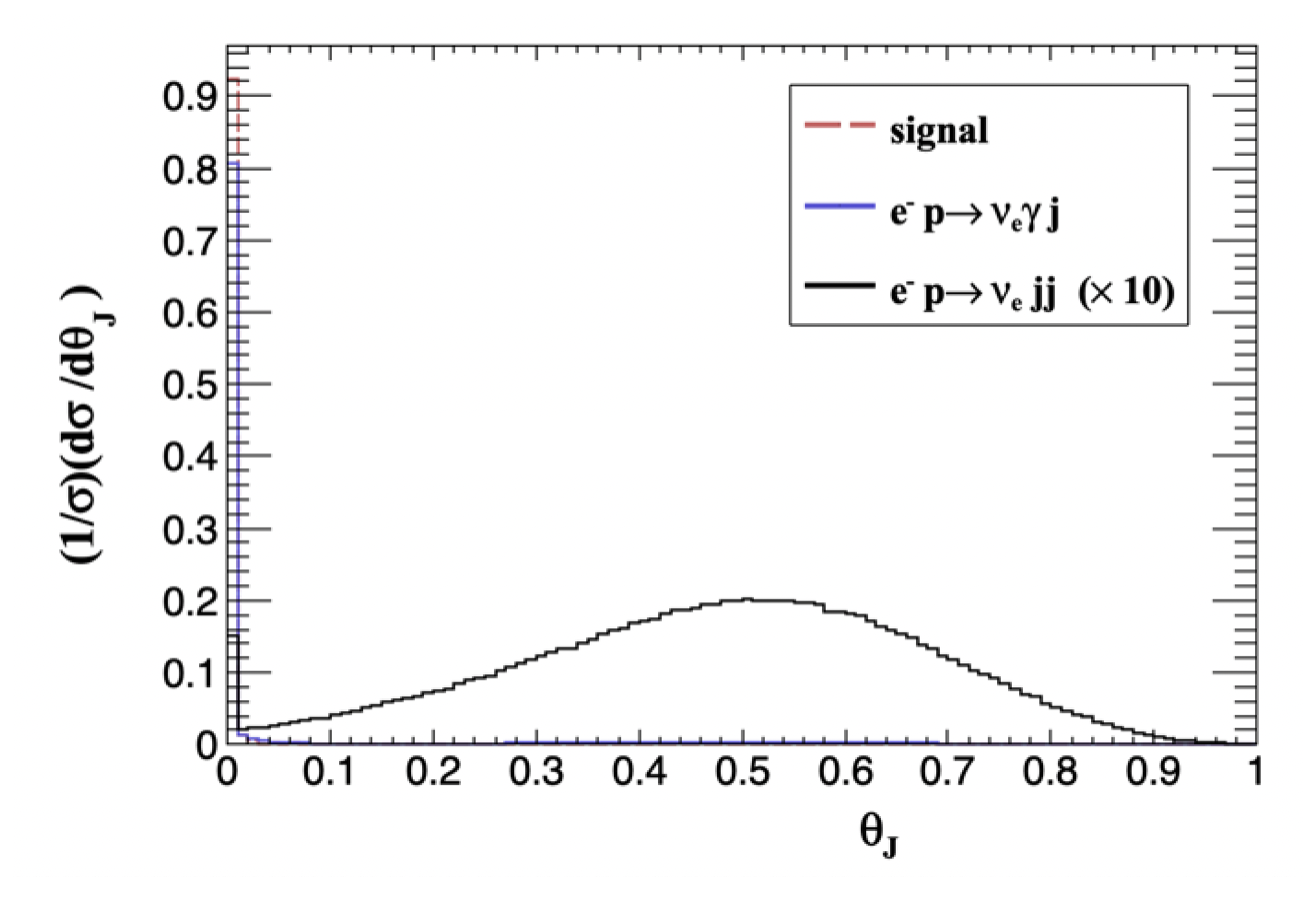}\qquad
\includegraphics[width=0.35\textwidth]{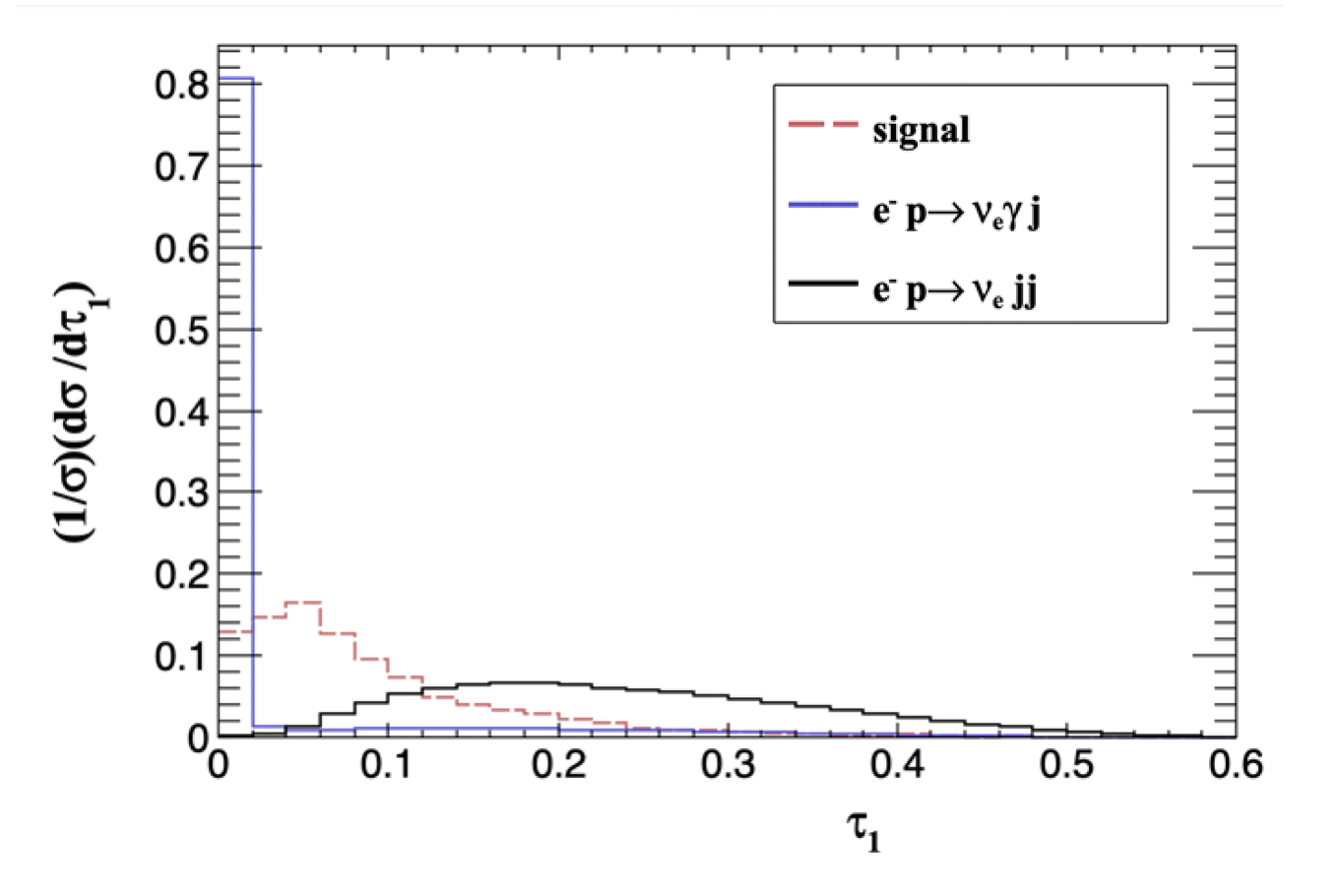}\qquad
\includegraphics[width=0.35\textwidth]{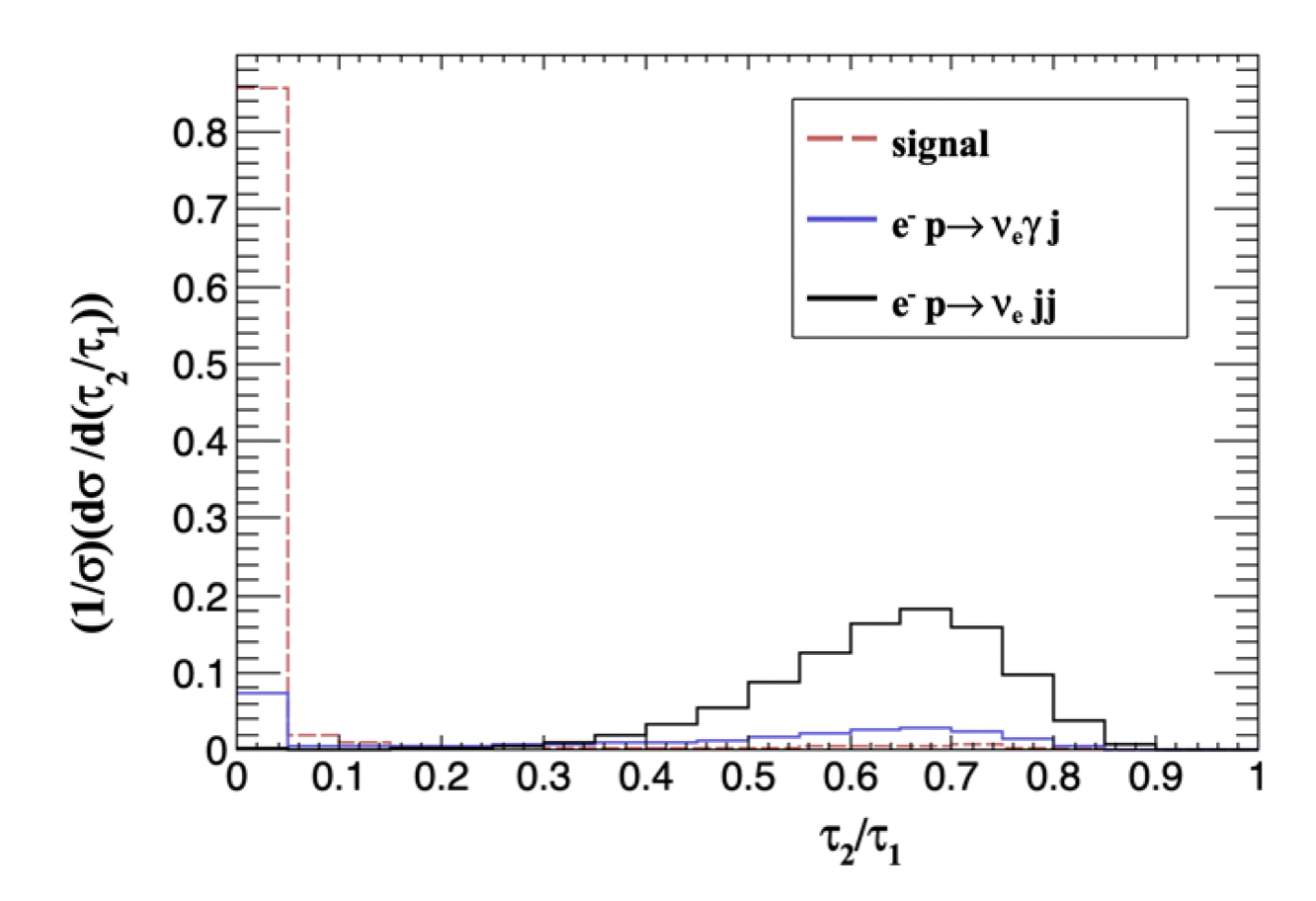}\qquad
\includegraphics[width=0.35\textwidth]{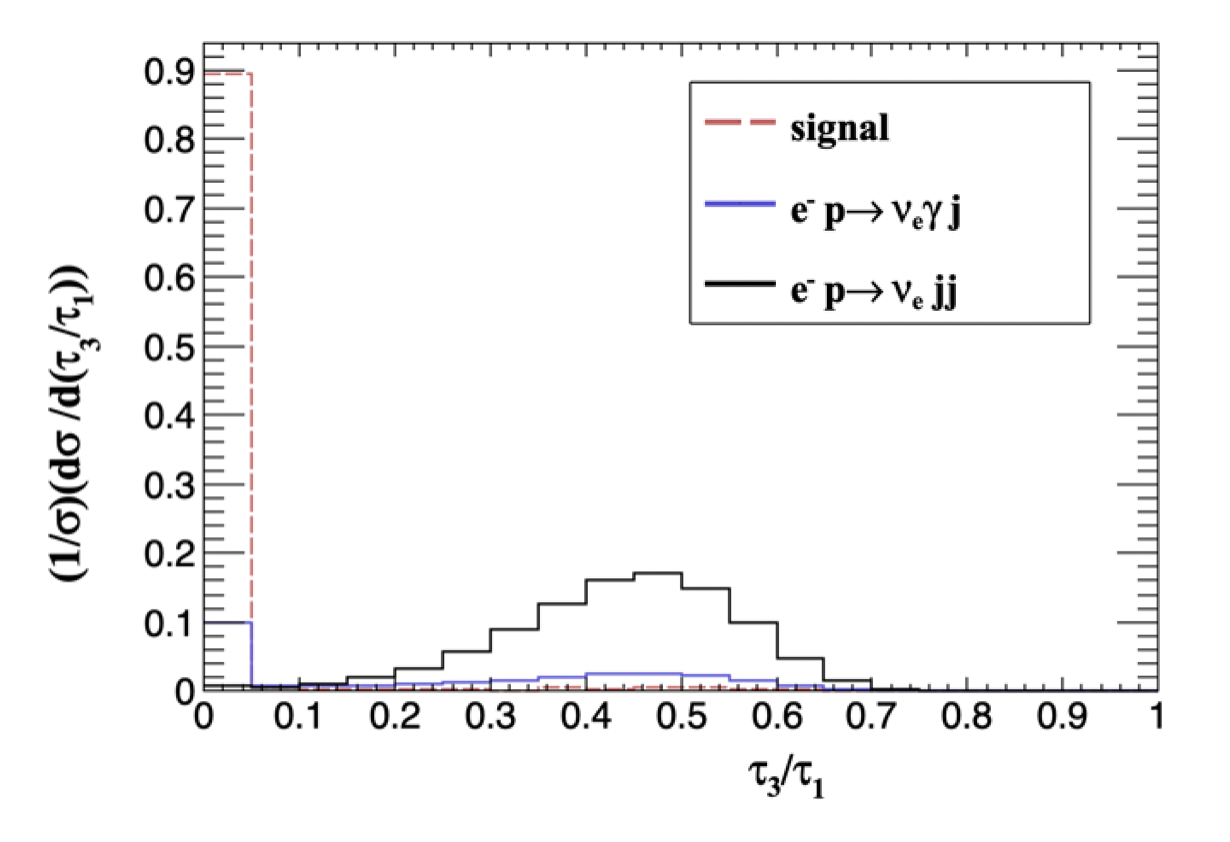}\qquad
\includegraphics[width=0.35\textwidth]{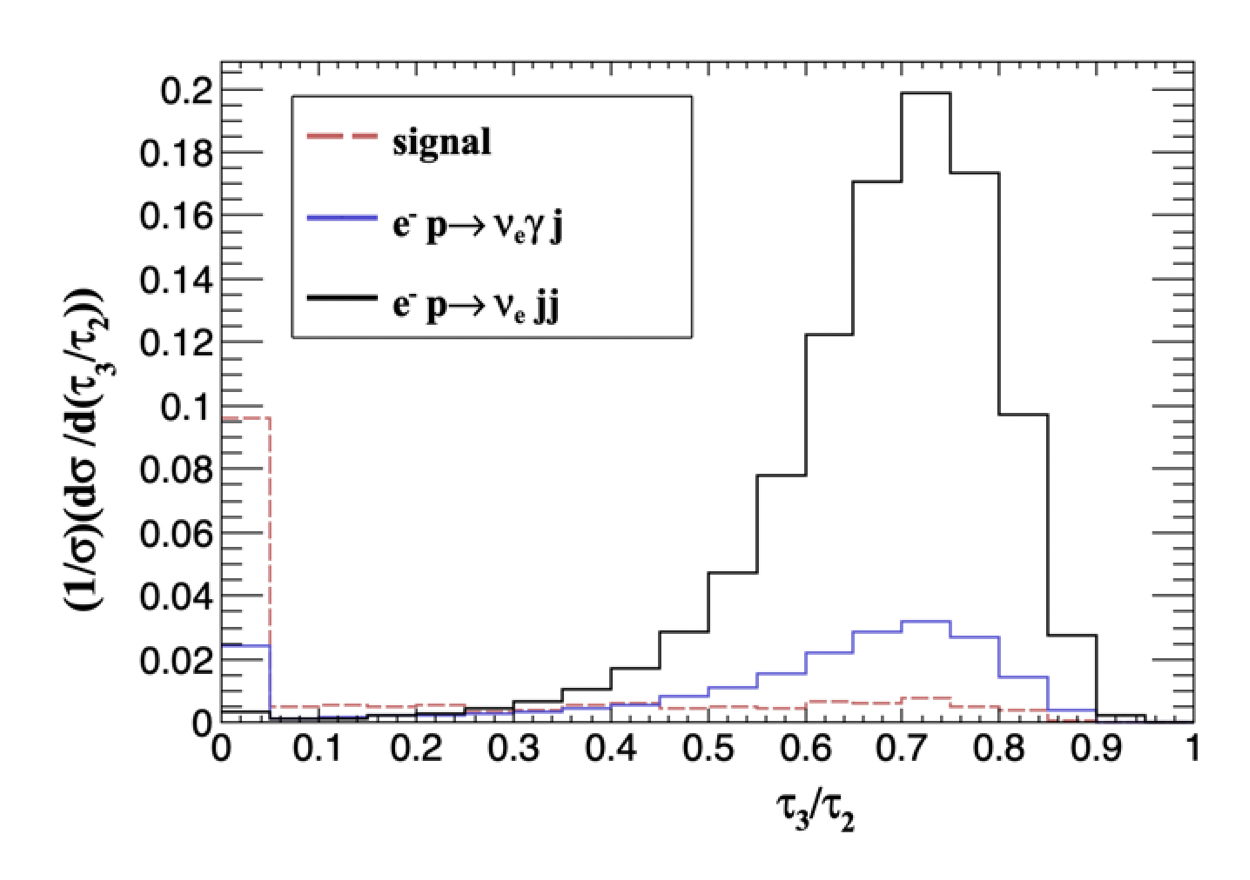}
\caption{Some signal and background kinematic distributions for the signature of a $J_{\gamma}$, a backward jet plus ${\:/\!\!\!\! E_T}$ at LHeC for $m_a = 5$ GeV with $c^A_e / \Lambda = 1$ TeV$^{-1}$.}
\label{fig:ep_5}
\end{figure*} 

In this Appendix, we choose some representative kinematic distributions mentioned in Sec.~\ref{sec:filter} for both signals and backgrounds at CEPC and LHeC in the following : 
\begin{itemize}
\item For the signature of two isolated photons plus ${\:/\!\!\!\! E}$ at CEPC, $E_{\gamma_{1,2}}$, $\eta_{\gamma_{1,2}}$, $M_{\:/\!\!\!\! E}$, $M_{\gamma_1\gamma_2}$, $\Delta\phi_{\gamma_2,{\:/\!\!\!\! E}}$ and ${\:/\!\!\!\! E}/M_{\gamma_1\gamma_2}$ distributions for $m_a = 50$ GeV with $c^A_e / \Lambda = 1$ TeV$^{-1}$ are shown in Fig.~\ref{fig:ee_50}. 
\item For the signature of a prompt $J_{\gamma}$ plus ${\:/\!\!\!\! E}$  at CEPC, $E_{J_{\gamma}}$, $\eta_{J_{\gamma}}$, $M_{\:/\!\!\!\! E}$, ${\:/\!\!\!\! E}/E_{J_{\gamma}}$, $M_{J_{\gamma}}$ and $\tau_1 (J_{\gamma})$ distributions for $m_a = 5$ GeV with $c^A_e / \Lambda = 1$ TeV$^{-1}$ are shown in Fig.~\ref{fig:ee_5}. 
\item For the signature of two isolated photons, a backward jet plus ${\:/\!\!\!\! E_T}$ at LHeC, $M_{\gamma_1\gamma_2}$, $({\:/\!\!\!\! E_T}+P^{j_1}_T)/M_{\gamma_1\gamma_2}$, $\Delta\phi_{\gamma_1,{\:/\!\!\!\! E_T}}$ and $\Delta\phi_{\gamma_2,{\:/\!\!\!\! E_T}}$ distributions for $m_a = 50$ GeV with $c^A_e / \Lambda = 1$ TeV$^{-1}$  are shown in Fig.~\ref{fig:ep_50}. 
\item For the signature of a prompt $J_{\gamma}$, a backward jet plus ${\:/\!\!\!\! E_T}$ at LHeC, $P^{J_{\gamma}}_T$, ${\:/\!\!\!\! E_T}$, $M_{J_{\gamma}}$, $\theta_J$, $\tau_1$, $\tau_2 / \tau_1$, $\tau_3 / \tau_1$, $\tau_3 / \tau_2$, distributions for $m_a = 5$ GeV with $c^A_e / \Lambda = 1$ TeV$^{-1}$ are shown in Fig.~\ref{fig:ep_5}.
\end{itemize}
Note that only physical quantities are shown in these figures. The $M_J$ distribution for the single photon inside $J_{\gamma}$ suffers from detector smearing effects and fluctuations which make parts of events with $M_J$ slightly less than zero. Besides, there are some events in signals and backgrounds with $\tau_1$, $\tau_2$, $\tau_3$ are equal or close to zero, which cause $\tau_2 / \tau_1$, $\tau_3 / \tau_1$ and $\tau_3 / \tau_2$ are ill-defined. Finally, in the $\theta_J$ distribution of Fig.~\ref{fig:ep_5}, we enlarge the event shape of $e^{-}p\rightarrow\nu_e jj$ by ten times to make it easily be visualized. 

\begin{table}[htp]
\begin{center}\begin{tabular}{|c|c|c|c|c|c|}\hline $m_a$ [GeV] & ${\:/\!\!\!\! E} $ [GeV] & $\Delta M_{\gamma_1\gamma_2}$ [GeV] & $\Delta\phi_{\gamma_1,{\:/\!\!\!\! E}}$ & $\Delta\phi_{\gamma_2,{\:/\!\!\!\! E}}$ & ${\:/\!\!\!\! E}/M_{\gamma_1\gamma_2}$ \\
\hline $10$ &  $> 125$  &  $< 1.0$  & $> 3.0$ &  $> 2.5$  &  $(13,21)$ \\ 
\hline $20$ &  $> 125$  &  $< 1.0$  &  $> 2.9$  &  $> 2.4$  &  $(6,9)$ \\ 
\hline $30$ &  same  &  $< 1.5$  &  $> 2.8$  &  $> 2.2$  &  $(4,6.2)$ \\ 
\hline $40$ &  same  &  $< 2.0$  &  $> 2.7$  &  same  &  $(3,4.5)$ \\ 
\hline $60$ &  same  &  same  &  same  &  same  &  $(2,2.6)$ \\ 
\hline $70$ &  $> 110$  &  same  &  $> 2.3$  &  $> 1.5$  &  $(1.6,2.2)$ \\ 
\hline $80$ &  $> 110$  &  same  &  $> 2.3$  &  $> 1.5$  &  $(1.4,1.8)$ \\ 
\hline \end{tabular} \caption{The changes of some event selections with various $m_a$ benchmark points for two isolated photons plus ${\:/\!\!\!\! E}$ at CEPC where $\Delta M_{\gamma_1\gamma_2}\equiv \lvert M_{\gamma_1\gamma_2}-m_a\rvert$ and "same" means the same event selection as the benchmark point $m_a = 50$ GeV in the main text.}
\label{tab:other}
\end{center}
\end{table}

On the other hand, we dynamically optimize some event selections for two isolated photons plus ${\:/\!\!\!\! E}$ with different $m_a$ at CEPC and these changes for some benchmark examples are listed in Table.~\ref{tab:other}. 
Note we also changed the setting of $e$ALP mass window selection, $\lvert M_{\gamma_1\gamma_2}-m_a\rvert < 2 (1.5)$ GeV for $m_a \lesssim 40 (20)$ GeV to optimize our analysis for two isolated photons, a backward jet plus ${\:/\!\!\!\! E_T}$ at LHeC. Similarly, we changed the settings of $e$ALP mass window selection, $\lvert M_{J_{\gamma}}-m_a\rvert < 1$ GeV, $0.02 < \tau_1 < 0.15$ and $\tau_3 / \tau_2 < 0.6$ for $m_a \lesssim 2$ GeV to optimize our analysis for a $J_{\gamma}$, a backward jet plus ${\:/\!\!\!\! E_T}$ at LHeC. 

\bibliographystyle{utphys}
\bibliography{eALPs}

\providecommand{\href}[2]{#2}\begingroup\raggedright\begin{thebibliography}{999}
\bibitem{Peccei:1977hh}
R.~D.~Peccei and H.~R.~Quinn,
Phys. Rev. Lett. \textbf{38}, 1440-1443 (1977)
doi:10.1103/PhysRevLett.38.1440

\bibitem{Weinberg:1977ma}
S.~Weinberg,
Phys. Rev. Lett. \textbf{40}, 223-226 (1978)
doi:10.1103/PhysRevLett.40.223

\bibitem{Wilczek:1977pj}
F.~Wilczek,
Phys. Rev. Lett. \textbf{40}, 279-282 (1978)
doi:10.1103/PhysRevLett.40.279

\bibitem{Preskill:1982cy}
J.~Preskill, M.~B.~Wise and F.~Wilczek,
Phys. Lett. B \textbf{120}, 127-132 (1983)
doi:10.1016/0370-2693(83)90637-8

\bibitem{Abbott:1982af}
L.~F.~Abbott and P.~Sikivie,
Phys. Lett. B \textbf{120}, 133-136 (1983)
doi:10.1016/0370-2693(83)90638-X

\bibitem{Dine:1982ah}
M.~Dine and W.~Fischler,
Phys. Lett. B \textbf{120}, 137-141 (1983)
doi:10.1016/0370-2693(83)90639-1

\bibitem{Kim:1979if}
J.~E.~Kim,
Phys. Rev. Lett. \textbf{43}, 103 (1979)
doi:10.1103/PhysRevLett.43.103

\bibitem{Bagger:1994hh}
J.~Bagger, E.~Poppitz and L.~Randall,
Nucl. Phys. B \textbf{426}, 3-18 (1994)
doi:10.1016/0550-3213(94)90123-6
[arXiv:hep-ph/9405345 [hep-ph]].

\bibitem{Svrcek:2006yi}
P.~Svrcek and E.~Witten,
JHEP \textbf{06}, 051 (2006)
doi:10.1088/1126-6708/2006/06/051
[arXiv:hep-th/0605206 [hep-th]].

\bibitem{Arvanitaki:2009fg}
A.~Arvanitaki, S.~Dimopoulos, S.~Dubovsky, N.~Kaloper and J.~March-Russell,
Phys. Rev. D \textbf{81}, 123530 (2010)
doi:10.1103/PhysRevD.81.123530
[arXiv:0905.4720 [hep-th]].

\bibitem{Cicoli:2012sz}
M.~Cicoli, M.~Goodsell and A.~Ringwald,
JHEP \textbf{10}, 146 (2012)
doi:10.1007/JHEP10(2012)146
[arXiv:1206.0819 [hep-th]].

\bibitem{Jeong:2018jqe}
K.~S.~Jeong, T.~H.~Jung and C.~S.~Shin,
Phys. Rev. D \textbf{101}, no.3, 035009 (2020)
doi:10.1103/PhysRevD.101.035009
[arXiv:1811.03294 [hep-ph]].

\bibitem{Im:2021xoy}
S.~H.~Im, K.~S.~Jeong and Y.~Lee,
Phys. Rev. D \textbf{105}, no.3, 035028 (2022)
doi:10.1103/PhysRevD.105.035028
[arXiv:2111.01327 [hep-ph]].

\bibitem{Graham:2015cka}
P.~W.~Graham, D.~E.~Kaplan and S.~Rajendran,
Phys. Rev. Lett. \textbf{115}, no.22, 221801 (2015)
doi:10.1103/PhysRevLett.115.221801
[arXiv:1504.07551 [hep-ph]].

\bibitem{Arias:2012az}
P.~Arias, D.~Cadamuro, M.~Goodsell, J.~Jaeckel, J.~Redondo and A.~Ringwald,
JCAP \textbf{06}, 013 (2012)
doi:10.1088/1475-7516/2012/06/013
[arXiv:1201.5902 [hep-ph]].

\bibitem{Brivio:2017ije}
I.~Brivio, M.~B.~Gavela, L.~Merlo, K.~Mimasu, J.~M.~No, R.~del Rey and V.~Sanz,
Eur. Phys. J. C \textbf{77}, no.8, 572 (2017)
doi:10.1140/epjc/s10052-017-5111-3
[arXiv:1701.05379 [hep-ph]].

\bibitem{Bauer:2017ris}
M.~Bauer, M.~Neubert and A.~Thamm,
JHEP \textbf{12}, 044 (2017)
doi:10.1007/JHEP12(2017)044
[arXiv:1708.00443 [hep-ph]].

\bibitem{Bauer:2018uxu}
M.~Bauer, M.~Heiles, M.~Neubert and A.~Thamm,
Eur. Phys. J. C \textbf{79}, no.1, 74 (2019)
doi:10.1140/epjc/s10052-019-6587-9
[arXiv:1808.10323 [hep-ph]].

\bibitem{Ebadi:2019gij}
J.~Ebadi, S.~Khatibi and M.~Mohammadi Najafabadi,
Phys. Rev. D \textbf{100}, no.1, 015016 (2019)
doi:10.1103/PhysRevD.100.015016
[arXiv:1901.03061 [hep-ph]].

\bibitem{Bauer:2020jbp}
M.~Bauer, M.~Neubert, S.~Renner, M.~Schnubel and A.~Thamm,
JHEP \textbf{04}, 063 (2021)
doi:10.1007/JHEP04(2021)063
[arXiv:2012.12272 [hep-ph]].

\bibitem{Bauer:2021mvw}
M.~Bauer, M.~Neubert, S.~Renner, M.~Schnubel and A.~Thamm,
JHEP \textbf{09}, 056 (2022)
doi:10.1007/JHEP09(2022)056
[arXiv:2110.10698 [hep-ph]].

\bibitem{Raffelt:1990yz}
G.~G.~Raffelt,
Phys. Rept. \textbf{198}, 1-113 (1990)
doi:10.1016/0370-1573(90)90054-6

\bibitem{Marsh:2015xka}
D.~J.~E.~Marsh,
Phys. Rept. \textbf{643}, 1-79 (2016)
doi:10.1016/j.physrep.2016.06.005
[arXiv:1510.07633 [astro-ph.CO]].

\bibitem{Bjorken:1988as}
J.~D.~Bjorken, S.~Ecklund, W.~R.~Nelson, A.~Abashian, C.~Church, B.~Lu, L.~W.~Mo, T.~A.~Nunamaker and P.~Rassmann,
Phys. Rev. D \textbf{38}, 3375 (1988)
doi:10.1103/PhysRevD.38.3375

\bibitem{Dobrich:2015jyk}
B.~D\"obrich, J.~Jaeckel, F.~Kahlhoefer, A.~Ringwald and K.~Schmidt-Hoberg,
JHEP \textbf{02}, 018 (2016)
doi:10.1007/JHEP02(2016)018
[arXiv:1512.03069 [hep-ph]].

\bibitem{Dobrich:2019dxc}
B.~D\"obrich, J.~Jaeckel and T.~Spadaro,
JHEP \textbf{05}, 213 (2019)
[erratum: JHEP \textbf{10}, 046 (2020)]
doi:10.1007/JHEP05(2019)213
[arXiv:1904.02091 [hep-ph]].

\bibitem{Izaguirre:2016dfi}
E.~Izaguirre, T.~Lin and B.~Shuve,
Phys. Rev. Lett. \textbf{118}, no.11, 111802 (2017)
doi:10.1103/PhysRevLett.118.111802
[arXiv:1611.09355 [hep-ph]].

\bibitem{Benson:2018vya}
S.~Benson and A.~Puig Navarro,
LHCb-PUB-2018-006.

\bibitem{Gori:2020xvq}
S.~Gori, G.~Perez and K.~Tobioka,
JHEP \textbf{08}, 110 (2020)
doi:10.1007/JHEP08(2020)110
[arXiv:2005.05170 [hep-ph]].

\bibitem{Altmannshofer:2022izm}
W.~Altmannshofer, J.~A.~Dror and S.~Gori,
Phys. Rev. Lett. \textbf{130}, no.24, 241801 (2023)
doi:10.1103/PhysRevLett.130.241801
[arXiv:2209.00665 [hep-ph]].

\bibitem{BaBar:2011kau}
J.~P.~Lees \textit{et al.} [BaBar],
Phys. Rev. Lett. \textbf{107}, 221803 (2011)
doi:10.1103/PhysRevLett.107.221803
[arXiv:1108.3549 [hep-ex]].

\bibitem{Belle-II:2020jti}
F.~Abudin\'en \textit{et al.} [Belle-II],
Phys. Rev. Lett. \textbf{125}, no.16, 161806 (2020)
doi:10.1103/PhysRevLett.125.161806
[arXiv:2007.13071 [hep-ex]].

\bibitem{OPAL:2002vhf}
G.~Abbiendi \textit{et al.} [OPAL],
Eur. Phys. J. C \textbf{26}, 331-344 (2003)
doi:10.1140/epjc/s2002-01074-5
[arXiv:hep-ex/0210016 [hep-ex]].

\bibitem{Mimasu:2014nea}
K.~Mimasu and V.~Sanz,
JHEP \textbf{06}, 173 (2015)
doi:10.1007/JHEP06(2015)173
[arXiv:1409.4792 [hep-ph]].

\bibitem{Jaeckel:2015jla}
J.~Jaeckel and M.~Spannowsky,
Phys. Lett. B \textbf{753}, 482-487 (2016)
doi:10.1016/j.physletb.2015.12.037
[arXiv:1509.00476 [hep-ph]].

\bibitem{ATLAS:2014jdv}
G.~Aad \textit{et al.} [ATLAS],
Phys. Rev. Lett. \textbf{113}, no.17, 171801 (2014)
doi:10.1103/PhysRevLett.113.171801
[arXiv:1407.6583 [hep-ex]].

\bibitem{ATLAS:2015rsn}
G.~Aad \textit{et al.} [ATLAS],
Eur. Phys. J. C \textbf{76}, no.4, 210 (2016)
doi:10.1140/epjc/s10052-016-4034-8
[arXiv:1509.05051 [hep-ex]].

\bibitem{Knapen:2016moh}
S.~Knapen, T.~Lin, H.~K.~Lou and T.~Melia,
Phys. Rev. Lett. \textbf{118}, no.17, 171801 (2017)
doi:10.1103/PhysRevLett.118.171801
[arXiv:1607.06083 [hep-ph]].

\bibitem{Kirpichnikov:2020lws}
D.~V.~Kirpichnikov, V.~E.~Lyubovitskij and A.~S.~Zhevlakov,
Particles \textbf{3}, no.4, 719-728 (2020)
doi:10.3390/particles3040047
[arXiv:2004.13656 [hep-ph]].

\bibitem{Han:2020dwo}
C.~Han, M.~L.~L\'opez-Ib\'a\~nez, A.~Melis, O.~Vives and J.~M.~Yang,
Phys. Rev. D \textbf{103}, no.3, 035028 (2021)
doi:10.1103/PhysRevD.103.035028
[arXiv:2007.08834 [hep-ph]].

\bibitem{Chang:2021myh}
C.~H.~V.~Chang, C.~R.~Chen, S.~Y.~Ho and S.~Y.~Tseng,
Phys. Rev. D \textbf{104}, no.1, 015030 (2021)
doi:10.1103/PhysRevD.104.015030
[arXiv:2102.05012 [hep-ph]].

\bibitem{Cheung:2021mol}
K.~Cheung, A.~Soffer, Z.~S.~Wang and Y.~H.~Wu,
JHEP \textbf{11}, 218 (2021)
doi:10.1007/JHEP11(2021)218
[arXiv:2108.11094 [hep-ph]].

\bibitem{Bertuzzo:2022fcm}
E.~Bertuzzo, A.~L.~Foguel, G.~M.~Salla and R.~Z.~Funchal,
Phys. Rev. Lett. \textbf{130}, no.17, 171801 (2023)
doi:10.1103/PhysRevLett.130.171801
[arXiv:2202.12317 [hep-ph]].

\bibitem{Cheung:2022umw}
K.~Cheung, J.~L.~Kuo, P.~Y.~Tseng and Z.~S.~Wang,
Phys. Rev. D \textbf{106}, no.9, 095029 (2022)
doi:10.1103/PhysRevD.106.095029
[arXiv:2208.05111 [hep-ph]].

\bibitem{Lucente:2022esm}
G.~Lucente, N.~Nath, F.~Capozzi, M.~Giannotti and A.~Mirizzi,
Phys. Rev. D \textbf{106}, no.12, 123007 (2022)
doi:10.1103/PhysRevD.106.123007
[arXiv:2209.11780 [hep-ph]].

\bibitem{Dobrescu:2000jt}
B.~A.~Dobrescu, G.~L.~Landsberg and K.~T.~Matchev,
Phys. Rev. D \textbf{63}, 075003 (2001)
doi:10.1103/PhysRevD.63.075003
[arXiv:hep-ph/0005308 [hep-ph]].

\bibitem{Toro:2012sv}
N.~Toro and I.~Yavin,
Phys. Rev. D \textbf{86}, 055005 (2012)
doi:10.1103/PhysRevD.86.055005
[arXiv:1202.6377 [hep-ph]].

\bibitem{Draper:2012xt}
P.~Draper and D.~McKeen,
Phys. Rev. D \textbf{85}, 115023 (2012)
doi:10.1103/PhysRevD.85.115023
[arXiv:1204.1061 [hep-ph]].

\bibitem{ATLAS:2012soa}
 [ATLAS],
ATLAS-CONF-2012-079.

\bibitem{Ellis:2012sd}
S.~D.~Ellis, T.~S.~Roy and J.~Scholtz,
Phys. Rev. Lett. \textbf{110}, no.12, 122003 (2013)
doi:10.1103/PhysRevLett.110.122003
[arXiv:1210.1855 [hep-ph]].

\bibitem{Ellis:2012zp}
S.~D.~Ellis, T.~S.~Roy and J.~Scholtz,
Phys. Rev. D \textbf{87}, no.1, 014015 (2013)
doi:10.1103/PhysRevD.87.014015
[arXiv:1210.3657 [hep-ph]].

\bibitem{Knapen:2015dap}
S.~Knapen, T.~Melia, M.~Papucci and K.~Zurek,
Phys. Rev. D \textbf{93}, no.7, 075020 (2016)
doi:10.1103/PhysRevD.93.075020
[arXiv:1512.04928 [hep-ph]].

\bibitem{Agrawal:2015dbf}
P.~Agrawal, J.~Fan, B.~Heidenreich, M.~Reece and M.~Strassler,
JHEP \textbf{06}, 082 (2016)
doi:10.1007/JHEP06(2016)082
[arXiv:1512.05775 [hep-ph]].

\bibitem{Chang:2015sdy}
J.~Chang, K.~Cheung and C.~T.~Lu,
Phys. Rev. D \textbf{93}, no.7, 075013 (2016)
doi:10.1103/PhysRevD.93.075013
[arXiv:1512.06671 [hep-ph]].

\bibitem{Aparicio:2016iwr}
L.~Aparicio, A.~Azatov, E.~Hardy and A.~Romanino,
JHEP \textbf{05}, 077 (2016)
doi:10.1007/JHEP05(2016)077
[arXiv:1602.00949 [hep-ph]].

\bibitem{Dasgupta:2016wxw}
B.~Dasgupta, J.~Kopp and P.~Schwaller,
Eur. Phys. J. C \textbf{76}, no.5, 277 (2016)
doi:10.1140/epjc/s10052-016-4127-4
[arXiv:1602.04692 [hep-ph]].

\bibitem{Domingo:2016unq}
F.~Domingo, S.~Heinemeyer, J.~S.~Kim and K.~Rolbiecki,
Eur. Phys. J. C \textbf{76}, no.5, 249 (2016)
doi:10.1140/epjc/s10052-016-4080-2
[arXiv:1602.07691 [hep-ph]].

\bibitem{Chiang:2016eav}
C.~W.~Chiang, H.~Fukuda, M.~Ibe and T.~T.~Yanagida,
Phys. Rev. D \textbf{93}, no.9, 095016 (2016)
doi:10.1103/PhysRevD.93.095016
[arXiv:1602.07909 [hep-ph]].

\bibitem{Dillon:2016tqp}
B.~M.~Dillon, C.~Han, H.~M.~Lee and M.~Park,
Int. J. Mod. Phys. A \textbf{32}, no.33, 1745006 (2017)
doi:10.1142/S0217751X17450063
[arXiv:1606.07171 [hep-ph]].

\bibitem{Allanach:2017qbs}
B.~C.~Allanach, D.~Bhatia and A.~M.~Iyer,
Eur. Phys. J. C \textbf{77}, no.9, 595 (2017)
doi:10.1140/epjc/s10052-017-5162-5
[arXiv:1706.09039 [hep-ph]].

\bibitem{Chakraborty:2017mbz}
A.~Chakraborty, A.~M.~Iyer and T.~S.~Roy,
Nucl. Phys. B \textbf{932}, 439-470 (2018)
doi:10.1016/j.nuclphysb.2018.05.019
[arXiv:1707.07084 [hep-ph]].

\bibitem{ATLAS:2018dfo}
M.~Aaboud \textit{et al.} [ATLAS],
Phys. Rev. D \textbf{99}, no.1, 012008 (2019)
doi:10.1103/PhysRevD.99.012008
[arXiv:1808.10515 [hep-ex]].

\bibitem{Sheff:2020jyw}
B.~Sheff, N.~Steinberg and J.~D.~Wells,
Phys. Rev. D \textbf{104}, no.3, 036009 (2021)
doi:10.1103/PhysRevD.104.036009
[arXiv:2008.10568 [hep-ph]].

\bibitem{Wang:2021uyb}
D.~Wang, L.~Wu, J.~M.~Yang and M.~Zhang,
Phys. Rev. D \textbf{104}, no.9, 095016 (2021)
doi:10.1103/PhysRevD.104.095016
[arXiv:2102.01532 [hep-ph]].

\bibitem{Ren:2021prq}
J.~Ren, D.~Wang, L.~Wu, J.~M.~Yang and M.~Zhang,
JHEP \textbf{11}, 138 (2021)
doi:10.1007/JHEP11(2021)138
[arXiv:2106.07018 [hep-ph]].

\bibitem{Ai:2022qvs}
X.~Ai, S.~C.~Hsu, K.~Li and C.~T.~Lu,
J. Phys. Conf. Ser. \textbf{2438}, no.1, 012114 (2023)
doi:10.1088/1742-6596/2438/1/012114
[arXiv:2203.16703 [hep-ph]].

\bibitem{CMS:2022fyt}
 [CMS],
[arXiv:2209.06197 [hep-ex]].

\bibitem{BDX:2016akw}
M.~Battaglieri \textit{et al.} [BDX],
[arXiv:1607.01390 [hep-ex]].

\bibitem{Liu:2021lan}
Y.~Liu and B.~Yan,
Chin. Phys. C \textbf{47}, no.4, 043113 (2023)
doi:10.1088/1674-1137/acbbc0
[arXiv:2112.02477 [hep-ph]].

\bibitem{Raffelt:1987yt}
G.~Raffelt and D.~Seckel,
Phys. Rev. Lett. \textbf{60}, 1793 (1988)
doi:10.1103/PhysRevLett.60.1793

\bibitem{Alloul:2013bka}
A.~Alloul, N.~D.~Christensen, C.~Degrande, C.~Duhr and B.~Fuks,
Comput. Phys. Commun. \textbf{185}, 2250-2300 (2014)
doi:10.1016/j.cpc.2014.04.012
[arXiv:1310.1921 [hep-ph]].

\bibitem{Alwall:2014hca}
J.~Alwall, R.~Frederix, S.~Frixione, V.~Hirschi, F.~Maltoni, O.~Mattelaer, H.~S.~Shao, T.~Stelzer, P.~Torrielli and M.~Zaro,
JHEP \textbf{07}, 079 (2014)
doi:10.1007/JHEP07(2014)079
[arXiv:1405.0301 [hep-ph]].

\bibitem{CEPCStudyGroup:2018ghi}
J.~B.~Guimar\~aes da Costa \textit{et al.} [CEPC Study Group],
[arXiv:1811.10545 [hep-ex]].

\bibitem{TLEPDesignStudyWorkingGroup:2013myl}
M.~Bicer \textit{et al.} [TLEP Design Study Working Group],
JHEP \textbf{01}, 164 (2014)
doi:10.1007/JHEP01(2014)164
[arXiv:1308.6176 [hep-ex]].

\bibitem{Baer:2013cma}
H.~Baer \textit{et al.} [ILC],
[arXiv:1306.6352 [hep-ph]].

\bibitem{CLICDetector:2013tfe}
H.~Abramowicz \textit{et al.} [CLIC Detector and Physics Study],
[arXiv:1307.5288 [hep-ex]].

\bibitem{Sjostrand:2007gs}
T.~Sjostrand, S.~Mrenna and P.~Z.~Skands,
Comput. Phys. Commun. \textbf{178}, 852-867 (2008)
doi:10.1016/j.cpc.2008.01.036
[arXiv:0710.3820 [hep-ph]].

\bibitem{deFavereau:2013fsa}
J.~de Favereau \textit{et al.} [DELPHES 3],
JHEP \textbf{02}, 057 (2014)
doi:10.1007/JHEP02(2014)057
[arXiv:1307.6346 [hep-ex]].

\bibitem{Cobal:2020hmk}
M.~Cobal, C.~De Dominicis, M.~Fabbrichesi, E.~Gabrielli, J.~Magro, B.~Mele and G.~Panizzo,
Phys. Rev. D \textbf{102}, no.3, 035027 (2020)
doi:10.1103/PhysRevD.102.035027
[arXiv:2006.15945 [hep-ph]].

\bibitem{Ahmed:2022ude}
W.~Ahmed, I.~Khan, T.~Li, S.~Raza and W.~Zhang,
Phys. Lett. B \textbf{832}, 137216 (2022)
doi:10.1016/j.physletb.2022.137216
[arXiv:2202.11011 [hep-ph]].

\bibitem{OPAL:1997zll}
K.~Ackerstaff \textit{et al.} [OPAL],
Eur. Phys. J. C \textbf{1}, 31-43 (1998)
doi:10.1007/s100520050060
[arXiv:hep-ex/9709022 [hep-ex]].

\bibitem{OPAL:1998jka}
K.~Ackerstaff \textit{et al.} [OPAL],
Phys. Lett. B \textbf{437}, 218-230 (1998)
doi:10.1016/S0370-2693(98)01095-8
[arXiv:hep-ex/9808014 [hep-ex]].

\bibitem{OPAL:1999tmg}
G.~Abbiendi \textit{et al.} [OPAL],
Phys. Lett. B \textbf{464}, 311-322 (1999)
doi:10.1016/S0370-2693(99)00861-8
[arXiv:hep-ex/9907060 [hep-ex]].

\bibitem{Cowan:2010js}
G.~Cowan, K.~Cranmer, E.~Gross and O.~Vitells,
Eur. Phys. J. C \textbf{71}, 1554 (2011)
[erratum: Eur. Phys. J. C \textbf{73}, 2501 (2013)]
doi:10.1140/epjc/s10052-011-1554-0
[arXiv:1007.1727 [physics.data-an]].

\bibitem{Dokshitzer:1997in}
Y.~L.~Dokshitzer, G.~D.~Leder, S.~Moretti and B.~R.~Webber,
JHEP \textbf{08}, 001 (1997)
doi:10.1088/1126-6708/1997/08/001
[arXiv:hep-ph/9707323 [hep-ph]].

\bibitem{Wobisch:1998wt}
M.~Wobisch and T.~Wengler,
[arXiv:hep-ph/9907280 [hep-ph]].

\bibitem{Zhang:2021orr}
M.~Zhang,
Phys. Rev. D \textbf{104}, no.5, 055008 (2021)
doi:10.1103/PhysRevD.104.055008
[arXiv:2104.06988 [hep-ph]].

\bibitem{Thaler:2010tr}
J.~Thaler and K.~Van Tilburg,
JHEP \textbf{03}, 015 (2011)
doi:10.1007/JHEP03(2011)015
[arXiv:1011.2268 [hep-ph]].

\bibitem{Thaler:2011gf}
J.~Thaler and K.~Van Tilburg,
JHEP \textbf{02}, 093 (2012)
doi:10.1007/JHEP02(2012)093
[arXiv:1108.2701 [hep-ph]].

\bibitem{LHeC:2020van}
P.~Agostini \textit{et al.} [LHeC and FCC-he Study Group],
J. Phys. G \textbf{48}, no.11, 110501 (2021)
doi:10.1088/1361-6471/abf3ba
[arXiv:2007.14491 [hep-ex]].

\bibitem{FCC:2018byv}
A.~Abada \textit{et al.} [FCC],
Eur. Phys. J. C \textbf{79}, no.6, 474 (2019)
doi:10.1140/epjc/s10052-019-6904-3

\bibitem{Hesari:2018ssq}
H.~Hesari, H.~Khanpour and M.~Mohammadi Najafabadi,
Phys. Rev. D \textbf{97}, no.9, 095041 (2018)
doi:10.1103/PhysRevD.97.095041
[arXiv:1805.04697 [hep-ph]].

\bibitem{Antusch:2019eiz}
S.~Antusch, O.~Fischer and A.~Hammad,
JHEP \textbf{03}, 110 (2020)
doi:10.1007/JHEP03(2020)110
[arXiv:1908.02852 [hep-ph]].

\bibitem{Andre:2022xeh}
K.~D.~J.~Andr\'e, L.~Aperio Bella, N.~Armesto, S.~A.~Bogacz, D.~Britzger, O.~S.~Br\"uning, M.~D'Onofrio, E.~G.~Ferreiro, O.~Fischer and C.~Gwenlan, \textit{et al.}
Eur. Phys. J. C \textbf{82}, no.1, 40 (2022)
doi:10.1140/epjc/s10052-021-09967-z
[arXiv:2201.02436 [hep-ex]].

\bibitem{ATLAS:2017muo}
 [ATLAS],
ATL-PHYS-PUB-2017-001.

\bibitem{Gutierrez-Rodriguez:2020gsi}
A.~Guti\'errez-Rodr\'\i{}guez, M.~A.~Hern\'andez-Ru\'\i{}z, E.~Gurkanli, V.~Ari and M.~K\"oksal,
Eur. Phys. J. C \textbf{81}, no.3, 210 (2021)
doi:10.1140/epjc/s10052-021-08991-3
[arXiv:2005.11509 [hep-ph]].

\bibitem{LHeCStudyGroup:2012zhm}
J.~L.~Abelleira Fernandez \textit{et al.} [LHeC Study Group],
J. Phys. G \textbf{39}, 075001 (2012)
doi:10.1088/0954-3899/39/7/075001
[arXiv:1206.2913 [physics.acc-ph]].

\bibitem{dEnterria:2021ljz}
D.~d'Enterria,
[arXiv:2102.08971 [hep-ex]].

\bibitem{CidVidal:2018blh}
X.~Cid Vidal, A.~Mariotti, D.~Redigolo, F.~Sala and K.~Tobioka,
JHEP \textbf{01}, 113 (2019)
[erratum: JHEP \textbf{06}, 141 (2020)]
doi:10.1007/JHEP01(2019)113
[arXiv:1810.09452 [hep-ph]].

\bibitem{BaBar:2021ich}
J.~P.~Lees \textit{et al.} [BaBar],
Phys. Rev. Lett. \textbf{128}, no.13, 131802 (2022)
doi:10.1103/PhysRevLett.128.131802
[arXiv:2111.01800 [hep-ex]].

\bibitem{KTeV:2003sls}
A.~Alavi-Harati \textit{et al.} [KTeV],
Phys. Rev. Lett. \textbf{93}, 021805 (2004)
doi:10.1103/PhysRevLett.93.021805
[arXiv:hep-ex/0309072 [hep-ex]].

\bibitem{LHCb:2015ycz}
R.~Aaij \textit{et al.} [LHCb],
JHEP \textbf{04}, 064 (2015)
doi:10.1007/JHEP04(2015)064
[arXiv:1501.03038 [hep-ex]].

\bibitem{CHARM:1985anb}
F.~Bergsma \textit{et al.} [CHARM],
Phys. Lett. B \textbf{157}, 458-462 (1985)
doi:10.1016/0370-2693(85)90400-9

\bibitem{Racco:2015dxa}
D.~Racco, A.~Wulzer and F.~Zwirner,
JHEP \textbf{05}, 009 (2015)
doi:10.1007/JHEP05(2015)009
[arXiv:1502.04701 [hep-ph]].

\bibitem{DeSimone:2016fbz}
A.~De Simone and T.~Jacques,
Eur. Phys. J. C \textbf{76}, no.7, 367 (2016)
doi:10.1140/epjc/s10052-016-4208-4
[arXiv:1603.08002 [hep-ph]].

\bibitem{Roe:2004na}
B.~P.~Roe, H.~J.~Yang, J.~Zhu, Y.~Liu, I.~Stancu and G.~McGregor,
Nucl. Instrum. Meth. A \textbf{543}, no.2-3, 577-584 (2005)
doi:10.1016/j.nima.2004.12.018
[arXiv:physics/0408124 [physics]].

\bibitem{Ayyar:2020ijy}
V.~Ayyar, W.~Bhimji, L.~Gerhardt, S.~Robertson and Z.~Ronaghi,
EPJ Web Conf. \textbf{245}, 06003 (2020)
doi:10.1051/epjconf/202024506003
[arXiv:2002.05761 [hep-ex]].

\bibitem{Lu:2023ryd}
C.~T.~Lu, X.~Luo and X.~Wei,
Chin. Phys. C \textbf{47}, no.10, 103102 (2023)
doi:10.1088/1674-1137/ace424
[arXiv:2303.03110 [hep-ph]].
\end{thebibliography}\endgroup

\end{document}